\documentclass[onecolumn,amsmath,amssymb,nofootinbib,12pt]{article}
\usepackage{jheppub}
\usepackage{ifpdf}
\usepackage{graphicx}
\usepackage{amsfonts}
\usepackage{amsmath}
\usepackage{amssymb}
\usepackage{epsfig}
\usepackage{pdfpages}
\usepackage{graphicx,epstopdf}

\usepackage{caption}
\usepackage[position=b]{subcaption}

\usepackage[makeroom]{cancel}
\usepackage{hyperref}
\hypersetup{pdftex,colorlinks=true,allcolors=blue}
\usepackage{array}

\newcommand{\RN}[1]{%
  \textup{\uppercase\expandafter{\romannumeral#1}}%
}
\newcommand{\del}{\partial}
\newcommand{\vac}{\mt{vac}} 

\newcommand{\BH}{\mt{BH}}
\newcommand{\BHx}{\mt{BH,sing}}
\newcommand{\bulk}{\text{bulk}}
\newcommand{\jnt}{\text{jnt}}

\newcommand{\be}{\begin{equation}}
\newcommand{\ee}{\end{equation}}
\newcommand{\bea}{\begin{eqnarray}}
\newcommand{\eea}{\end{eqnarray}}
\newcommand{\beq}{\begin{equation}}
\newcommand{\eeq}{\end{equation}}
\newcommand{\beqa}{\begin{eqnarray}}
\newcommand{\eeqa}{\end{eqnarray}}
\newcommand{\beqar}{\begin{eqnarray*}}
\newcommand{\eeqar}{\end{eqnarray*}}

\newcommand{\labell}[1]{\label{#1}} 

\newcommand{\eg}{{\it e.g.,}\ }
\newcommand{\ie}{{\it i.e.,}\ }
\newcommand{\reef}[1]{(\ref{#1})}

\newcommand{\mt}[1]{\textrm{\tiny #1}}

\newcommand{\mC}{\mathcal{C}}
\newcommand{\mV}{\mathcal{V}}

\newcommand{\Gn}{G_\mt{N}}
\newcommand{\eps}{\epsilon}
\newcommand{\tk}{{\tilde k}}

\begin{document}

\title{Complexity of Formation in Holography}

\author[a]{Shira Chapman,}
\author[a, b]{Hugo Marrochio}
\author[a]{and Robert C. Myers}
\affiliation[a]{Perimeter Institute for Theoretical Physics, Waterloo, ON N2L 2Y5, Canada}
\affiliation[b]{Department of Physics $\&$ Astronomy and Guelph-Waterloo Physics Institute,\\
University of Waterloo, Waterloo, ON N2L 3G1, Canada}

\emailAdd{schapman@perimeterinstitute.ca}
\emailAdd{hmarrochio@perimeterinstitute.ca}
\emailAdd{rmyers@perimeterinstitute.ca}

\date{\today}

\abstract{It was recently conjectured that the quantum complexity of a holographic boundary state can be computed by evaluating the gravitational action on a bulk region known as the Wheeler-DeWitt patch. We apply this complexity=action duality to evaluate the `complexity of formation' \cite{Brown1,Brown2}, \ie the additional complexity arising in preparing the entangled thermofield double state with two copies of the boundary CFT compared to preparing the individual vacuum states of the two copies.  We find that for boundary dimensions $d>2$, the difference in the complexities grows linearly with the thermal entropy at high temperatures. For the special case $d=2$, the complexity of formation is a fixed constant, independent of the temperature. We compare these results to those found using the complexity=volume duality.}

\maketitle

\section{Introduction}

In recent years, it has become widely appreciated that quantum information theory is a fruitful lens with which to examine the conundrums of quantum gravity. While most of the ongoing research has focused on holographic entanglement entropy \cite{rt1,rt2}, `quantum complexity' (\eg see \cite{johnw,tobias,four}) is another concept from quantum information theory that has recently found a place in this discussion. These ideas emerged from studies aimed at understanding the growth of the Einstein-Rosen bridge for AdS black holes in terms of quantum complexity in the dual boundary CFT \cite{suss1,CompVolume,suss2,suss3}.

Loosely speaking, the complexity $\mC$ of a particular state $|\psi\rangle$ is the minimum number of quantum gates required to produce this state from a particular reference state $|\psi_0 \rangle$.\footnote{See \cite{johnw,suss2} for further details. A more refined definition is still required for application to continuum quantum field theories \cite{Brown2}, perhaps using the geometric perspective of \cite{miken1,miken2,miken3}  --- see also \cite{prep9}.} Now in the context of the AdS/CFT correspondence, two proposals have been made to evaluate the complexity of a boundary state: The first is that the complexity should be dual to the volume of the extremal codimension-one bulk hypersurface which meets the asymptotic boundary on the time slice where the boundary state is defined \cite{CompVolume} --- see section \ref{sec:CompVol}.  The second conjecture states \cite{Brown1,Brown2}
\begin{equation}\label{Complexity}
\mathcal{C}=\frac{I}{\pi \,\hbar}\,,
\end{equation}
where $I$ is the gravitational action evaluated on a particular spacetime region  in the bulk, known as the `Wheeler-DeWitt (WDW) patch.' In particular, the WDW patch is the region enclosed by past and future light sheets sent into the bulk spacetime from the time slice on the boundary, \eg see figure \ref{PenroseBHa}.

Both of these holographic conjectures satisfy a number of properties expected of complexity, \eg they continue to grow (linearly with time) after the boundary theory reaches thermal equilibrium. However, the second conjecture has certain advantages. In particular, the complexity=volume duality requires introducing an additional length scale in relating the bulk geometric quantity to the complexity --- see eq.~\reef{volver}. However, the complexity=action duality faced the obstacle that when the conjecture was originally proposed, there was no rigorous method for evaluating the gravitational action on spacetime regions with null boundaries. However, this problem was recently overcome with a careful analysis of the boundary terms which must be added to the gravitational action for null boundary surfaces and for joints where such null boundaries intersect with other boundary surfaces \cite{LuisRob}.

The new boundary terms developed in \cite{LuisRob} have opened up the possibility of investigating the complexity=action proposal in a variety of new situations and this is the focus of the present paper. In particular, we use the new boundary terms to answer a question posed in \cite{Brown1,Brown2}: What is the `complexity of formation' for a thermal state of temperature $T$? That is, the full geometry of an eternal AdS black hole can be interpreted as being dual to the thermofield double state  \cite{eternal},
\begin{equation}
| {\rm TFD} \rangle = Z^{-1/2} \sum_\alpha e^{-{E_\alpha}/({2T})} |E_\alpha\rangle_L |E_\alpha\rangle_R\,,
\label{TFD}
\end{equation}
where we have two copies of the CFT, which are associated with the left (L) and right (R) asymptotic boundaries --- see the Penrose diagram in figure \ref{PenroseBHa}. Of course, integrating out either the left or right copy in the above state leaves us with the thermal density matrix for the CFT at temperature $T$. The Einstein-Rosen bridge emerges in the bulk geometry through the entanglement between these two sets of degrees of freedom \cite{eternal,EPR}.
The question, which we investigate here, is what is the additional complexity involved in forming this entangled thermofield double state \reef{TFD} compared to preparing each of the two individual CFTs in their vacuum state, \ie the complexity of formation.

Hence applying the complexity=action proposal \cite{Brown1,Brown2}, we begin by evaluating the action of the WDW patch in the AdS black hole spacetime to determine the complexity of the thermofield double state \reef{TFD}. Then for comparison to the CFT vacuum, we evaluate the action of the WDW patch in (two copies of) the vacuum AdS spacetime. One of our key results is that when the spacetime dimension $d$ of the boundary theory is three or higher, the complexity of formation grows linearly with the thermal entropy at high temperatures, \ie $\Delta\mC=k_d S$, where the proportionality coefficient depends only on $d$. For the special case of $d=2$, we find that the complexity of formation is a fixed constant, independent of the temperature.

The paper is organized as follows: In section \ref{sec:genset}, we review the details of our action computation including the various null surface and joint terms. In section \ref{sec:Cform}, we evaluate the complexity of formation for black holes in five and four bulk dimensions, where we consider various possible horizon geometries -- spherical, planar and hyperbolic. This section also presents the result for planar black holes with general $d$. In section \ref{sec:BTZ}, we study the complexity of formation for the special case of $d=2$, where the bulk geometry is described by a BTZ black hole. Section \ref{sec:CompVol} compares our results for the complexity of formation using the complexity=action duality to those found with the complexity=volume approach. Finally, we close with a brief discussion in section \ref{discuss}. A number of technical details are left to four appendices: Appendix \ref{App:FeGra}  presents some details about the choice of the UV cutoff surfaces, which are needed to regulate the action. In appendix \ref{App:Vac}, we describe certain subtle differences in the calculation of the vacuum complexity that arise for the different spatial geometries. Appendix \ref{app:Hypers} describes the calculation of the complexity of formation for `small' hyperbolic black holes, \ie with a negative mass. Appendix \ref{AppendixC} demonstrates that our results for the complexity of formations are robust against ambiguities in the definition of the gravitational action found in \cite{LuisRob}. In appendix \ref{lastA}, we examine the description of the analogous CFT states in terms of MERA tensor networks to gain some insight into our results.

\section{General Framework}\label{sec:genset}

\begin{figure}
\centering
\includegraphics[scale=0.35]{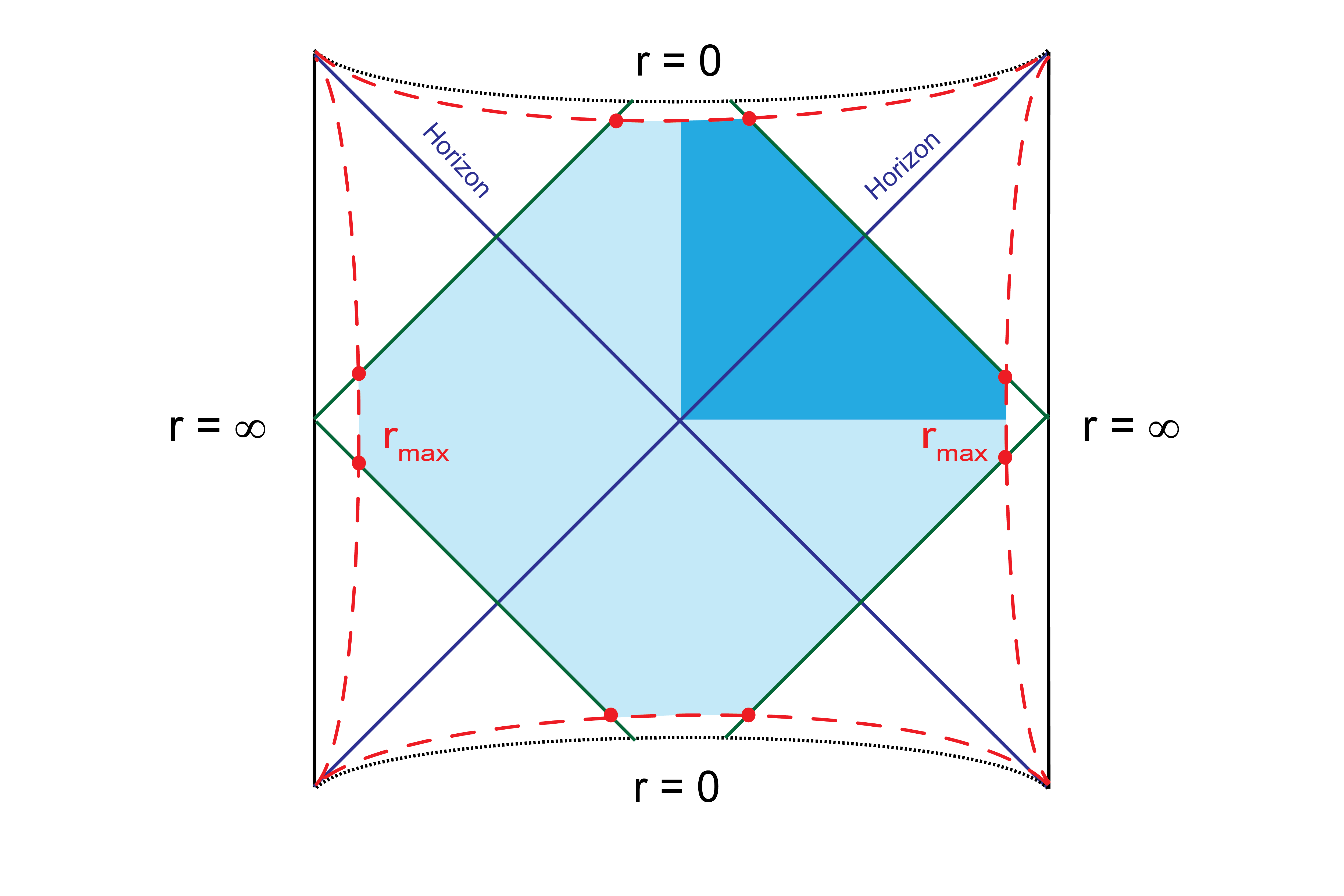}
\caption{Penrose diagram for black holes in more than three bulk dimensions ($d>2$).
We define surfaces of constant $r$ to regulate the action both  near the asymptotic boundary ($r=r_\mt{max}$) and near the past and future singularities ($r=\eps_0$). We identify the Wheeler-DeWitt patch as the area of the bulk bounded by the four null sheets which originate from the boundary at $t=0$. The joints between the null sheets and the regulating surfaces are indicated by red dots.}
\label{PenroseBHa}
\end{figure}

In this section, we describe the evaluation of the (regulated) gravitational action for the Wheeler-DeWitt (WDW) patch in various asymptotically locally AdS spacetimes. In particular, we focus on the AdS black holes in $d+1$ dimensions, whose metric takes the general form:\footnote{Here, we will assume that the boundary dimension satisfies $d>2$. The special case of the BTZ black hole \cite{btz1,btz2} with $d=2$ will be treated in section \ref{sec:BTZ}.}
\begin{equation}\label{HigherDMetric}
d s^{2} = - f(r)\, d t^{2} + \frac{d r^{2}}{f(r)} + r^{2}\, d \Sigma^{2}_{k,d-1}\,,
\end{equation}
with
\begin{equation}\label{BlackeningFactor}
f(r) = \frac{r^2}{L^2}+k -
\frac{\omega^{d-2}}{r^{d-2}}\,.
\end{equation}
Here, $L$ denotes the AdS curvature scale while $k=\lbrace+1,0,-1\rbrace$ indicates the curvature of the ($d$--1)-dimensional line element $d \Sigma^{2}_{k,d-1}$,
which is given by
\begin{equation}
d\Sigma^2_{k,d-1}=\left\lbrace\begin{matrix}
d\Omega^2_{d-1}&=d\theta^2+\sin^2\theta\, d\Omega^2_{d-2}\ \ \ &\ \  {\rm for\ }k=+1\ \,,\cr
d\ell^2_{d-1}&=\sum_{i=1}^{d-1} dx_i^2/L^2\ \ \qquad
\qquad&\ \ {\rm for\ }k=0\ \ \ \,,\cr
d\Xi^2_{d-1}&=d\theta^2+\sinh^2\theta\, d\Omega^2_{d-2}&\ \  {\rm for\ }k=-1\,.
\end{matrix}\right. \label{geometries}
\end{equation}
Hence, with $k=+1$, we have $d\Omega^2_{d-1}$, the standard round metric on a unit ($d$--1)-sphere; while for $k=0$,
$d\ell^2_{d-1}$ is the flat metric on $R^{d-1}$ (normalized by $1/L^2$); and for $k=-1$, $d\Xi^2_{d-1}$ is the metric on a ($d$--1)-dimensional
hyperbolic `plane' with unit curvature. In particular then, the black
holes corresponding to $k=\{+1, 0, -1\}$ have spherical, planar, and
hyperbolic horizons, respectively. The position of the horizon $r_h$ is determined by the `mass' parameter $\omega$ with
\beq
\omega^{d-2}= r_h^{d-2}\left(\frac{r_h^2}{L^2}+k\right)\,.
\label{horiz}
\eeq
Each of these solutions \reef{HigherDMetric} of the ($d$+1)-dimensional Einstein equations can be represented by the same Penrose diagram shown in figure \ref{PenroseBHa}.\footnote{`Small' hyperbolic black holes (with $k=-1$) require some extra consideration --- see comments below, as well as appendix \ref{app:Hypers}.} Of course, these geometries are
also static with the Killing vector $\partial_t$.

In the context of the AdS/CFT correspondence, these AdS black holes provide a holographic description of a uniform thermal bath in the dual CFT in the corresponding $d$-dimensional boundary geometry:
\beq
d s^{2}_\mt{boundary} = - d t^{2} +L^{2}\, d \Sigma^{2}_{k,d-1}\,.
\label{bound}
\eeq
The temperature of the thermal ensemble is given by
\beq
T=\frac{1}{4\pi}\left.\frac{\partial f}{\partial r}\right|_{r=r_h}=\frac{1}{4\pi r_h}\left(d\,\frac{r_h^2}{L^2} + (d-2)\,k \right)\,,
\label{temp}
\eeq
where we have used eq.~\reef{horiz} to substitute for $\omega^{d-2}$ in the above expression. The total energy is given by \cite{energy,count}
\begin{equation}\label{eq:BHMass}
M = \frac{(d-1)\Omega_{k, d-1}}{16\pi \Gn}\, \omega^{d-2}\,,
\end{equation}
where $\Omega_{k,d-1}$  denotes the dimensionless volume of the relevant spatial geometry in eq.~\reef{geometries}.
For instance, for $k=+1$, we have the volume of a ($d$--1)-dimensional unit sphere:  $\Omega_{1,d-1} = 2 \pi^{d/2}/\Gamma\left({d}/{2}\right)$. For the hyperbolic and planar geometries, we must introduce an infrared regulator to produce a finite volume, \eg for $k=0$, we could simply identify the spatial coordinates with $x_i\sim x_i + R_i$.\footnote{This choice then yields the dimensionless volume $\Omega_{0,d-1} =\prod_{i=1}^{d-1}\! R_i/L^{d-1}$, while the dimensionful spatial volume of the boundary geometry \reef{bound} would be simply ${\mathcal{V}}_{\,0,d-1} =L^{d-1}\,\Omega_{0,d-1} = \prod_{i=1}^{d-1}\! R_i$.}
The entropy of the system is determined by  the  usual Bekenstein-Hawking entropy of the event horizon:
\begin{equation}\label{eq:EntropyBH}
S = \frac{\mathcal {A}_\mt{horizon}}{4\Gn} = \frac{\Omega_{k,d-1} }{4\Gn}\,r_h^{d-1}\,.		
\end{equation}

In using the language of a thermal ensemble, we are describing the physics of the CFT dual to a single asymptotic boundary of the black hole geometry \reef{HigherDMetric}. As described in the introduction, the full geometry illustrated in figure \ref{PenroseBHa} can be interpreted as the dual of the thermofield double state \reef{TFD}, which provides a purification of the thermal ensemble with the second asymptotic boundary being dual to the thermofield double of the original CFT \cite{eternal}. Now the central question, which we wish to address here, is what is the additional complexity involved in forming this entangled thermofield double state
\reef{TFD} compared to preparing each of the two individual CFTs in their vacuum state.  In the nomenclature of \cite{Brown1,Brown2}, we wish to evaluate the `complexity of formation.'

\begin{figure}
   \centering
\vspace{0.5cm}
        \begin{subfigure}[b]{0.26\textwidth}
       \includegraphics[width=\textwidth]{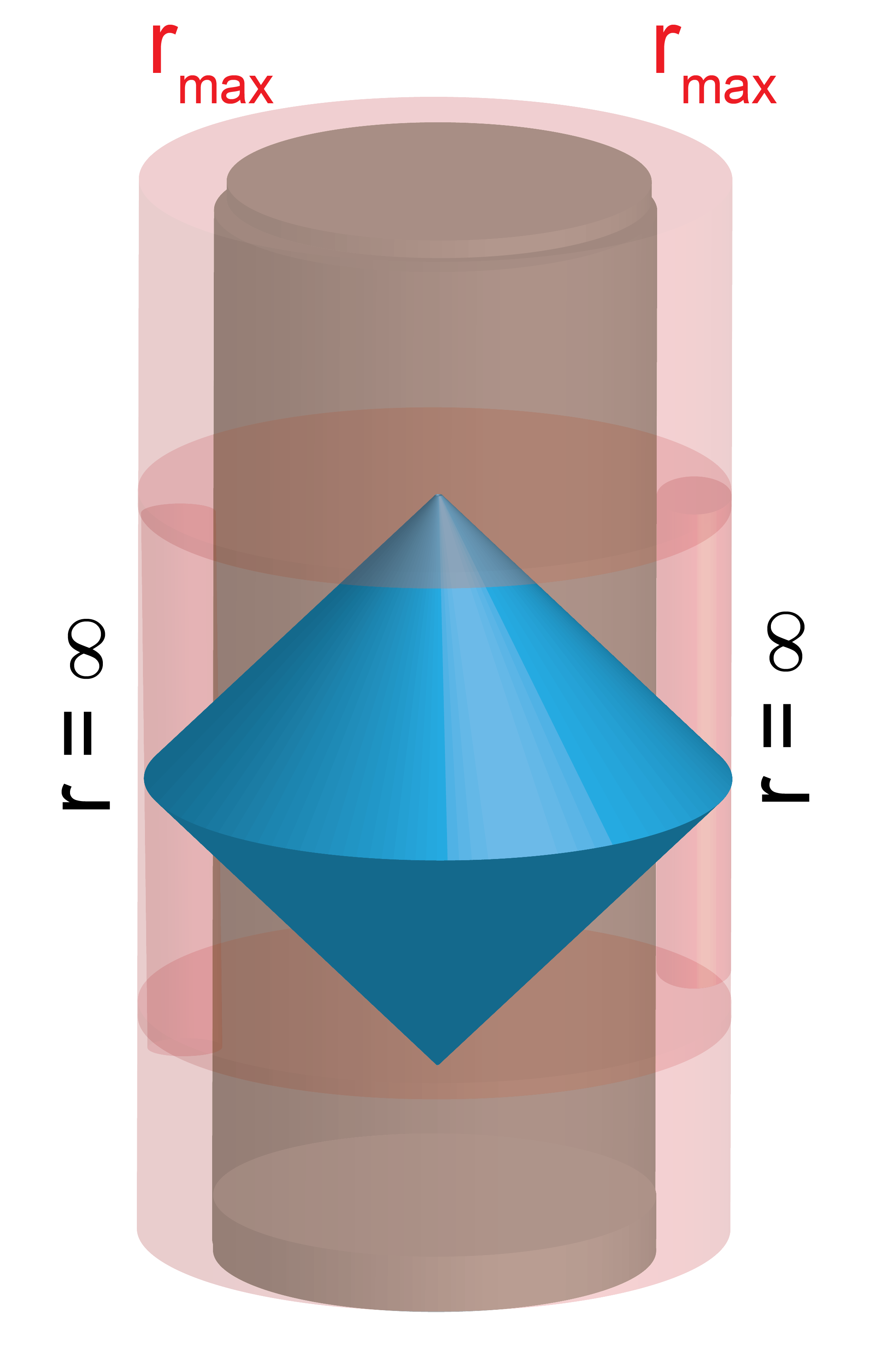}
                \caption{$k=+1$}
\label{EmptyAdSSpherical}
        \end{subfigure}
~
        \begin{subfigure}[b]{0.25\textwidth}
                \includegraphics[width=\textwidth]{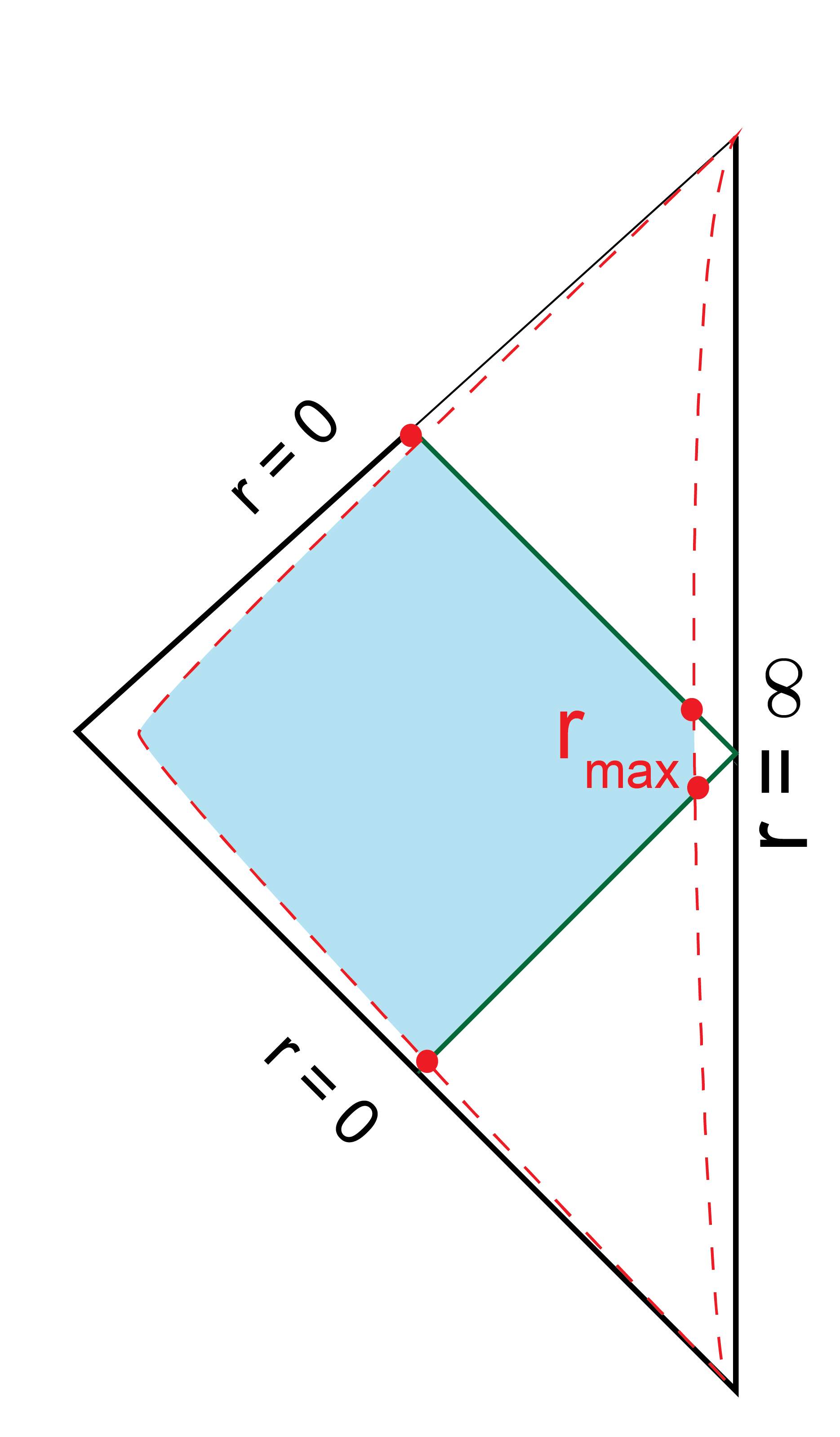}
                \caption{$k=0$}
\label{EmptyAdSPlanar}
        \end{subfigure}
~~~~
        \begin{subfigure}[b]{0.40\textwidth}
                \includegraphics[width=\textwidth]{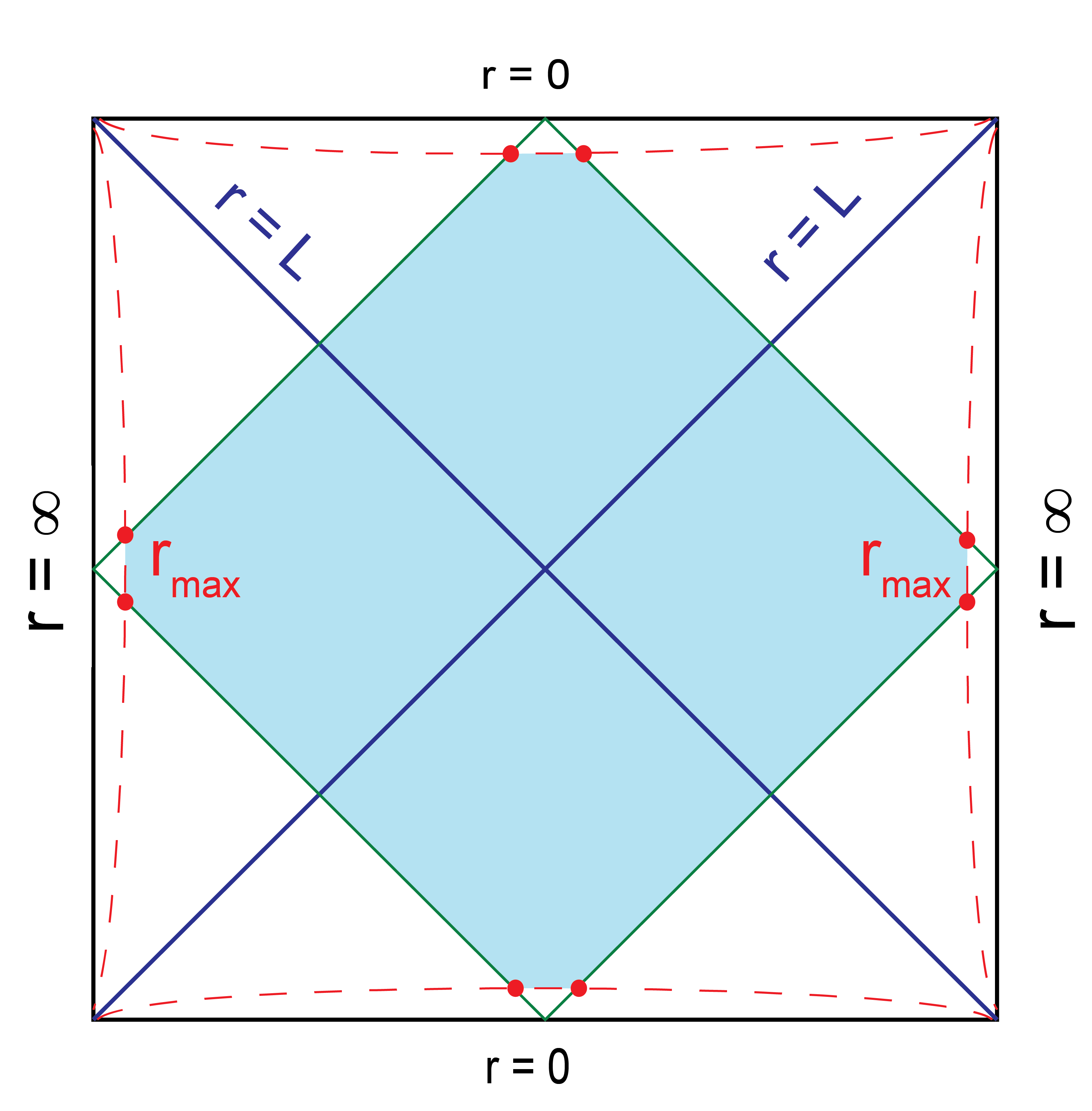}
                \caption{$k=-1$}
\label{EmptyAdSHyperbolic}
        \end{subfigure}
        \caption{Penrose diagrams of the Wheeler-DeWitt patch in vacuum AdS for the different values $k=\{+1, 0, -1\}$. }
\label{EmptyAdSDrawing}
\end{figure}

Hence applying the complexity=action proposal \cite{Brown1,Brown2}, we begin by evaluating the action of the WDW patch with $t_L=0=t_R$, shown in figure \ref{PenroseBHa}, to determine the complexity of the thermofield double state \reef{TFD}. Then for comparison, we evaluate the action of the WDW patch in the vacuum AdS spacetime, corresponding to the metric \reef{HigherDMetric} with $\omega=0$, \ie replacing $f(r)$ with
\begin{equation}\label{EmptyBlack}
f_0(r) = \frac{r^2}{L^2}+k \,.
\end{equation}
While the evaluation of the action in the black hole backgrounds is essentially the same for the three different geometries corresponding to $k=\{+1,0,-1\}$, there are small differences for the vacuum geometries which should be accounted for. We will describe these subtleties here, \ie various singularities in the geometry. However, we defer evaluating their contributions to the gravitational action to appendix \ref{App:Vac}, because our final conclusion will be that in fact these singularities do {\it not} affect the final value of the WDW action in the vacuum spacetimes. The WDW patches for the AdS vacua are shown in figure \ref{EmptyAdSDrawing}. 

\noindent{\bf a) Spherical geometry:} With $k=+1$, the vacuum metric in the bulk is the AdS geometry in global coordinates. In particular then, these coordinates cover the entire AdS spacetime. Choosing a constant time slice, the corresponding WDW patch is the causal diamond shown in figure \ref{EmptyAdSSpherical}. The only point to note here is that the past and future tips of the causal diamond are caustics, \ie all of the null rays in the associated null boundaries cross each other at these points. Singular features like this were not considered in the recent discussion of boundary terms for the gravitational action \cite{LuisRob} and so will require some special attention.\vspace{.5em}

\noindent{\bf b) Planar geometry:} With $k=0$, the vacuum bulk metric is the AdS space in Poincar\'e coordinates, which only cover a portion of the full AdS geometry.
We note that in these coordinates, the $t=0$ time slice covers the entire Cauchy surface at $t=0$ in global coordinates.\footnote{We must include an extra point at infinity, \ie at $x_i,\, r\to \infty$.} In the present context, however, we are  compactifying the spatial coordinates --- as described below eq.~\reef{eq:BHMass} --- and as a result, the Poincar\'e horizon becomes a null orbifold or `conical' singularity. That is, the proper volume of the spatial geometry shrinks to zero along this null line. Further, the null generators of the corresponding WDW patch all intersect when they hit this null spacetime singularity, as shown in figure \ref{EmptyAdSPlanar}. Hence both this caustic and the null singularity will require special attention in evaluating the action of the WDW patch.\vspace{.5em}

\noindent{\bf c) Hyperbolic geometry:} With $k=-1$, the vacuum bulk metric is AdS space in the so-called `AdS-Rindler' coordinates. Again these coordinates only cover a portion of the full AdS geometry, and in particular,  the infinite hyperbolic geometry only covers a portion of the boundary time slice at $t=0$ and $r\to\infty$ --- see figure 2 in \cite{Mark1}. Examining $f_0(r)$ in eq.~\reef{EmptyBlack} with $k=-1$, we see that there is still a horizon at $r_h=L$ even when $\omega=0$. Further, eq.~\reef{temp} then yields a finite temperature $T=1/(2\pi L)$ in this case. Hence the vacuum metric still has the form of an AdS black hole and it can be interpreted in terms of an entangled state of two copies of the CFT on a hyperbolic hyperplane. This curious interpretation of the `AdS vacuum' can be understood from the discussion of \cite{CHM}. The hyperbolic boundary geometry, \ie eq.~\reef{bound} with $k=-1$, can be mapped to the spherical or planar boundary geometry with a conformal transformation. However, this mapping takes the $t=0$ time slice in the $k=-1$ geometry to the interior of a finite spherical region in either of the other two geometries. While the CFT vacuum is a pure state on the global time slice  of either of the latter backgrounds, it becomes a mixed state when reduced to this spherical region. The entangled state of two copies of the CFT on a hyperbolic plane appearing above can then be understood as a conformally transformed description of the global vacuum state which entangles the CFT degrees of freedom on the interior with those on the exterior of the sphere. Since the vacuum  already contains two copies of the CFT in the hyperbolic case, we only need to consider a single copy of the vacuum AdS geometry when evaluating the complexity of formation.

We should also add that since we are compactifying the spatial geometry, the volume of the spatial sections shrinks to zero at $r=0$ producing an orbifold singularity. However, for $k=-1$, this singularity lies behind the horizon and as shown in figure \ref{EmptyAdSHyperbolic}, the tips of the WDW patch just touch this singular surface. Again this singularity requires special attention in evaluating the action of the WDW patch.\vspace{.5em}

At this point, we might also mention that with $k=-1$, the event horizon persists when $\omega^{d-2}$ takes on negative values and the black hole mass \reef{eq:BHMass} becomes negative \cite{roberto}. In this case, eq.~\reef{horiz} yields two  real positive solutions for $r_h$ and the causal structure of the geometry takes a form similar to that of a charged black hole --- see figure \ref{PenroseInOut}. Hence the evaluation of the action in this case demands some extra attention, as described in appendix \ref{app:Hypers}.

Finally, let us close here by observing that we can follow the procedure outlined below to evaluate the complexity of formation for any value of $r_h$. However, in the case of spherical horizons, we should recall the Hawking-Page phase transition \cite{HP,witten1,Cho2}, which occurs for small black holes.\footnote{Recall that there is no analogous phase transition for the planar or hyperbolic black holes \cite{Cho2}.} That is, when $r_h<L$,  the saddle point which dominates the bulk partition function is still vacuum AdS space. This implies then that the complexity of formation is only an order one quantity in the large $N$ (or large central charge) expansion of the boundary CFT --- see section \ref{discuss}.

\subsection{Evaluating the Action}\label{subsec:actionsetup}

Next, we describe in detail the evaluation of the gravitational action for the Wheeler-DeWitt patch. Including all of the various boundary terms, the gravitational action
can be written as \cite{LuisRob},\footnote{We will be using slightly modified conventions from those given in \cite{LuisRob} --- see \cite{Pratik}.}
\begin{equation}
\begin{split} \label{ActionGeneral}
I=\ & \frac1{16\pi G_N}\int_\mathcal{M} d^{d+1}x \sqrt{-g} \left(R +   \frac{d(d-1)}{L^2}\right)+\frac1{8\pi G_N}\int_\mathcal{B} d^{d}x \sqrt{|h|} \,K
 \\
&
- \frac1{8\pi G_N}\int_{\mathcal{B}'}d\lambda\, d^{d-1}\theta \sqrt{\gamma}\, \kappa
+ \frac1{8\pi G_N}\int_{\Sigma} d^{d-1}x \sqrt{\sigma}\, \eta
+ \frac1{8\pi G_N}\int_{\Sigma'} d^{d-1}x \sqrt{\sigma}\, a \,  .
\end{split}
\end{equation}
The various terms include: the Einstein-Hilbert and cosmological constant terms (with $\Lambda=-d(d-1)/(2L^2)$) integrated over the $d+1$-dimensional volume $\mathcal{M}$; the Gibbons-Hawking-York extrinsic curvature term \cite{York,GH} integrated over the timelike and spacelike boundary surfaces, denoted  by $\mathcal{B}$; the $\kappa$ boundary contribution \cite{LuisRob} (see also \cite{Pad}) integrated over $d$-dimensional null boundary surfaces, denoted by $\mathcal{B'}$; the Hayward joint terms \cite{Hay1,Hay2} which are included at the intersections $\Sigma$ of two boundaries which are either timelike or spacelike; and finally the $a$ joint terms \cite{LuisRob} which are included at the intersections $\Sigma'$ of two boundary surfaces where either or both are null surfaces. In the following, we consider the contribution of each of these terms to the action of the WDW patch in the static black hole background \reef{HigherDMetric} at $t_L=0=t_R$, as well as in the corresponding AdS vacuum geometries. We will examine the full time evolution of the complexity elsewhere \cite{prep5}.

However, before proceeding with these calculations, we first observe that the action of the WDW patch is divergent because this spacetime region extends all the way to the asymptotic boundary of the bulk geometry. This divergence would naturally be associated with a UV divergence in the complexity related to establishing correlations between the CFT degrees of freedom at arbitrarily short distance scales, \eg see \cite{Pratik}. Hence to make sense of the calculation, we regulate with the standard approach of truncating the region on which the action is evaluated by a cutoff surface at $r=r_\mt{max}$, see, \eg \cite{count,sken1,sken2}.\footnote{The standard approach is to eliminate the corresponding divergences in the regulated action by adding surface counterterms, \eg \cite{count,sken1,sken2}. The `renormalized' action is then evaluated by taking the limit $r_\mt{max}\to\infty$ (or $\delta\to0$). We do not apply this approach here in evaluating the action of the WDW patch for two reasons: First, the same surface counterterms simply do not remove the divergences in the WDW action \cite{Pratik}. Second, the UV divergences have a physical interpretation in terms of the complexity, as described above. Note, however, that the UV divergences will cancel below in the difference of the WDW actions for the black hole and vacuum, and hence the complexity of formation is finite, \ie independent of the details of the UV regulator.} A potential subtlety here is that we wish to compare the WDW actions in the two different spacetimes (\ie the AdS black hole and vacuum AdS space) and so we need to choose our cutoff surfaces in a consistent way. As described in appendix \ref{App:FeGra}, we do so by describing both geometries with the corresponding metric in the canonical form given by the Fefferman-Graham expansion \cite{FG1,FG2} and then we set the radial cutoff surface at $z=\delta$ in both cases. As usual, $\delta$ plays the role of a short-distance cutoff in the dual boundary theory. The final result is given by
\begin{equation}
r^\mt{BH}_{\mt{max}}-r^\mt{\vac}_{\mt{max}}= \frac{\omega^{d-2}  }{2 d L^{2 (d-2)}}\ \delta ^{d-1}\ +\ \mathcal{O}\!\left(\delta^{d+1}\right)\, .
\label{differ}
\end{equation}
It turns out that this difference appears at a sufficiently high order that, in fact, the complexity of formation is not affected --- see appendix \ref{App:FeGra}.
Note that the (timelike) UV cutoff surfaces at $r=r_\mt{max}$ are shown in the Penrose diagrams in figures \ref{PenroseBHa} and \ref{EmptyAdSDrawing}.\footnote{We have also shown various other regulator surfaces, \eg near the spacetime singularity in the black hole geometry. These will appear in the discussion below and in appendix \ref{App:Vac}.}

Next we need to define the (null) boundaries of the WDW patch. For this purpose,
it will be useful to define the tortoise coordinate,
\begin{equation}\label{tortoise}
r^* = \int \frac{dr}{f(r)}\,,
\end{equation}
with which we construct the Eddington-Finkelstein outgoing and infalling coordinates,
\beq\label{EddFinkCoords}
u  = t - r^*(r)\qquad{\rm and}\qquad
v  = t + r^*(r)\,,
\eeq
respectively.
In terms of these coordinates, the metric \eqref{HigherDMetric}  becomes
\beqa
d s^{2} &=& - f(r)\, d u^{2} - 2\, du\, dr + r^{2} d \Sigma^{2}_{k,d-1}\label{newmetric}\\
&=&- f(r)\, d v^{2} + 2\, dv \, dr + r^{2} d \Sigma^{2}_{k,d-1}\,,
\nonumber
\eeqa
which are well-behaved on the past and future event horizons, respectively. Now let us focus our attention on the right-hand side of the Penrose  diagram in figure \ref{PenroseBHa}. We are interested in the WDW patch corresponding to the time slice $t=0$ (\ie $t_R=0$) and so the past null boundary can be defined as
\begin{equation}\label{eq:uinf}
u=u_\infty \qquad{\rm with}\ \ u_\infty = -\lim_{r\rightarrow\infty} r^{*}(r)\,.
\end{equation}
Similarly, the future null boundary is given by
\begin{equation}\label{eq:vinf}
v=v_\infty \qquad{\rm with}\ \ v_\infty = \lim_{r\rightarrow\infty} r^{*}(r)\,.
\end{equation}
Note that the two constants are the same up to a sign, \ie $u_{\infty} = -v_{\infty}$.

Analogous boundaries can be constructed for the left-hand side of the Penrose diagram, however, the details for these will not be needed. In particular, there is a four-fold symmetry in the case of interest (\ie the WDW patch corresponding to $t_R =0=t_L$) consisting of the left-right symmetry in the Penrose diagram and the time reflection symmetry, \ie $t\to-t$. Hence for simplicity, our calculations of the action focus only on the upper right quadrant in figure \ref{PenroseBHa}, \ie the region between $t=0$ and $v=v_\infty$.

As a final note here, it will be useful for the following calculations to evaluate the tortoise coordinate \eqref{tortoise}. In general, the blackening factor can be written in the form:
\begin{equation}
f(r) = (r-r_h)\, F(r)
\end{equation}
where $F(r)$ has no positive real roots.\footnote{The only exception is the case of small hyperbolic black holes, where the blackening factor has two positive real roots. We will deal with this case separately in appendix \ref{app:Hypers}.}
Hence the inverse of $f(r)$ can be decomposed as:
\begin{equation}\label{eq:factorize}
\frac{1}{f(r)} =
\frac{1}{F(r_h) (r-r_h)} +
\frac{F(r_h)-F(r)}{F(r)F(r_h)(r-r_h)}\,.
\end{equation}
Note that while the first term contains a pole at $r=r_h$, the second term  above is regular at the horizon. Integrating with respect to $r$, we obtain the tortoise coordinate:
\begin{equation}
r^{*}(r) = \frac{\log{|r-r_{h}|}}{F(r_h)}  + G(r)\qquad
{\rm where}\ \ \ G(r) = \int \frac{F(r_h)-F(r)}{F(r)F(r_h)(r-r_h)} dr\,.
\label{ggg}
\end{equation}
Again, the function $G(r)$ is completely regular at $r=r_h$. Eq.~\reef{ggg} will be useful to explicitly evaluate the tortoise coordinate \eqref{tortoise} for the specific examples presented in the following section.

We now turn to the evaluation of each of the contributions in the gravitational action \reef{ActionGeneral} for the WDW patch shown in figure \ref{PenroseBHa}.

\subsubsection{Bulk Contribution}
We start by evaluating the Einstein-Hilbert and cosmological constant terms in eq.~\eqref{ActionGeneral}:
\begin{equation}\label{bulky}
I_{\textrm{bulk}} =  \frac{1}{16\pi G_N} \int_{\mathcal{M}} d^{d+1} x \sqrt{- g} \left(R +   \frac{d(d-1)}{L^2}\right)\,.
\end{equation}
Einstein's equations yield $R = -d(d+1)/L^{2}$ and so the above can be written as\footnote{We are evaluating eq.~\reef{bulky} using the original $(t,r)$ coordinates in eq.~\reef{HigherDMetric}. For the upper right quadrant described above, we integrate over the time coordinate as: $\int_{0}^{v_\infty -r^*(r)} dt = v_\infty -r^*(r)$.}
\begin{align}\label{ActionBulk}
I_{\textrm{bulk}} =  - \frac{\Omega_{k,d-1} \, d}{2\pi G_N L^2} \, \int_0^{r_{\mt{max}}} dr\, r^{d-1}\,\big(v_\infty -r^*(r)\big)
\,,
\end{align}
where $v_\infty$ is the constant defining the null boundary for this quadrant, as in eq.~\reef{eq:vinf}. Further, as described above, we have multiplied by a factor of 4 and we are only performing the integral over the upper right quadrant of the WDW patch.

We might note that the same expression can be applied for the vacuum AdS spacetime. The latter only requires that we replace $f(r)$ by $f_0(r)$ from eq.~\reef{EmptyBlack}, which is used in the definition of $r^*(r)$ in eq.~\reef{tortoise} --- as well as $v_\infty$ then in eq.~\reef{eq:vinf}. In this case, the factor of 4 in eq.~\reef{ActionBulk} automatically includes the contribution of two vacuum AdS geometries.

\subsubsection{Surface Contributions}

\begin{figure}
\centering
\includegraphics[scale=0.30]{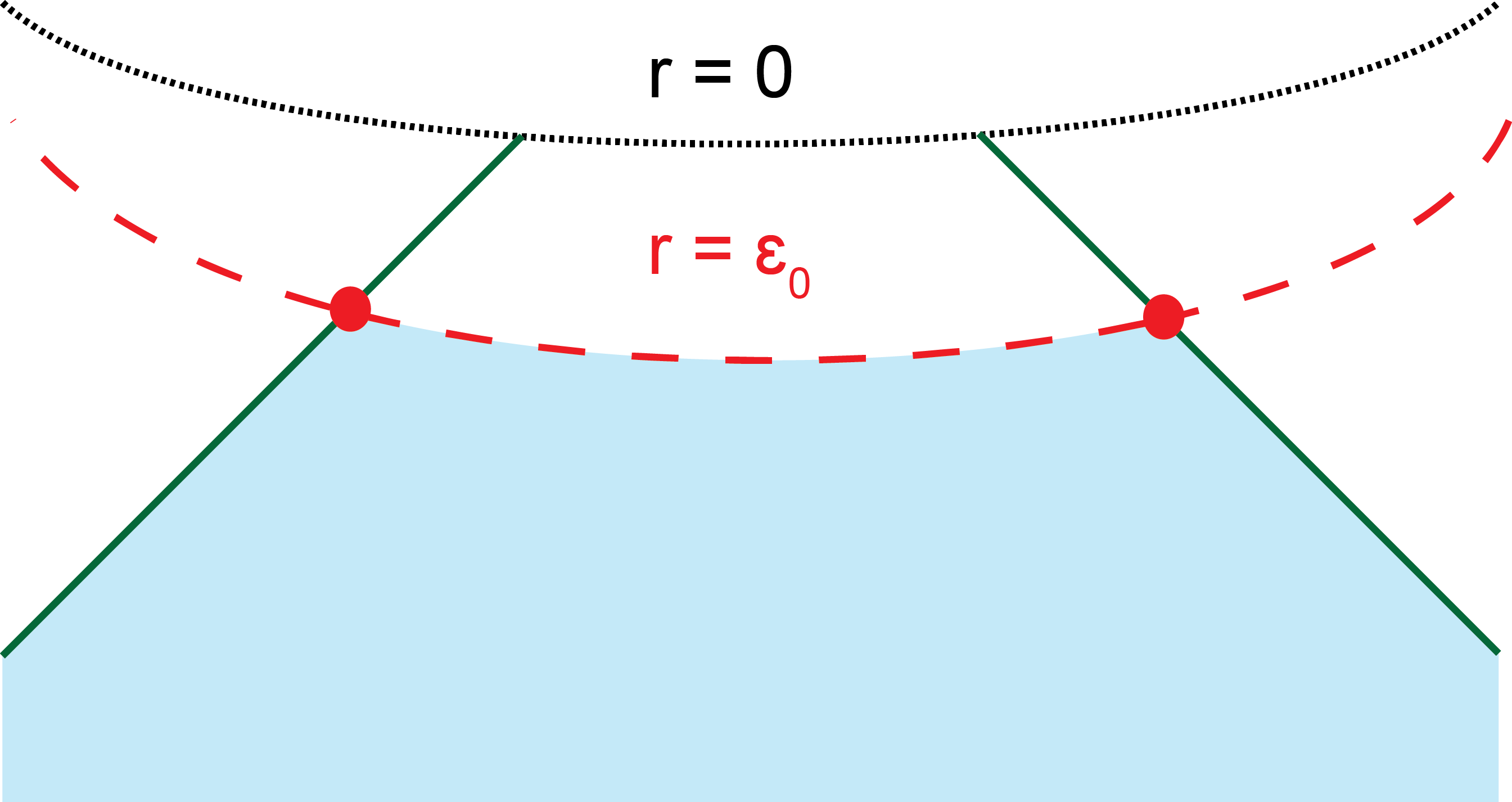}
\caption{The top of the WDW patch for black holes in $d>2$. The GHY surface term evaluated on the regulator surface at $r=\epsilon_0$ makes a finite contribution to the action.}
\label{ZoomTopSurface}
\end{figure}

Next we have the GHY extrinsic curvature term \cite{York,GH}, which is integrated over the timelike or spacelike boundary surfaces,
\beq\label{ActSurf}
I_\mt{GHY}=\frac1{8\pi G_N}\int_{\mathcal{B}} d^{d}x \sqrt{|h|} \,K\,.
\eeq
There are two pairs of such surfaces for the WDW patch in figure \ref{PenroseBHa}: the timelike  surfaces at $r=r_{\mt{max}}$, which are introduced in both of the asymptotic regions to provide a UV cutoff, as discussed above eq.~\reef{differ}; and the spacelike surfaces at $r=\epsilon_0$, which are inserted to regulate the geometry of the WDW patch where it touches the future and past curvature singularities behind the horizon, following \cite{Brown2} --- see figure \ref{ZoomTopSurface}.
As described above, we will only focus on the contribution of the GHY terms in the upper right quadrant. We can write the unit normal vectors to these surfaces as
\begin{equation}\label{VecBH}
\begin{split}
r=r_{\mt{max}}\ :\qquad&{\bf s}=s_{\mu}\,dx^\mu =   \frac{dr}{\sqrt{f(r_{\mt{max}})}} \, ,
\\
r=\epsilon_0\ :\qquad&{\bf t}=t_{\mu}\,dx^\mu =  -\frac{dr}{\sqrt{-f(\epsilon_0)}}\, .
\end{split}
\end{equation}
Note that our convention here is that these normals (as one-forms) point outward from the spacetime volume of interest. The trace of the extrinsic curvature is then given by
\begin{equation}
K= \frac{n_r}{2} \left( {\del_r f(r)}+ \frac{2(d-1)}{r}\,f(r) \right) \, ,
\end{equation}
where $n_\mu$ denotes the unit normal of interest.
Substituting the appropriate normals from eq.~\reef{VecBH} into this expression then yields for \eqref{ActSurf}:
\beqa
I_\mt{GHY}(r=\epsilon_0) & =& -  \frac{\Omega_{k,d-1}\,r^{d-1}}{4 \pi G_N}   \left( \del_r f(r) + \frac{2(d-1)}{r}\,f(r) \right) \bigg( v_\infty-r^*(r) \bigg)\bigg|_{r=\epsilon_0},~~~~~~~~~~
\label{extrinsic1}\\
I_\mt{GHY}(r=r_{\mt{max}}) & = & \frac{\Omega_{k,d-1}\,r^{d-1}}{4 \pi G_N}   \left( \del_r f(r) + \frac{2(d-1)}{r}\,f(r) \right) \bigg( v_\infty-r^*(r) \bigg)\bigg|_{r=r_{\mt{max}}},~~~
\label{extrinsic2}
\eeqa
where we have included an additional factor of 4 to include the contributions from all four quadrants of the Penrose diagram.

Note that eq.~\reef{extrinsic2} for the contribution of the UV cutoff surface can also be used for the vacuum AdS spaces upon replacing $f(r)$ with $f_0(r)$ from eq.~\reef{EmptyBlack}. Recall that there is a small difference in the value of $r_{\mt{max}}$ for the black hole and vacuum AdS geometries, as shown in eq.~\reef{differ}. However, a detailed analysis shows that the difference between these surface contributions in the two geometries vanishes. That is, when the corresponding contribution for vacuum AdS is subtracted from eq.~\reef{extrinsic2} for the black hole geometry, the result is proportional to a single power of $\delta$ and so vanishes in the limit $\delta \rightarrow 0$ --- see appendix \ref{app:Fegra2} for details.

On the other hand, the contribution \reef{extrinsic1} coming from the singularity has no counterpart in the vacuum AdS geometry.
Examining this expression in more detail, we find that the black hole geometry yields a finite result,
\beq
I_\mt{GHY}(r=\epsilon_0) =   \frac{d\,\Omega_{k,d-1}\,\omega^{d-2}}{4 \pi G_N}  \big( v_\infty-r^*(0) \big)+ O(\eps_0)\,.
\label{extrinsic1a}
\eeq
Hence this is the only contribution which the GHY surface terms make to the complexity of formation.\footnote{We note that the computation for small hyperbolic black holes is slightly different and there is no contribution from the spacetime singularity inside the event horizon --- see appendix \ref{app:Hypers} for details.}

We also have the surface term introduced in \cite{LuisRob} for null boundary surfaces,
\beq\label{ActSurf2}
I_\mt{null surface}=- \frac1{8\pi G_N}\int_{\mathcal{B}'}d\lambda\, d^{d-1}\theta \sqrt{\gamma}\, \kappa\,,
\eeq
where the hypersurface is described parametrically by $x^\mu=x^\mu(\lambda,\theta^A)$ with $\lambda$, the parameter along the null generators spanning the boundary surface and $\theta^A$ constant on each generator.
Then $\kappa$ measures the failure of $\lambda$
to be an affine parameter on the null generators of the surface, \ie
\beq\label{kappa}
k^{\mu} \, \nabla_{\mu} k_{\nu}= \kappa \, k_{\nu} \qquad{\rm with}
\ \ \ k^\mu=\frac{\del x^\mu}{\del \lambda}\,.
\eeq
Hence this contribution can be easily dismissed by using the ambiguity in the null normals to choose them to be affinely parameterized and so setting $\kappa=0$, as discussed in \cite{LuisRob}.\footnote{In appendix \ref{AppendixC}, we consider a different parameterization of the null surfaces yielding $\kappa=$ constant and we find that our results for the complexity of formation do not change.}
This is easily achieved here using the definition of the null boundaries in terms of the Eddington-Finkelstein coordinates \reef{eq:uinf} and \reef{eq:vinf}. In particular for the null boundary in the upper right quadrant, we set
\beq
{\bf k}=dv\big|_{v=v_\infty}=\left(dt + \frac{dr}{f(r)}\right)\bigg|_{v=v_\infty}\,.
\labell{niceNull}
\eeq
Implicitly, we have normalized this null normal at the asymptotic AdS boundary such that ${\bf k} \cdot {\bf{\hat t}} = 1$ where ${\bf \hat t}=\del_t$, as suggested in \cite{LuisRob}.

\subsubsection{Joint Contributions}

This leaves the joint terms in the gravitational action \reef{ActionGeneral}
which are needed where two of the boundary surfaces intersect. First, we have the Hayward terms \cite{Hay1,Hay2}
\beq
I_\mt{Hay}=\frac1{8\pi G_N}\int_{\Sigma} d^{d-1}x \sqrt{\sigma}\, \eta\  ,
\eeq
but these are not relevant here since all of the joints in figure \ref{PenroseBHa} involve at least one null surface. Hence we only need to consider the last term in the gravitational action \reef{ActionGeneral}
\beq
I_{\jnt}= \frac1{8\pi G_N}\int_{\Sigma'} d^{d-1}x \sqrt{\sigma}\, a \, ,
\label{ActJ2}
\eeq
where $a$ is defined as \cite{LuisRob},
\begin{equation}
a= \begin{cases}
 \epsilon \log{ |k \cdot t| } \qquad\text{for  spacelike-null joint with }\epsilon = -\mbox{sign}(k \cdot t)\, \mbox{sign}(k \cdot \hat s)\,, \\
\epsilon \log{| k \cdot s |}\qquad \text{for  timelike-null joint with }\epsilon = -\mbox{sign}(k \cdot s) \, \mbox{sign}(k \cdot \hat t)\,. 
\end{cases}\label{ActJ2a}
\end{equation}
and $\hat s$ and $\hat t$ are auxiliary unit \emph{vectors} in the tangent space of the spacelike/timelike boundary surface, which are orthogonal to the junction and point outwards from the boundary region of interest --- see figure  \ref{cornerdrawing} and reference \cite{Pratik}.

Again focusing our attention on the upper right quadrant of the WDW patch, we have a spacelike-null joint where the null boundary \reef{eq:vinf} meets the regulator surface $r=\epsilon_0$.\footnote{Again, we note that the computation for small hyperbolic black holes is slightly different --- see appendix \ref{app:Hypers} for details.} Using the corresponding normals in eqs.~\reef{VecBH} and \reef{niceNull} and ${\bf \hat s}=\hat s^\mu\,\del_\mu = \del_t/\sqrt{-f(r)}$, the null joint term \reef{ActJ2} yields
\beqa\label{CornerSingSP}
I_{\jnt,\text{sing}} &=&  - \frac{\Omega_{k, d-1}}{4 \pi G_N}\,  r^{d-1}  \log|f(r)|\,\Big|_{r=\epsilon_0}\\
&\simeq&\   \frac{\Omega_{k, d-1}}{4 \pi G_N}\, \epsilon_0^{d-1}  \log(\epsilon_0^{d-2}/\omega^{d-2}) \,,\nonumber
\eeqa
where as usual we have included a factor of 4 to include the contributions of all of the joints near the future and past singularities. However, the key observation about this result is that this contribution vanishes in the limit $\epsilon_0\to0$.

We also have a timelike-null joint where the null boundary \reef{eq:vinf} meets the cut-off surface $r=r_{\mt{max}}$. In this case ${\bf \hat t}=\hat t^\mu\,\del_\mu = \del_t/\sqrt{f(r)}$  and the corresponding contribution (including the usual factor of 4) is
\begin{equation}\label{CornerCut}
I_{\jnt,\text{cut}} =  \frac{\Omega_{k, d-1}}{4 \pi G_N}\,  r^{d-1}\log{f(r)}\,\Big|_{r=r_{\mt{max}}} \,.
\end{equation}
Again, this contribution from the UV cutoff surface appears in the vacuum AdS calculation as well but with $f(r)$ replaced by $f_0(r)$, given in eq.~\reef{EmptyBlack}. Further analysis shows that the difference between these contributions in the black hole and vacuum AdS geometries again vanishes in the limit $\delta\to0$ --- see appendix \ref{app:Fegra2} for details.

At this point, let us reiterate that there are certain subtleties, \eg caustics and orbifold singularities, in the case of the AdS vacua, which should be accounted for in evaluating the gravitational action. However, as described in appendix \ref{App:Vac}, we find that in fact they do not produce any additional nonvanishing contributions to the gravitational action of the vacuum WDW patch.

\section{Complexity of Formation} \label{sec:Cform}

We can combine the various results of section \ref{sec:genset} to evaluate the desired complexity of formation,
\beq
\Delta\mC = \frac{1}{\pi}\big[\, I(\text{BH}) - 2\,I(\text{AdS})\,\big]\,.
\label{Cform}
\eeq
We already established that the surface and joint contributions associated with the cutoff surface at $r=r_{\mt{max}}$ precisely cancel between the two geometries. Hence the only nonvanishing contributions that need to be considered are the bulk contributions \reef{ActionBulk} for both geometries and the GHY surface contribution \reef{extrinsic1a} at the black hole singularity. Combining the various results above then, we arrive at
\begin{equation}\label{Cform1}
\Delta\mC = \frac{1}{\pi}\big[\,\Delta I_{\bulk} + I_{\mt{GHY}}^{\BHx} \,\big]
\end{equation}
where
\beqa
\Delta I_{\bulk}&=& - \frac{\Omega_{k,d-1} \, d}{2\pi G_N L^2} \, \int_0^{r_{\mt{max}}} dr\, r^{d-1} \,\Big[v_\infty -v'_\infty-\big(r^*(r) -r_0^*(r)\big)\Big]\,,
\nonumber\\
I_{\mt{GHY}}^{\BHx} &=& \frac{\Omega_{k,d-1}\,d}{4 \pi G_N}\,\omega^{d-2}\,  \big( v_\infty-r^*(0) \big)\,.\label{Cform1a}
\eeqa
In the expression for $\Delta I_{\bulk}$, we use $r_0^*(r)$ and $v'_{\infty}$ to denote the tortoise coordinate and the null boundary in the AdS vacuum --- see eq.~\reef{vacvac} below.

As mentioned previously, the computation for ``small' hyperbolic black holes with $k=-1$ and  $r_h<L$ is slightly different. The full details are described in appendix  \ref{app:Hypers}. The essential difference is that $f(r)$ has two positive real roots and the Penrose diagram resembles that of a charged black hole --- see figure \ref{PenroseInOut}. In this case, the null boundaries from the two asymptotic regions meet between the two horizons at $r^*(r_{\text{meet}})=v_{\infty}$  and hence the surface term near the singularity is replaced by a null joint term \reef{ActJ2}. The complexity of formation then becomes:
\begin{equation}\label{Cform2}
\text{small hyperbolic BH:}\qquad\Delta\mC =  \frac{1}{\pi}\big[\,\Delta I_{\bulk} + I_{\text{jnt}} \,\big]
\end{equation}
where
\beqa
\Delta I_{\bulk}&= & - \frac{\Omega_{-1,d-1} \, d}{2\pi G_N L^2} \, \left[\int_{r_{\text{meet}}}^{r_{\mt{max}}} dr\, r^{d-1} \,\Big[v_\infty -r^*(r)\Big]-\int_0^{r_{\mt{max}}} dr\, r^{d-1} \Big[v'_\infty-r_0^*(r)\Big]\right]\,,
\nonumber\\
I_{\text{jnt}}&=& - \frac{\Omega_{-1,d-1}}{4 \,\pi \, G_N} \,r_{\text{meet}}^{d-1} \, \log{|f(r_{\text{meet}})|} \, .
\label{Cform2a}
\eeqa

However, we should add that this result will change with redefinitions allowed by the ambiguities in the definition of the gravitational action \cite{LuisRob} --- for further discussion, see section \ref{discuss} and appendix \ref{AppendixC}.

We now evaluate the above results for some specific examples:

\subsection{$d=4$}\label{sec:AdS5}

For $d=4$, \ie a five-dimensional AdS black hole, the blackening factor  \eqref{BlackeningFactor} becomes
\begin{equation}\label{SphericalHorizonFunction5}
f(r) = \frac{r^2}{L^2} + k -
\frac{r_h^{2}}{r^{2}}\left(\frac{r_h^2}{L^2} + k\right)  \, ,
\end{equation}
while for vacuum AdS, we have $f_0(r)$ in eq.~\eqref{EmptyBlack}. From the results in appendix \ref{App:FeGra}, we fix the UV cutoff surface at
\beq
r_{\mt{max}} = \frac{L^{2}}{\delta} -  \frac{k}{4}\,\delta + \frac{r_h^2 (r_h^2+k\,L^2) }{8 L^6}\,\delta^{3} + \mathcal{O}(\delta^{5}) \, .
\eeq
Setting $r_h=0$ in the above expression yields the cutoff for the vacuum spacetime, but in accord with eq.~\reef{differ}, we see the difference is $O(\delta^3)$. To evaluate the tortoise coordinate \reef{tortoise}, we use eq.~\eqref{eq:factorize}
to first write:
\begin{equation}
\frac{1}{f(r)} =
\frac{L^2\, r_h}{2 \left(2 r_h^2+k\,L^2\right)(r-r_h)}
-\frac{L^2 }{ 2 r_h^2+k\,L^2}
	\left( \frac{r_h}{2(r+r_h)} - \frac{ r_h^2+k\,L^2}{ r^2+r_h^2+k\,L^2} \right) \, .
\end{equation}
Eq.~\reef{tortoise} then yields
\begin{equation}\label{rstarEqD5}
r^{*}(r) =\frac{L^2\sqrt{r_h^2+k\,L^2}}{ \left(2 r_h^2+k\,L^2\right)}  \, \tan ^{-1}\!\left[ \frac{r}{\sqrt{r_h^2+k\,L^2}}\right] + \frac{L^2\, r_h}{2 \left(2 r_h^2+k\,L^2\right)} \,\log{\frac{| r-r_{h} |}{r+r_h}} \, ,
\end{equation}
which leads to:\footnote{Note that we have chosen an (arbitrary) integration constant in eq.~\reef{rstarEqD5} but this choice cancels in the difference $v_\infty-r^*(r)$ appearing, \eg in eq.~\reef{Cform1a}.}
\begin{equation}\label{vinfD5}
v_{\infty} =  \frac{\pi  L^2}2\,\frac{ \sqrt{r_h^2+k\,L^2}}{ 2 r_h^2+k\,L^2} \, .
\end{equation}
For $k=+1$, it is straightforward to substitute $r_h=0$ into the above expressions to recover the vacuum results, \ie $r_0^*(r)$ and $v'_{\infty}$ as given in eqs.~\reef{TortoiseEmpty} and \reef{vinfsphempt}. Unfortunately, this substitution is more subtle for $k=0$ and $-1$ but one can calculate the desired quantities directly. From appendix \ref{App:Vac}, the results are
\beqa \label{TortoiseVacuum}
k=+1\ :&\qquad r^{*}_0(r) = L \ \tan ^{-1}\! \left({r}/{L}\right) \, ,&\ \
v'_{\infty} =  {L\,\pi}/{2} \,,
\nonumber\\
k=0\ :&\qquad r_0^*(r) = -{L^2}/{r}\, ,\quad\ \qquad&\ \ v'_\infty=0\,,
\label{vacvac}\\
k=-1\ :&\qquad r^{*}_0(r) = \frac{L}{2}\, \log\!  \frac{  |r-L| }{r+L}\, ,\qquad
&\ \ v'_\infty=0\,.
\nonumber
\eeqa

Now it is straightforward to evaluate the expressions in eq.~\reef{Cform1a}:
\beqa\label{BulkActiond4k}
\Delta I_{\textrm{bulk}} &=  &
 - \frac{\Omega_{k,3}}{4G_N}\left[ \frac{ (r_h^2 +k\, L^2)^{5/2}}{ 2 r_h^2 +k\, L^2}  - L^3\,\delta_{k,1} \right] \, ,\\
I^{\BHx}_{\text{GHY}}
 &=& \frac{\Omega_{k, 3}  }{2 \, G_N}\,  \frac{r_h^2 \, (r_h^2 +k\, L^2)^{3/2} }{2 r_h^2 +k\, L^2} \, .
\eeqa
Combining these results in eq.~\reef{Cform1} then yields
\begin{equation}\label{TotalAction5Spherical}
\Delta\mC =
\frac{\Omega_{k,3}}{4\pi G_N} \left[ \, \frac{ (r_h^2+k L^2)^{3/2} \left(r_h^2-k L^2	\right)}{\left(2 r_h^2+k L^2\right)} +
L^3\,\delta_{k,1}    \right]\,.
\end{equation}
With an expansion in large horizon radius, this result becomes
\begin{equation}
\Delta\mC = \frac{\Omega_{k, 3}\,L^3 }{8\pi G_N}\,\left[\frac{r_h^3}{L^3} + 2\delta_{k,1}-\frac{9 \,  k^2 }{8 }\,\frac{L}{r_h} +\frac{k^3 }{8}\,\frac{L^3}{r_h^3} + \mathcal{O}(L^5/r_h^{5}) \right] \, . \label{largeR}
\end{equation}
or expressed as a function of entropy \eqref{eq:EntropyBH}:
\begin{equation}\label{EntFormD5}
\Delta\mC =  \frac{S}{2\pi} + \frac{\Omega_{k, 3}\,\pi^2 }{20}\,C_T\,\delta_{k,1} -9\pi^3 k^2  \left(\frac{\Omega_{k, 3}}{160}\right)^{4/3} \frac{C_T^{4/3}}{S^{1/3}}+ \pi^{5} k\left( \frac{\Omega_{k,3}}{80}   \right)^{2} \frac{C_T^2}{S}+\mathcal{O}(S^{-5/3}) \, .
\end{equation}
where we have introduced the central charge in the boundary theory \cite{hologb}: $C_T=\frac{5}{\pi^3}\,\frac{L^3}{G_N}$. Hence we see that to leading order in this large entropy expansion (\ie implicitly a high temperature expansion), the complexity of formation grows linearly with the entropy.
Further, eq.~\reef{EntFormD5} shows that this expansion is an expansion for large values of $S/C_T$. Finally, the coefficient of the leading behavior in $\Delta\mC$ is independent of the spatial geometry. In section \ref{plane}, we derive an analytic expression for this leading coefficient as a function of the boundary dimension $d$.

The ``small'' hyperbolic black holes are discussed in detail in appendix \ref{app:Hypers}. Using the results presented there, eq.~\reef{Cform2} yields the following complexity of formation
\begin{align}\label{car2}
\begin{split}
\Delta\mC = & - \frac{ \,  \Omega_{-1,3}}{4 \pi^2 G_N }
\left(
r_2 \,\frac{r_{\text{meet}}^4-r_2^4}{2r_h^2-L^2} \log \left[\frac{r_{\text{meet}}+r_2}{r_{\text{meet}}-r_2}\right]
+ r_h \,\frac{r_h^4-r_{\text{meet}}^4}{2r_h^2-L^2} \log \left[\frac{r_h+r_{\text{meet}}}{r_h-r_{\text{meet}}}\right]
\right.
\\
&
\left.\qquad\qquad\quad - \frac{2}3\, r_{\text{meet}} \left(3 L^2+r_{\text{meet}}^2\right)+ r_{\text{meet}}^{3} \, \log{|f(r_{\text{meet}})|}\right)
\, .
\end{split}
\end{align}
for small hyperbolic black holes, with
\begin{equation}
r_2=\sqrt{L^2 - r_h^2}\qquad{\rm and}\qquad
r^{*}(r_{\text{meet}})
= 0\, .
\end{equation}
Here, $r_2$ is the second root of $f(r)=0$, which defines the position of the inner horizon --- see figure \ref{PenroseInOut}. Further, $r_{\text{meet}}$ is the coordinate radius where the null surfaces from the left and right asymptotic regions meet behind the horizon. Since $r_{\text{meet}}$ is the solution to a transcendental equation, evaluating the expression in eq.~\reef{car2} requires some numerical treatment. Finally, as we mentioned above, this result is also ambiguous --- see further discussion in section \ref{discuss} and appendix \ref{AppendixC}.

\begin{figure}
\centering
\includegraphics[scale=0.8]{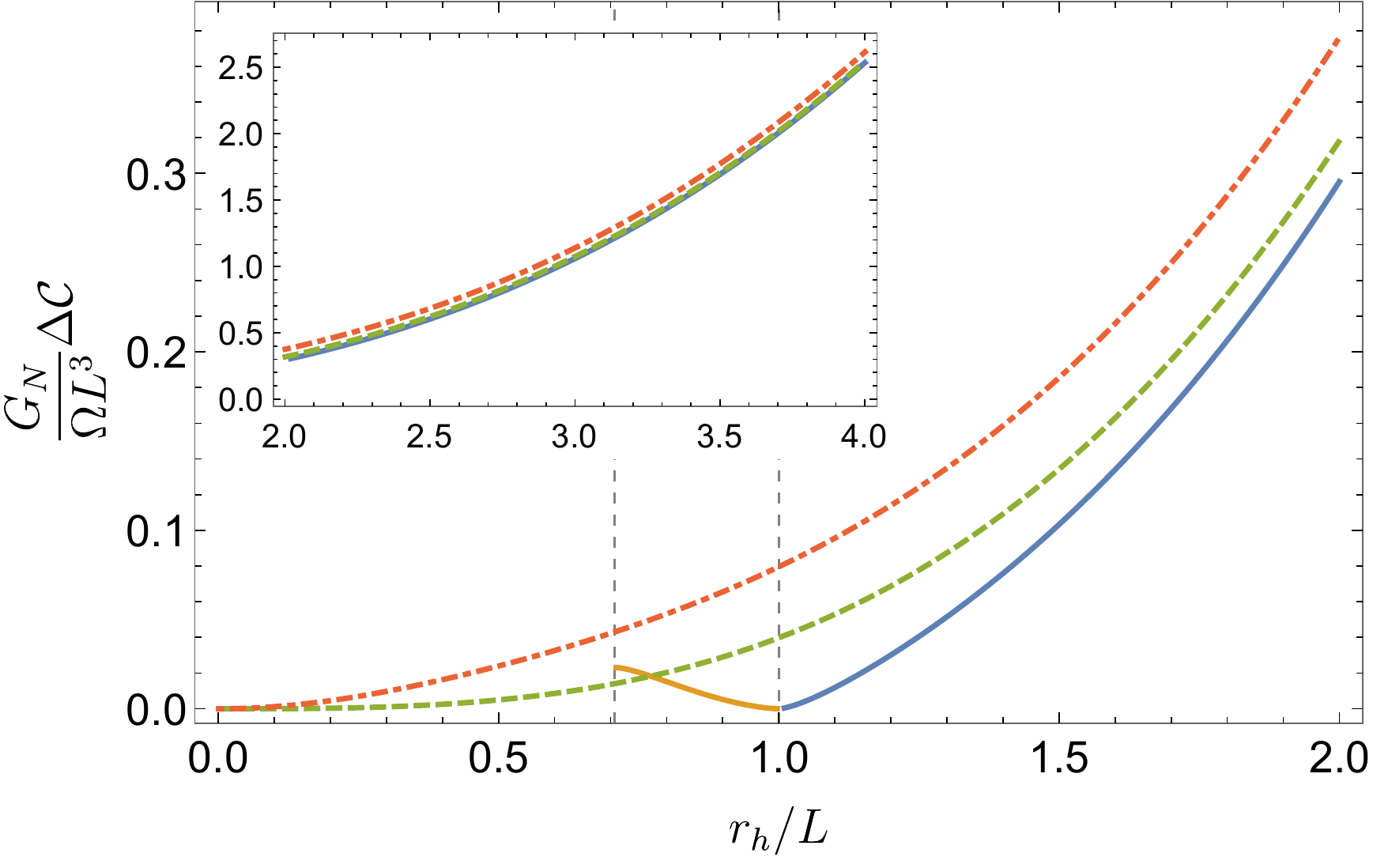}
\caption{Complexity of formation for the different geometries in four boundary (five bulk) dimensions: large hyperbolic (blue), small hyperbolic (orange), planar (dashed green) and spherical (dot-dashed red). In the inset, a larger range of horizon radii is presented demonstrating that the leading behavior at large $r_h$ is the same for the three different horizon geometries.
The two vertical dashed lines are: (1) $r_h=L/\sqrt{2}$, where the (small) hyperbolic black holes become extremal; (2) $r_h=L$, where the Hawking-Page phase transition takes place for the spherical black holes (planar and hyperbolic black holes do not admit a similar transition).}
\label{CompareAdS5}
\end{figure}

Figure \ref{CompareAdS5} summarizes the results of this subsection.

\subsection{$d=3$}\label{sec:AdS4}

For $d=3$ (four-dimensional bulk), the blackening factor  \eqref{BlackeningFactor} becomes
\begin{equation}\label{SphericalHorizonFunction4}
f(r) = \frac{r^2}{L^2} + k -
\frac{r_h}{r}\left(\frac{r_h^2}{L^2} + k\right)  \, ,
\end{equation}
and as before for vacuum AdS, we have $f_0(r)$ in eq.~\eqref{EmptyBlack}. In appendix \ref{App:FeGra}, the UV cutoff surface is set at
\beq
r_{\mt{max}} = \frac{L^{2}}{\delta} -  \frac{k}{4}\,\delta + \frac{r_h}{6 L^2}\,\left(\frac{r_h^2}{L^2}+k\right)  \, \delta^{2} + \mathcal{O}(\delta^{4}) \, ,
\eeq
which also fixes the vacuum cutoff with $r_h =0$.
To evaluate the tortoise coordinate \reef{tortoise}, we first use eq.~\eqref{eq:factorize} to write
\begin{equation}
\frac{1}{f(r)} =\frac{L^2 r_h}{\left(r-r_h\right) \left(3 r_h^2+k L^2\right)}
+
\frac{L^2 \left(r_h^2-r r_h+k L^2\right)}{\left(r^2+r r_h+r_h^2+k L^2\right) \left(3 r_h^2+k L^2\right)} \, .
\end{equation}
Eq.~\reef{tortoise} then yields\footnote{Here and below, we assume $r_h>{2L}/{\sqrt{3}}$ for the hyperbolic case with $k=-1$. In the range $L\le r_h\le{2L}/{\sqrt{3}}$ , $f(r)$ has two additional negative real roots. While these do not signify the presence of  additional horizons, this case has some similarities to that of small hyperbolic black holes, \ie $r_h<L$. Hence it will be treated separately in appendix \ref{app:Hypers}.}
\begin{equation}\label{rstarEqD4}
r^{*}(r) =\frac{L^2\,r_h}{3 r_h^2+k L^2} \, \left(\log\! \left[\frac{\bigl| r-r_h \bigl| }{\sqrt{r^2+r r_h+r_h^2+k L^2}}\right]+\frac{3 r_h^2+2 k L^2 }{r_h \sqrt{3 r_h^2+4 k L^2}}\,\tan ^{-1}\!\left[\frac{2 r+r_h}{\sqrt{3 r_h^2+4 k L^2}}\right]\right) \, ,
\end{equation}
which in eq.~\reef{eq:vinf} yields
\begin{equation}\label{vinfD4}
v_{\infty} = \frac{\pi  L^2 \left(3 r_h^2+2 k L^2\right)}{2 \left(3 r_h^2+k L^2\right) \sqrt{3 r_h^2+4 k L^2}} \, .
\end{equation}
The vacuum results, $r_0^*(r)$ and $v'_{\infty}$, are identical to those shown in eq.~\eqref{TortoiseVacuum}.

Using these results to evaluate the expressions in eq.~\reef{Cform1a}, the complexity of formation \reef{Cform1} becomes
\beqa
\Delta\mC &=&   \frac{\Omega_{k,2} }{8 \pi^2  G_N \left(3 r_h^2+k L^2\right) \sqrt{3 r_h^2+4 k L^2}} \nonumber\\
&&\ \ \times\ \bigg[ 2 r_h \left(3 r_h^4+5 k L^2 r_h^2+4 k^2 L^4\right)\left( \frac{\pi}2-\tan ^{-1}\!\left[\frac{r_h}{\sqrt{3 r_h^2+4 k L^2}}\right]\right) +  \label{TotalAction4Spherical} \\
&&\sqrt{3 r_h^2+4 k L^2} \left(\left(r_h^4-3 k L^2 r_h^2-2 k^2 L^4\right) \log \left[\frac{r_h^2}{L^2}+k\right]-2 \left(r_h^4+3 k L^2 r_h^2\right) \log \frac{r_h}{L}\right)  \bigg] \, .  \nonumber
\eeqa
An expansion in large $r_h/L$ then yields
\begin{equation}
\Delta\mC = \frac{\Omega_{k, 2}\,L^2 }{8\pi^2 G_N}\,\left[ \frac{2\pi}{3\sqrt{3}} \frac{r_h^2}{L^2} - 4 \, k\, \log{\frac{r_h}{L}}  + \frac{2 \, k \, (9 + 2 \sqrt{3} \pi)}{27 } - \frac{4 \, k^2}{27 } \left(9 -\sqrt{3} \pi \right) \frac{L^2}{r_h^2} + \mathcal{O}(r_h^{-4})  \right]  \, ,\label{largeR2}
\end{equation}
or alternatively, an expansion for large entropy \eqref{eq:EntropyBH} gives
\begin{align}
&\Delta\mC = \frac{S}{3 \sqrt{3}\pi}- \frac{ k \pi\, \Omega_{k,2} }{12}\,C_T \, \log{\left[ \frac{12 }{\Omega_{k,2} \, \pi^3} \,\frac{S}{ C_T} \right]} +
 \label{EntFormD4} \\
&\qquad\qquad+ \frac{k \pi \, \Omega_{k,2}}{324}\,C_{T}  \left(9+2 \sqrt{3} \pi \right)- \frac{k^2 \, \pi^4 \, \Omega_{k,2}^{2}}{1944}\,\frac{C_T^2 }{S}\, (9-\sqrt{3} \pi)+\mathcal{O}(S^{-2}) \, ,    \nonumber
\end{align}
where we used $C_T=3L^2/(\pi^3 G_N)$ for the boundary central charge. As in the previous case, the coefficient of the leading order term matches with the general $d$ argument in section \ref{plane}.

The ``small'' hyperbolic black holes for $d=3$ are discussed in detail in appendix \ref{app:Hypers}, and the complexity of formation is given by
\begin{equation}
\begin{split}
&\Delta\mC=
\frac{\Omega_{-1,2}  }{4 \pi^2  G_N (r_2-r_3) (r_h-r_2) (r_h-r_3)}
\left[ 2 r_h \,(r_2-r_3) \left(r_h^3-r_{\text{meet}}^3\right) \log \left(\frac{r_h-r_{\text{meet}}}{L}\right)
\right.
\\
&\left.
+2 r_2\, (r_h-r_3)\left(r_{\text{meet}}^3-r_2^3\right)  \log \left(\frac{r_{\text{meet}}-r_2}{L}\right)-2 r_3\, (r_h-r_2) \left(r_{\text{meet}}^3-r_3^3\right) \log \left(\frac{r_{\text{meet}}-r_3}{L}\right)
 \right]
\\
&
+\frac{\Omega_{-1,2}  }{4 \pi^2  G_N }\Big[r_{\text{meet}}\,(2r_h+2 r_2+2r_3+r_{\text{meet}})
-  r_{\text{meet}}^{2} \, \log{|f(r_{\text{meet}})|}\Big]
\end{split}
\end{equation}
where
\begin{equation}
r_2 = \frac{1}{2} \left(\sqrt{4 L^2-3 r_h^2}-r_h\right)\quad{\rm and}
\quad
r_3 = - \frac{1}{2} \left(\sqrt{4 L^2-3 r_h^2}+r_h\right)\,.
\end{equation}
Here $r_2$ denotes the second positive root of $f(r)=0$, which specifies the position of the inner horizon, while $r_3$ is a third real but negative root (which does not correspond to the location of a horizon).
As before, $r_{\text{meet}}$ is the radius of the meeting point of the null surfaces behind the horizon, which satisfies $r^{*}(r_{\text{meet}})= 0$. 

\begin{figure}
\centering
\includegraphics[scale=0.8]{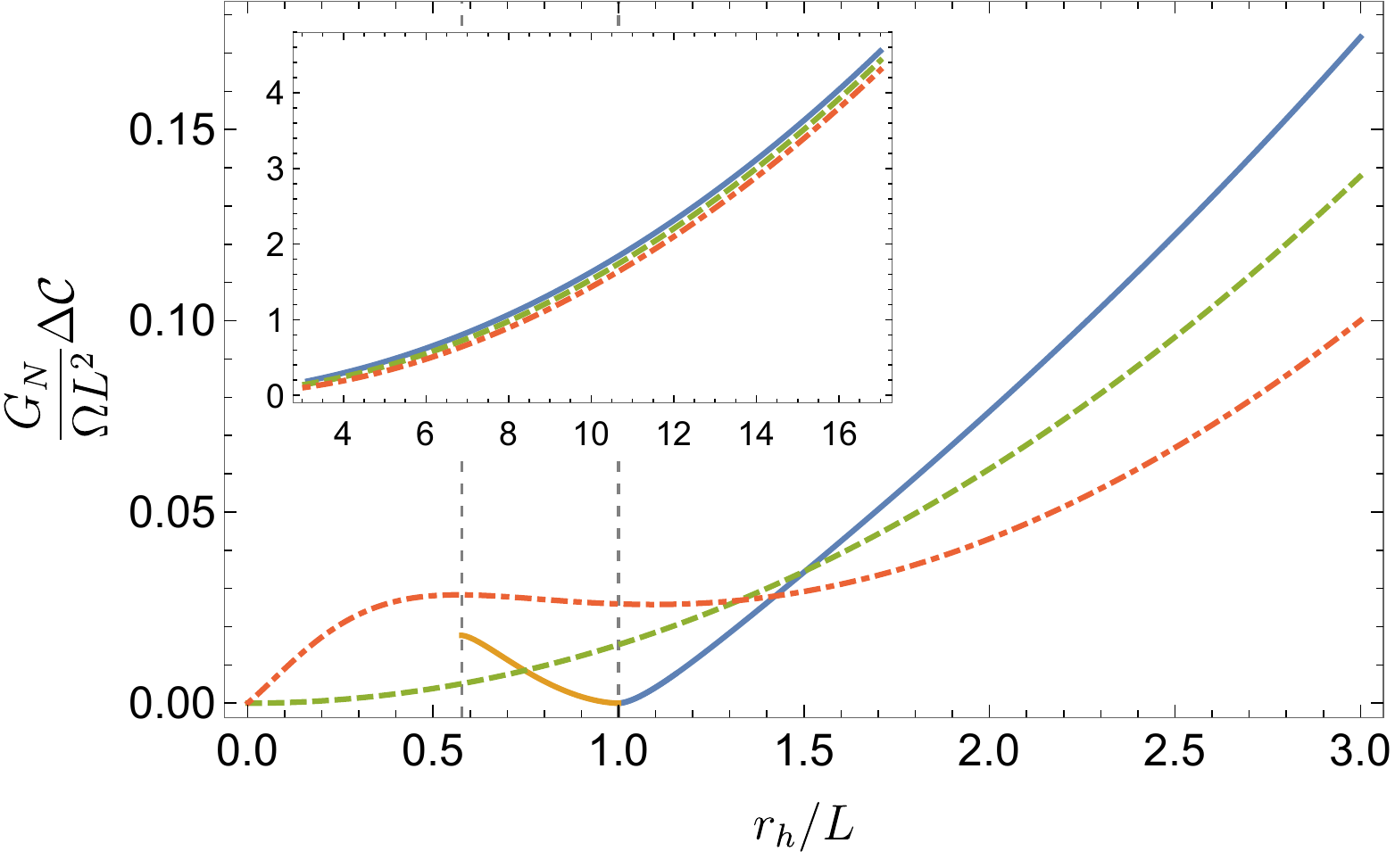}
\caption{Complexity of formation for the different geometries in three boundary (four bulk) dimensions: large hyperbolic (blue), small hyperbolic (orange), planar (dashed green) and spherical (dot-dashed red). In the inset, a larger range of horizon radii is presented demonstrating that the leading behavior at large $r_h$ is the same for the three different horizon geometries.
The two vertical dashed lines are: (1) $r_h=L/\sqrt{3}$, where the (small) hyperbolic black holes become extremal; (2) $r_h=L$, where the Hawking-Page phase transition takes place for the spherical black holes (planar and hyperbolic black holes do not admit a similar transition).}
\label{CompareAdS4}
\end{figure}

We show the results of this subsection in figure \ref{CompareAdS4}.

\subsection{Planar Case for General $d$}\label{plane}

In the previous subsections, we saw that our results for $d=4$ and $3$ reduce to a single term proportional to $r_h^{d-1}$ for the planar black holes. Therefore, up to an overall coefficient, the complexity of formation is given by the horizon entropy in these cases. Further, for hyperbolic and spherical black holes this same term appears as the leading behavior for large black holes, irrespective of the sign of $k$.
In this section, we compute the complexity of formation for planar black holes in general dimension ($d>2$) and find a similar result, \ie $\Delta\mC = k_d\,S$ where the proportionality constant $k_d$ is a relatively simple function of the dimension $d$.

\begin{figure}
\centering
\includegraphics[scale=0.65]{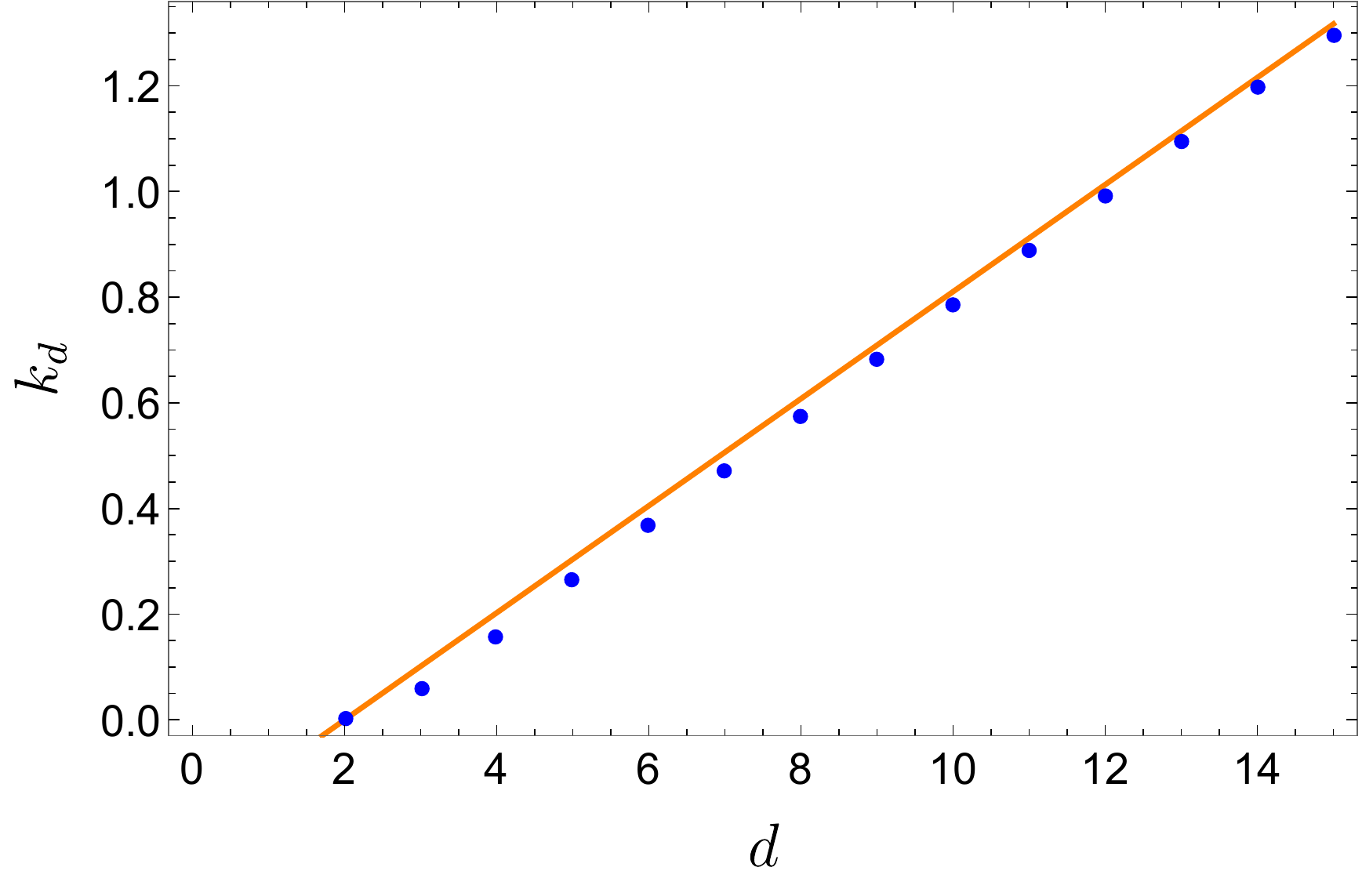}
\caption{Coefficient of entropy in eq.~\reef{PlanarGenerald}, $k_d\equiv\Delta\mC/S$, plotted as a function of the boundary dimension $d$, for planar AdS black holes. The orange line shows the linear approximation in eq.~\reef{line}.}
\label{coefgend}
\end{figure}

With $k=0$, the blacking factor \eqref{BlackeningFactor} reduces to
\begin{equation}
f(r) = \frac{r^2}{L^2}-\frac{r_h^d}{L^2 r^{d-2}} \,,
\end{equation}
and for vacuum AdS, we have $f_0(r)=r^2/L^2$. Following the calculations from appendix \ref{App:FeGra}, the UV cutoff surface is
\begin{equation}
r_{\mt{max}}=\frac{L^2}{\delta} + \frac{ r_h^{d}\ \delta ^{d-1} }{2 d\, L^{2 (d-1)}}+O\left(\delta^{d+1}\right)\,,
\end{equation}
while the cutoff for vacuum AdS is given by taking the limit $r_h\rightarrow 0$ in this expression.
As usual, the tortoise coordinate is given by integrating $1/f(r)$ with the result\footnote{Note that the inverse blackening factor can be decomposed as follows:
\begin{equation}
\frac{1}{f(r)} = \frac{L^2}{d \, r_h} \left( \frac{1}{r-r_h} + \frac{-r^{d-2}  + \sum_{n=0}^{d-3}(n+1) r^{n} r_h^{d-2-n} }{\sum_{m=0}^{d-1} r^{m} r_h^{d-1-m}} \right) \, ,
\end{equation}
where all of the terms in the sum are regular at the location of the horizon and the first term leads to a contribution in the tortoise coordinate of the form $r^*(r)=L^2 /(d r_h) \log|r-r_h|+\cdots$.}
\begin{align}\label{tortoise_planar}
\begin{split}
&r_{\text{in}}^{*}(r) = \frac{L^2}{r} \left[ \, _2F_1\left(1,-\frac{1}{d};1-\frac{1}{d};\left(\frac{r}{r_h}\right)^d\right) - 1\right] \, ,  \\
&r_{\text{out}}^{*}(r)=\frac{L^2}{r_h} \left[\frac{\pi}{d}  \cot \left(\frac{\pi }{d}\right)-
\left(\frac{r_h^d}{r^d-r_h^d}\right)^{1/d} \, _2F_1\left(\frac{1}{d},\frac{1}{d};1+\frac{1}{d};\frac{r_h^d}{r_h^d-r^d}\right)\right]\, .
\end{split}
\end{align}
where the subscripts in/out indicate inside ($r<r_h$) and outside ($r>r_h$) of the horizon, respectively. Note that we have to impose that the nonlogarithmic part of these functions is continuous across the horizon, \ie
\begin{equation}
\lim_{r\rightarrow r_h^+} \left(r_{\text{out}}^{*}(r)-\frac{L^2}{d r_h} \log(r-r_h)\right)=
\lim_{r \rightarrow r_h^-} \left(r_{\text{in}}^{*}(r)-\frac{L^2}{d r_h} \log(r_h-r)\right)
\end{equation}
 to fix the relative integration constant between the two hypergeometric functions. This condition is responsible for the appearance of the constant $(L^2\pi)/(d r_h)\,\cot(\pi/d)$ in $r_{\text{out}}^{*}(r)$. This also yields:
\begin{equation}
v_{\infty}= \frac{\pi L^2}{d r_h}  \cot \left(\frac{\pi }{d}\right)\, .
\end{equation}
The vacuum expressions are the same as in eq.~\reef{vacvac}. Now, the bulk contribution in eq.~\eqref{Cform1a} yields
\beqa
\Delta I_{\textrm{bulk}}&=& -\frac{\Omega_{0, d-1}}{2 \pi \, G_{N}}\,\frac{d}{d-1} \Bigg[ (r^d-r_h^d)^{\frac{d-1}{d}}  \, _2F_1\left(\frac{1}{d}-1,\frac{1}{d};1+\frac{1}{d};\frac{r_h^d}{r_h^d-r^d}\right) \bigg|_{r_h}^{r_{\mt{max}}}\nonumber\\
&&\ \ + r^{d-1}
 \left[\frac{\pi(d-1)}{d^2}\,\frac{ r}{r_h }\, \cot \left(\frac{\pi }{d}\right)-  {}_2F_1\left(1,-\frac{1}{d};2-\frac{1}{d};\left(\frac{r}{r_h}\right)^d\right)+1\right]_{0}^{r_h}
- r_{\mt{max}}^{d-1} \Bigg] \nonumber \\
&=& - \frac{\Omega_{0,d-1}}{2 \, d\, G_N} \, \cot{\bigg(\frac{\pi}{d} \bigg)} \, r_h^{d-1} \, . \label{Bulk_planar}
\eeqa
The corresponding surface contribution is also easily evaluated
\begin{equation}
I^\BHx_{\text{GHY}} =\frac{\Omega_{0,d-1}}{4 \, G_N} \, \cot{\bigg(\frac{\pi}{d} \bigg)} \, r_h^{d-1} \, ,
\end{equation}
and the total complexity of formation becomes
\begin{equation}
\Delta \mathcal{C} = \frac{1}{\pi}\left[\Delta I_{\textrm{bulk}}+I^\BHx_{\text{GHY}}\right]  = \frac{d-2}{d} \,  \cot{\bigg(\frac{\pi}{d} \bigg)} \, \frac{\Omega_{0,d-1} \, r_h^{d-1}}{4  \pi \, G_N} \, .
\end{equation}
Therefore, the complexity of formation has a simple form in terms of the horizon entropy \reef{eq:EntropyBH}
\begin{equation}\label{PlanarGenerald}
\Delta\mC   = \frac{d-2}{d\,\pi} \,  \cot{\!\bigg(\frac{\pi}{d} \bigg)} \, S \equiv  k_d \, S\, .
\end{equation}
Note that for large $d$, the coefficient $k_d$ approaches a linear function of $d$, \ie
\beq
k_d\simeq \frac{d-2}{\pi^2}+\mathcal{O}(1/d)\,.\label{line}
\eeq
In figure \ref{coefgend}, we plot the coefficient $k_d$ as a function of the dimension and show that it rapidly approaches the linear approximation above. Note that $k_d$ vanishes for $d=2$. Strictly speaking, however, our calculations above only apply for $d>2$ and  $d=2$ is a special case which we discuss in the next section. Nevertheless, we will confirm there that the complexity of formation is independent of the entropy for $d=2$.

\section{Complexity of BTZ Black Holes}\label{sec:BTZ}

\begin{figure}
\centering
\includegraphics[scale=0.35]{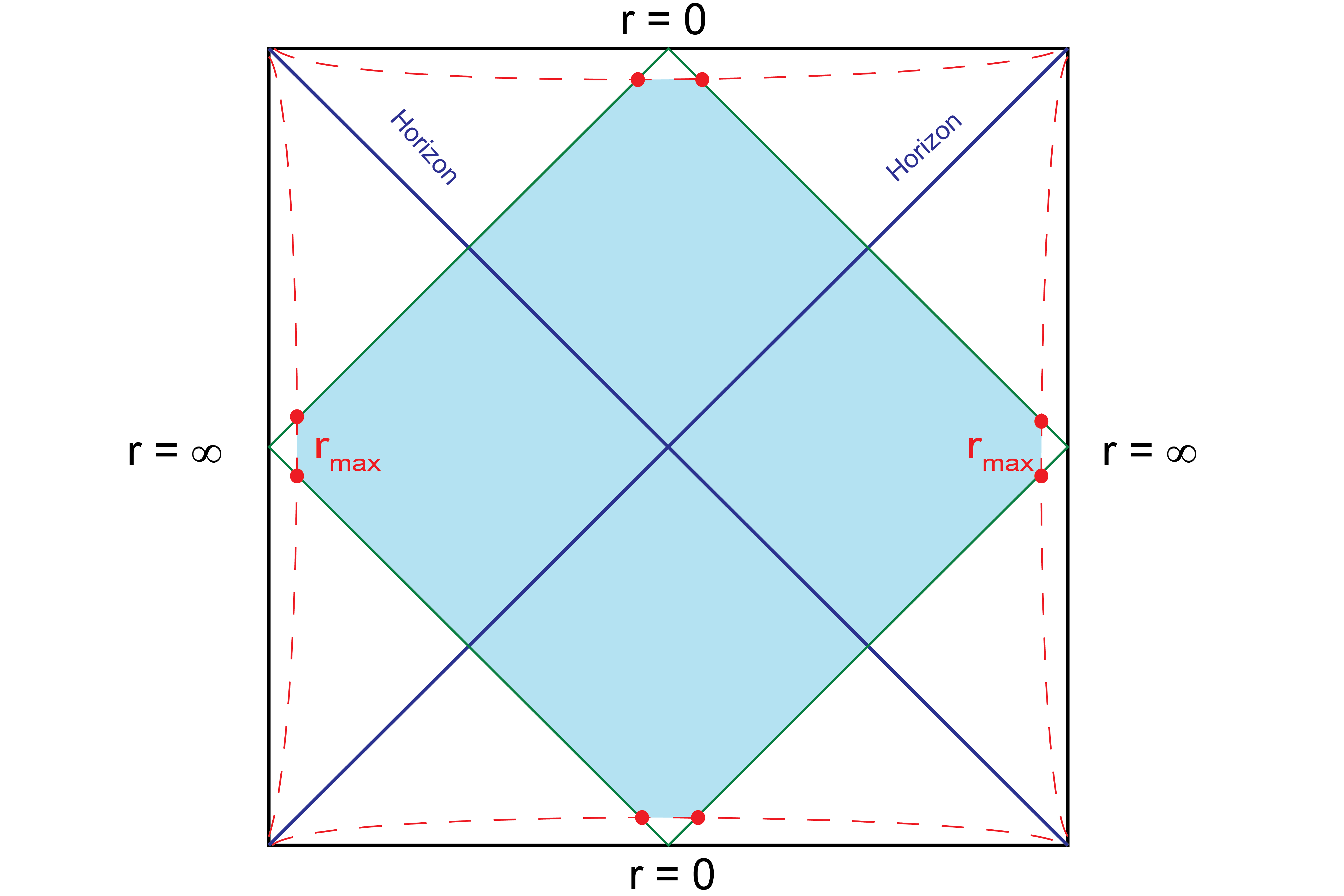}
\caption{Penrose diagram of the WDW patch in the BTZ black hole background (with zero angular momentum). Note that unlike the higher dimensional case, the null sheets originating from the $t=0$ slices on the left and right boundaries meet with each other in a joint, precisely at $r=0$.}
\label{BTZPenroseCorner}
\end{figure}

Of course, the case of two dimensions in the boundary theory is special. In this situation, the corresponding BTZ black hole \cite{btz1,btz2} can be seen as an orbifold of the vacuum AdS$_3$ solution. The corresponding calculation of the complexity of formation is slightly different from that for its higher dimensional counterparts. The main difference is that the null surfaces from the two asymptotic boundaries now meet each other  at a joint precisely on the (orbifold) singularity, instead of ending separately on the singularity.
The metric can still be written in the form given in eq.~\eqref{HigherDMetric} with $d\Sigma^2_{k,d-1}$ replaced by $d\phi^2$ and
 with the blackening factor
\begin{equation}\label{metricBTZ}
f(r) = \frac{r^{2} - r^{2}_{h}}{L^{2}}\,.
\end{equation}
For the vacuum solution, we take eq.~\reef{EmptyBlack} with $k=+1$,\footnote{We could also choose $k=-1$ or 0. However, the $k=-1$ solution is precisely the BTZ black hole with $r_h=L$ and the $k=0$ vacuum will be discussed at the end of this section.} \ie
\beq
f_0(r)=\frac{r^2}{L^2}+1\,.
\label{vac8}
\eeq
The Penrose diagram representing the BTZ black hole is shown in figure \ref{BTZPenroseCorner}. The corresponding mass, temperature and entropy are given by
\beq
M=\frac{r_h^2}{8G_NL^2}\,,\qquad T=\frac{r_h}{2\pi L^2}\,, \qquad{\rm and} \qquad S=\frac{\pi\,r_h^2}{2\,G_N}\,.
\label{props}
\eeq

The cutoff surfaces follow again from the near boundary expansion in appendix \ref{App:FeGra},
\beq\label{BTZcutoffs1}
r_{\mt{max}}^{\text{BTZ}}  =\frac{L^2}{\delta } + \frac{r_h^2}{4 L^2} \delta \quad{\rm and}\quad
r_{\mt{max}}^{\text{vac}}  =  \frac{L^2}{\delta }-\frac{\delta }{4} \, .
\eeq
The vacuum AdS calculation follows immediately from previous examples and the bulk integral \eqref{ActionBulk} becomes for $d=2$:
\begin{equation}\label{Emp3Action}
I_{\bulk}^{\vac} = -\frac{L^2}{G_N\,\delta }+\frac{\pi  L}{4G_N}+ \mathcal{O}(\delta) \, .
\end{equation}
Next, we evaluate the action for the BTZ black hole. The tortoise coordinate \reef{tortoise} is
\begin{equation}
r^{*}(r) = \frac{L^{2}}{2 \, r_{h}} \log{\frac{| r - r_{h}|}{r+r_{h}}}\, ,
\end{equation}
and from eq.~\reef{eq:vinf}, $v_\infty=0$.  The bulk integral
result is
\begin{equation}
I_{\bulk}^{\text{BTZ}} = -\frac{2L^2}{G_N\,\delta }+ \mathcal{O}(\delta) \, .
\end{equation}
Of course, the divergence cancels when subtracting twice the action of vacuum AdS. There is no contribution from a surface term near the singularity, as the null boundaries meet as described above, and the joint contribution there vanishes. Therefore, the complexity of formation is simply given by
\begin{equation}\label{cc1}
\Delta\mC = \frac{1}\pi\left[ I_{\bulk}^{\text{BTZ}} - 2I_{\bulk}^{\text{vac}}\right]= -\frac{  L}{2 G_N}=-\frac{c}3 \, ,
\end{equation}
where we have introduced the central charge of the boundary theory $c=3L/(2G_N)$ \cite{brown}.
Hence $\Delta\mC$ is a fixed constant, independent of the temperature or horizon radius.

One notable fact about the above expression is that it does not vanish when the mass (or $r_h$) vanishes. At a pragmatic level, this occurs because in the limit $r_h\to0$, the blackening factor \eqref{metricBTZ} does not become $f_0(r)$ in eq.~\reef{vac8} for the vacuum AdS$_{3}$ spacetime. Implicitly, in choosing eq.~\reef{vac8}, we are choosing  to consider the Neveu-Schwarz vacuum of the boundary theory \cite{couscous}. Alternatively, we could have chosen $f_0(r)=r^2/L^2$ (\ie the $r_h\to0$ limit of the BTZ blackening factor), but this choice would correspond to the Ramond vacuum of the boundary theory. In this case, we find that the complexity of formation vanishes, \ie $\Delta\mC=0$.

\section{Comparison with Complexity$=$Volume}\label{sec:CompVol}

In the previous sections, we were investigating the conjectured duality between complexity and action (CA) \cite{Brown1, Brown2}.
However, it was previously conjectured that the complexity of states in a holographic theory should be dual to the volume of the extremal codimension-one bulk hypersurface which meets the asymptotic boundary on the desired time slice \cite{CompVolume}.\footnote{An alternative proposal related to complexity=volume was recently put forward by \cite{willy}. See also \cite{ali,Ben-Ami:2016qex} for  proposed generalizations for subregions.}
More precisely, the  complexity $=$ volume (CV) duality states that the complexity of the state on a time slice denoted $\Omega$ is given by:
\begin{equation}\label{volver}
\mathcal{C}_{\text{V}}(\Omega) =\ \mathrel{\mathop {\rm
max}_{\scriptscriptstyle{\Omega=\partial \mathcal{B}}} {}\!\!}\left[\frac{\mathcal{V(B)}}{G_N \, \ell}\right] \, ,
\end{equation}
where $\mathcal B$ is the corresponding bulk surface and $\ell$ is some length scale associated with the bulk geometry, \eg the AdS radius for large
black holes and $r_h$ for small black holes, see, \eg \cite{Brown2}. The ambiguity in defining the latter is somewhat unsatisfactory and provided some motivation for developing the CA duality, since this choice is naturally eliminated in this framework. For simplicity, we will set $\ell=L$ in all of the following calculations. In this section, we compare our previous results for the complexity of formation obtained from the CA duality to those obtained by the CV duality.
\begin{figure}
\centering
\includegraphics[scale=0.25]{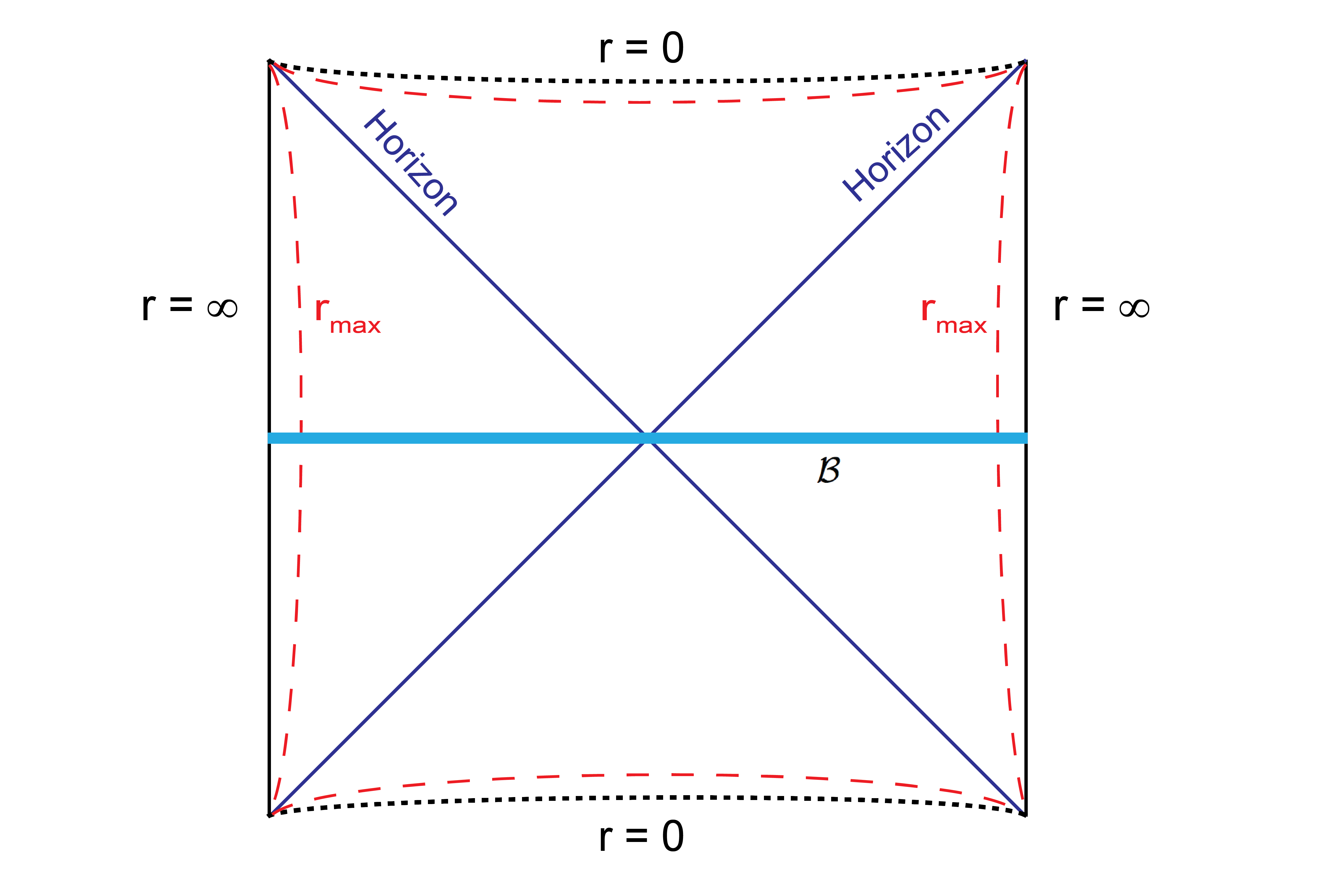}
\caption{The maximal volume slice $\mathcal{B}$ connecting the two boundaries at $t_L = t_R =0$ through the Einstein-Rosen bridge.}
\label{MaximalVol}
\end{figure}

We are interested in evaluating the complexity for the thermal state defined on the time slice at $t_L = t_R = 0$.
By symmetry, the maximal volume is given by the bulk $t=0$ slice, \ie  the straight line connecting the two boundaries through the bifurcation surface in the Penrose diagram shown in figure \ref{MaximalVol}. The volume integral then simplifies to:
\begin{equation}\label{wack1}
\mathcal{V} = 2  \Omega_{k, d-1} \int_{r_h}^{r_{\mt{max}}} \, \frac{r^{d-1}}{\sqrt{f(r)}}\, dr \, .
\end{equation}
To evaluate the complexity of formation, we will subtract from this integral, the corresponding contribution from (two copies of) the vacuum AdS background:
\begin{equation}\label{wack2}
\mathcal{V}_0 = 2  \Omega_{k, d-1} \int_{r_{\text{min}}}^{r_{\mt{max}}} \, \frac{r^{d-1}}{\sqrt{f_0(r)}}\, dr \, .
\end{equation}
Here we have introduced the minimum radius $r_{\text{min}}$ because while the integration starts at $r_{\text{min}}=0$ for $k=+1$ and 0, we must set $r_{\text{min}}=L$ for $k=-1$.
Hence in this framework, the complexity of formation becomes\footnote{Using the methods of appendix \ref{App:FeGra}, we confirmed that the difference between $r_\mt{max}$ evaluated for the vacuum AdS and the black hole backgrounds will not contribute to $\Delta\mC_{V}$ when we send $\delta \rightarrow 0$.}
\begin{equation}\label{wack0}
\Delta\mC_{V} =
\frac{2 \, \Omega_{k, d-1}}{G_N\,L} \left[
\int_{r_h}^{r_{\mt{max}}} \, \frac{r^{d-1}dr}{\sqrt{f(r)}}
-
\int_{r_{\text{min}}}^{r_{\mt{max}}} \, \frac{r^{d-1}dr}{\sqrt{f_0(r)}}
\right]\, .
\end{equation}

\subsection{Planar Geometry}

It is easiest to evaluate this expression \reef{wack0} for planar black holes with $k=0$. The volume integral \reef{wack1} can be evaluated analytically for any $d$:
\begin{equation}
\mathcal{V} =  \frac{4 \, \Omega_{0, d-1}  \,L}{d} \, r_h^{\frac{d}{2}-1} \sqrt{r^d-r_h^d} \, \, _2F_1\left(\frac{1}{2},\frac{1}{d}-\frac{1}{2};\frac{3}{2};1-\left({r}/{r_h}\right)^d\right) \, \biggr|_{r_h}^{r_{\mt{max}}}\,.
\end{equation}
The cutoff $r_{\mt{max}}$ is given in appendix \ref{App:FeGra} --- see eq.~\eqref{appeq:rmaxBH}. In the limit of a small short distance cutoff $\delta $, the volume integral  becomes
\begin{equation}\label{VacPlan}
\mathcal{V} = \frac{2 \, \Omega_{0, d-1}  \, L^{2 d -1}}{(d-1) \, \delta^{d-1}} + \frac{2 \, \Omega_{0, d-1}  \,\sqrt{\pi} \, \Gamma(-1 + \frac{1}{d})}{d \, \Gamma(-\frac{1}{2} + \frac{1}{d})} L \, r_h^{d-1} + \mathcal{O}(\delta) \, .
\end{equation}
The leading (divergent) term above is exactly canceled when subtracting the volume of the maximal slice in the vacuum AdS geometry. The complexity of formation \reef{wack0} can be written as
\begin{equation}\label{trac9}
\Delta\mC_{V} =   \frac{\sqrt{\pi}\, \Omega_{0, d-1}  }{G_N }\,\frac{\left(d-2\right)\,\Gamma(1+ \frac{1}{d})}{\left(d-1\right)\Gamma(\frac{1}{2} + \frac{1}{d})}  \, r_h^{d-1} \, .
\end{equation}

Again, this result for the complexity of formation can be expressed in terms of the entropy \reef{eq:EntropyBH}
\begin{equation} \label{VolSd}
\Delta\mC_{V} =   4\sqrt{\pi} \,\frac{\left(d-2\right)\,\Gamma(1+ \frac{1}{d})}{\left(d-1\right)\Gamma(\frac{1}{2} + \frac{1}{d})}  \, S\equiv {\tilde k}_d\,S \, .
\end{equation}
Note that in this case, the coefficient ${\tilde k}_d$ approaches a constant for large $d$, \ie
\beq
\tk_d\simeq 4+O(1/d)\,.\label{line2}
\eeq
It is interesting, of course, to compare these results to the analogous results found using the CA duality --- see eqs.~\reef{PlanarGenerald} and \reef{line}. It is perhaps notable that in both approaches, the coefficient vanishes for $d=2$. However, otherwise the coefficients $k_d$ and $\tk_d$ seem to bear little resemblance to each other. 
For example, we saw that for large $d$, the coefficient ${\tilde k}_d$ approaches a constant for the CV duality while $k_d$ grows linearly with $d$ for the CA duality. The two coefficients are roughly equal in the vicinity of $d=42$.

\begin{figure}
\centering
	\includegraphics[scale=0.7]{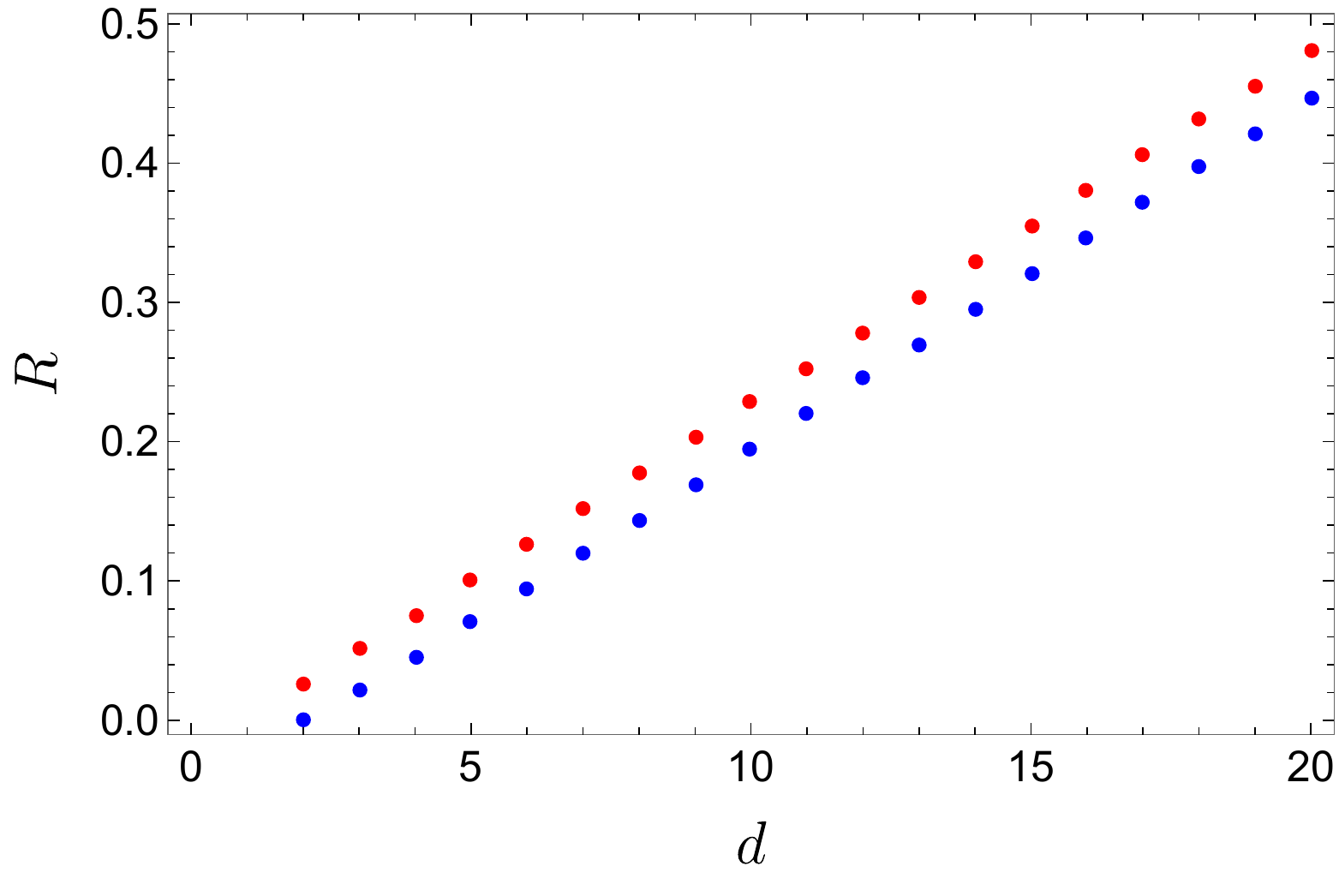}
\caption{The two ratios $R_{\text{form}}$ (blue) and $R_{\text{rate}}$ (red) as a function of $d$.}
\label{VolAct}
\end{figure}

However, one should be aware that the definition of complexity is not completely precise and different choices of, \eg the universal gate set may lead to changing the complexity of a given family of states by a multiplicative constant --- see discussions in \cite{Brown2,LuisRob}.  Hence an interesting approach is to combine the above comparison with a comparison of the late time growth of complexity from the CV and CA dualities. In particular, examining the growth of complexity for an uncharged AdS black hole using the two conjectures yields \cite{CompVolume,Brown1,Brown2}\footnote{Recall that it was suggested in \cite{Brown1,Brown2} that the late time limit of $d \mC_A/dt$ was related to Loyd's bound for the rate of computation for a system of energy $M$ \cite{Lloyd}. Recently, ref.~\cite{Yang:2016awy} considered conditions under which this bound is compatible with the CA duality conjecture.}
\begin{equation}\label{GrowthCompVol}
\frac{d \mathcal{C}_V}{d t} \bigg|_{t \rightarrow \infty} = \frac{8 \pi}{d -1} M \qquad{\rm and}\qquad \frac{d \mC_A}{dt} \bigg|_{t \rightarrow \infty}  = \frac{2 M}{\pi}\,.
\end{equation}
We note that the  late time growth rate above from the CV duality is only valid in the limit of large temperatures for $k=\pm1$ \cite{CompVolume}. Of course, our results for the complexity of formation $\Delta\mC_{A,V}$ only apply for high temperatures, as well. Now let us compare the two ratios\footnote{We have simplified the first ratio using
$\cot\! \left(\pi x\right)= \frac{\Gamma \left(1-x\right) \Gamma \left(x\right)}{\Gamma \left(\frac{1}{2}-x\right) \Gamma \left(\frac{1}{2}+x\right)}$.}
\beqa
R_{\rm form} &=& \frac{\Delta\mC_{A}}{\Delta\mC_{V}}
=\frac{d-1 }{4 \pi ^{3/2}}\,\frac{\Gamma \left(1-\frac{1}{d}\right)}{\Gamma \left(\frac{1}{2}-\frac{1}{d}\right)}
\,,\nonumber\\
R_{\rm rate} &=& \frac{d \mC_A/dt}{d \mC_V/dt}=\frac{d-1}{4\pi^2}\,.
\label{rat6}
\eeqa
Now it is straightforward to see that in the limit of large $d$, both ratios grow linearly with $d$ and further we may write
\beq\label{largeD}
R_{\rm rate}-R_{\rm form}=\frac{\log2}{2\pi^2}+{\cal O}(1/d)\,.
\eeq
However, from figure \ref{VolAct}, we can see that apart from the constant shift in eq.~\reef{largeD}, the two ratios agree very well for all values of $d$. This comparison then suggests that the two holographic approaches to complexity are more or less consistent up to an overall multiplicative factor.

\subsection{Spherical and Hyperbolic Geometries}

Here we evaluate the complexity of formation \reef{wack0} using the CV duality for the spherical and hyperbolic black holes. In the following, it is convenient to define the dimensionless coordinate $x \equiv r/L$, as well as $x_h \equiv r_h/L$ and $x_{\rm min}=0$ or 1 for $k=+1$ or $-1$, respectively. Then eq.~\reef{wack0} may be expressed as
\begin{equation}\label{romd}
\Delta\mC_{V} = 2 \, \Omega_{k, d-1}\,\frac{L^{d-1}}{G_N} \left[ \int_{x_h}^{\infty}   \left( \frac{x^{d-1}}{\sqrt{f(x)}} - \frac{x^{d-1}}{\sqrt{x^2+k}}\right) \, d x -   \int_{x_{\rm min}}^{x_h}  \frac{x^{d-1}\, dx}{\sqrt{x^2+k}}  \, \right]  \, ,
\end{equation}
where $f(x)$ is the usual blackening factor \eqref{BlackeningFactor}, \ie $f(x)=x^2+k-(x_h/x)^{d-2}(x_h^2+k)$. Note that in extending the upper limit of integration to infinity in the first integral, we have allowed for the cancellation of the divergences which appear individually in eqs.~\reef{wack1} and \reef{wack2}.
Again, the small hyperbolic black holes are a special case and eq.~\reef{romd} must be modified slightly in this situation since with $\omega^{d-2}<0$, one finds that $x_h<x_{\rm min}=1$. Hence for small hyperbolic black holes, we write instead
\begin{equation}\label{rome}
\Delta\mC_{V} = 2 \, \Omega_{-1, d-1}\,\frac{L^{d-1}}{G_N} \left[ \int_{1}^{\infty}   \left( \frac{x^{d-1}}{\sqrt{f(x)}} - \frac{x^{d-1}}{\sqrt{x^2-1}}\right) \, d x +   \int^1_{x_h} \frac{x^{d-1}\, dx}{\sqrt{f(x)}}  \, \right]  \, ,
\end{equation}

\begin{figure}[t]
\centering
\includegraphics[scale=0.8]{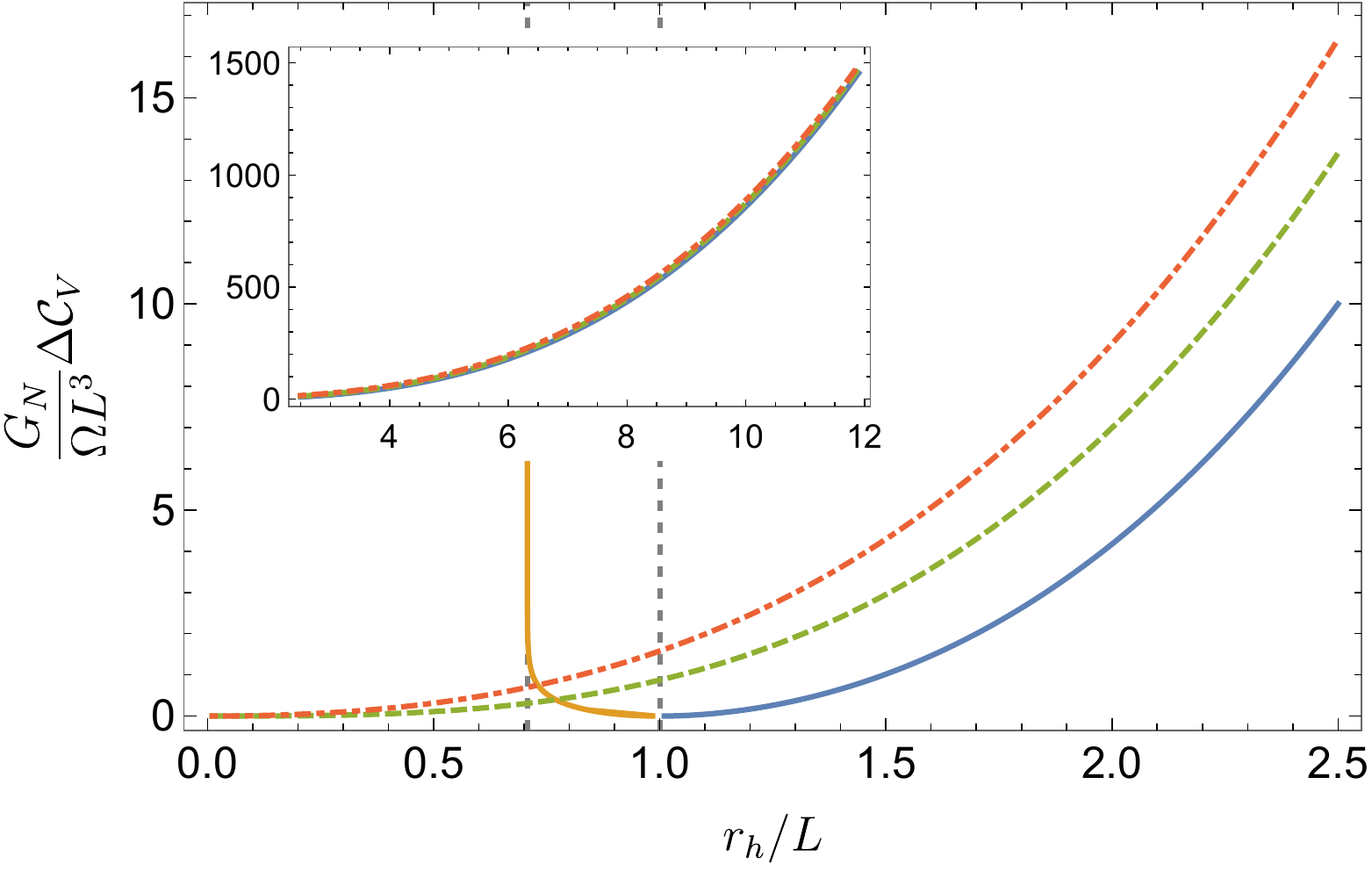}
\caption{Complexity calculated from the CV duality for the three different geometries in $d=4$ -- large hyperbolic (blue), small hyperbolic (orange), planar (dashed green) and spherical (dot-dashed red). In the inset, a larger range of horizon radii is presented demonstrating that the leading behavior at large $r_h$ is the same for the three different horizon geometries.  The  dashed vertical line at $r_h=L$ denotes the position of the Hawking-Page transition for the spherical black holes, while the one at $r_h=L/\sqrt{2}$ indicates where the (small) hyperbolic black holes become extremal. The volume, and hence the complexity of formation, diverges for these extremal black holes. }
\label{Cvol4}
\end{figure}
\begin{figure}[t]
\centering
\includegraphics[scale=0.8]{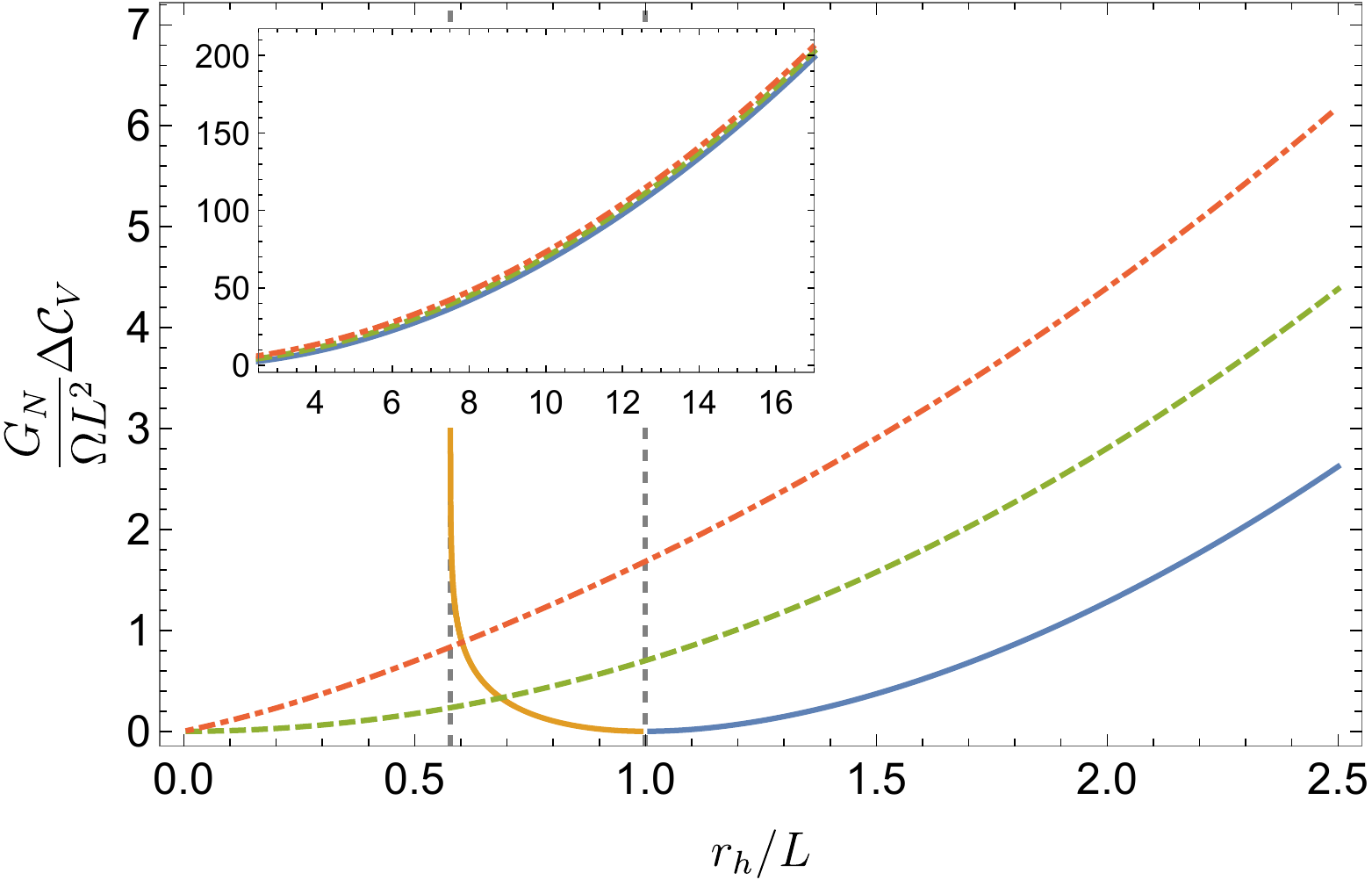}
\caption{Complexity calculated from the CV duality for the three different geometries in $d=3$ -- large hyperbolic (blue), small hyperbolic (orange), planar (dashed green) and spherical (dot-dashed red).
In the inset, a larger range of horizon radii is presented demonstrating that the leading behavior at large $r_h$ is the same for the three different horizon geometries. The  dashed vertical line at $r_h=L$ denotes the position of the Hawking-Page transition for the spherical black holes, while the one at $r_h=L/\sqrt{3}$ indicates where the (small) hyperbolic black holes become extremal. The volume, and hence the complexity of formation, diverges for these extremal black holes. }
\label{Cvol3}
\end{figure}
The above expressions can be evaluated in terms of elliptic integrals (at least for certain dimensions), however, these are not particularly enlightening.  Therefore, we evaluate these expressions numerically instead to study their behavior. Figures \ref{Cvol4} and \ref{Cvol3} show the results for $\Delta\mC_{V}$ in $d=4$ and $d=3$, respectively. There are many features found in common with the CA results shown in figures \ref{CompareAdS5} and \ref{CompareAdS4}, but there are some differences as well.

For instance, for large horizon radius, the result for the spherical and hyperbolic geometries approaches eq.~\reef{trac9} for the planar case, as expected, \ie $\Delta\mC_{V}\simeq \tk_d\,S$ as in eq.~\reef{VolSd}. We can subtract this leading behavior and fit the residual numerical result to find
\beqa
\begin{split}
&d=4,\quad k=+1:  \qquad\ \, \frac{G}{\Omega_{1,3} L^3}\left(\Delta\mC_{V}- \tk_4\,S\right) =\ \
1.55 \,\frac{r_h}{L}-1.33+0.55\frac{L}{r_h}+ \cdots\,,
\\
&d=4,\quad k=-1:  \qquad \frac{G}{\Omega_{-1,3} L^3}\left(\Delta\mC_{V}- \tk_4\,S\right) =
-1.55 \,\frac{r_h}{L}\, +0.00+ 0.55\frac{L}{r_h} +\cdots\,,
\label{gummybear}\\
&d=3,\quad k=+1: \qquad\ \ \frac{G}{\Omega_{1,2} L^2}\left(\Delta\mC_{V}- \tk_3\,S\right) =\ \ 1.00
 \log\!\left(\frac{r_h}{L}\right)+0.865 +0.14\frac{L^2}{r_h^2}+\cdots\,,
\\
&d=3,\quad k=-1:  \qquad\frac{G}{\Omega_{-1,2} L^2}\left(\Delta\mC_{V}- \tk_3\,S\right) =
\,-1.00\log\!\left(\frac{r_h}{L}\right) -0.865+ 0.14\frac{L^2}{r_h^2}+ \cdots\,.
\end{split}
\eeqa
We observe that in many respects, the structure here is very similar to that found in eqs.~\reef{largeR} and \reef{largeR2} for the CA duality. For example, there are clearly factors of $k$ multiplying the various terms; a special $\delta_{k,1}$ constant contribution appears in $d=4$;  and a logarithmic contribution appears in $d=3$. Note, however, that the first term for $d=4$ is proportional to $r_h/L$ above, whereas the term at this order vanishes for $\Delta\mC_A$ in eq.~\reef{largeR}. Further, note that the first term for $d=3$ above seems to be $k\,\log(r_h/L)$ whereas the first subleading contribution in eq.~\reef{largeR2} has the same form but the opposite sign. As a result, the curves in figure \ref{CompareAdS4} cross in the vicinity of $r_h/L\sim 1.4$, but no such crossing appears in figure \ref{Cvol3}.

Another interesting difference is that in both figures, $\Delta\mC_{V}$ diverges as $T$ approaches zero for $k=-1$, \ie as the small hyperbolic black holes approach the extremal limit.\footnote{Note that choosing $\ell=r_h$ (rather than $\ell=L$) in eq.~\reef{volver} for these `small' hyperbolic black holes does not remedy this divergence, since $r_h$ remains finite in the extremal limit, \eg $r_h = L/\sqrt{2}$ for $d=4$.} This divergence arises because the throat of the black hole grows to have infinite (proper) length in this limit.  In contrast with the CA duality, $\Delta\mC_{A}$ remains finite in this limit, but recall that the results for small hyperbolic black holes are ambiguous in this approach --- see appendix \ref{AppendixC}.

\subsection*{Special Case of $d=2$:}
Recall that $d=2$ is a special case, which is described by the BTZ black hole
in the bulk.  In this case, it is possible to evaluate the complexity of formation for the CV duality analytically.  We use the blackening factors $f(r)$ and $f_0(r)$ given by  eqs.~\eqref{metricBTZ} and \eqref{vac8}, respectively, as well as $r_{\mt{max}}$ given by eq.~\eqref{BTZcutoffs1} and $r_{\text{min}}=0$. The required volumes in eqs.~\eqref{wack1} and \eqref{wack2} are then given by
\begin{equation}
\mathcal{V} = \frac{4 \pi L^3}{\delta} +{\cal O} (\delta) \, , \qquad
\mathcal{V}_0
=\frac{4 \pi L^3  }{\delta }-4\pi L^2
+{\cal O} (\delta) \, .
\end{equation}
Hence the complexity of formation for the BTZ black hole becomes
\begin{equation}\label{eq:dcvBTZ}
\Delta \mathcal{C}_{V} = \frac{4 \pi L}{G_{N}} = \frac{8 \pi}{3}\, c \,,
\end{equation}
where $c=3L/(2G_N)$ is the boundary central charge. As before, we are implicitly considering the Neveu-Schwarz vacuum in choosing $f_0(r)$ in eq.~\reef{vac8}. If instead we consider the Ramond vacuum with $f_0(r)=r^2/L^2$, we find $\Delta\mC=0$.

In any event, we find that the complexity of formation is a fixed constant, independent of the temperature. Of course, this result for $d=2$ agrees with that found in section \ref{sec:BTZ} using the CA duality. One curious difference is that the sign of $\Delta \mathcal{C}_{V}$ in eq.~\reef{eq:dcvBTZ} is positive while the corresponding result for $\Delta \mathcal{C}_{A}$ in eq.~\reef{cc1} is negative.

\section{Discussion} \label{discuss}

In this paper, we considered the conjectured duality between complexity and action \cite{Brown1, Brown2} to evaluate  the complexity of formation, \ie the additional complexity involved in preparing an entangled thermofield double state between two boundary CFTs compared to preparing each of the individual CFTs in their vacuum state. Using the results of \cite{LuisRob} to account for the contributions of null hypersurfaces and joints  to the gravitational action, we were able to evaluate the action of the WDW patch for the dual AdS black holes and vacuum spacetimes. While the individual actions need to be regulated because of divergences coming from the asymptotic boundary, these divergences cancel in the difference of the actions in eq.~\reef{Cform} and hence the complexity of formation remains finite in the $\delta\to0$ limit.

We evaluated $\Delta\mC$ for three horizon geometries (\ie for the three different spatial geometries \reef{bound} in the boundary theory) --- spherical, planar and hyperbolic. For high temperatures, this geometry is unimportant and as indicated in eq.~\reef{PlanarGenerald}, the leading contribution is proportional to the entropy, \ie
\begin{equation}\label{PGd}
\Delta\mC   = k_d\,S +\cdots \qquad{\rm with}\quad k_d=\frac{d-2}{d\,\pi} \,  \cot{\!\bigg(\frac{\pi}{d} \bigg)} \, .
\end{equation}
The ellipsis indicates the presence of subleading terms for $k=\pm1$. From the explicit examples in eqs.~\reef{EntFormD5} and \reef{EntFormD4}, we can see that for curved horizons, eq.~\reef{PGd} is the leading term in an expansion for large values of $S/C_T$ where $C_T$ is the central charge in the boundary theory. Above, we referred to this as a high temperature expansion because up to numerical factors, $S/C_T \sim {\cal V}\,T^{d-1}$ at high temperatures, where $\cal V$ is the spatial volume in the boundary theory. This explains why the spatial curvature was unimportant in this limit and the leading result in eq.~\reef{PGd} is independent of the parameter $k$. Let us note that more generally, the results for $d=4$ and 3 show that we can write $\Delta\mC= C_T\,f(S/C_T)$.

Hence at least for high temperatures, the additional complexity required in preparing the entangled thermofield double state is proportional to the entanglement entropy between the two boundary CFTs in this state. It is perhaps useful to think of the description of analogous CFT states with MERA tensor networks \cite{mera1,mera2,mera3} to gain some insight into this result. The interested reader may find a more detailed discussion in appendix \ref{lastA}, however, we describe some of the salient points here. If we compare the tensor networks describing the individual ground states and the entangled thermofield double state, a large portion of the circuits are in fact identical and prepare the short range entanglements in the final UV state from a coarse-grained IR state. In the holographic context, this is reflected in the fact that the asymptotic AdS regions are nearly identical in both cases and make the same UV divergent contributions to the individual WDW actions. The difference between the MERA tensor networks at high temperatures is that the IR portion of the two vacuum circuits is removed and replaced with a layer of distinct tensors which entangles the two CFTs and introduces the appropriate thermal spectrum of eigenvalues --- see figure \ref{googooa} in appendix \ref{lastA} or figure 2c in \cite{tnr2}. Of course, this central layer of tensors can be thought of
as representing the Einstein-Rosen bridge connecting the two asymptotic AdS regions \cite{tnr2}. Hence in considering the complexity of formation, there is some competition between the additional complexity of preparing these bridge tensors and the simplification coming from removing the IR components of the vacuum circuits. At high temperatures, our holographic results indicate that the former dominates since $\Delta\mC>0$. One can also argue that the complexity of the bridge tensors and of the corresponding IR vacuum circuits should both be proportional to the entanglement entropy between the two copies of the CFT --- see appendix \ref{lastA}. Hence the complexity of formation should be proportional to this entropy, in accord with our holographic results in higher dimensions --- see comments below on $d=2$. 

In section \ref{sec:CompVol}, a similar result was obtained for the complexity of formation using the previously proposed duality relating complexity to the volume of an extremal bulk surface \cite{CompVolume}. For high temperatures, the leading contribution is independent of the geometry and given by eq.~\reef{VolSd},
\begin{equation}\label{PGdx}
\Delta\mC_{V} = \tk_d\,S +\cdots \qquad{\rm with}\quad \tk_d=4\sqrt{\pi} \,\frac{\left(d-2\right)\,\Gamma(1+ \frac{1}{d})}{\left(d-1\right)\Gamma(\frac{1}{2} + \frac{1}{d})} \, .
\end{equation}
Again this leading term is the complete result of the planar case while the ellipsis indicates subleading terms which appear with a spherical or hyperbolic horizon. Comparing eqs.~\reef{PGd} and \reef{PGdx} shows that both of the proposed dualities yield more or less the same complexity of formation up to an overall multiplicative factor. In fact, comparing the growth of complexity at late times found with the two different approaches yields essentially the same multiplicative factor --- see eq.~\reef{rat6}. Now as
emphasized in \cite{Brown1, Brown2}, the circuit complexity of a given quantum state can only be assigned a precise value once the algorithm for constructing the state is defined. For example, the value will depend on the choice of an initial reference state and the specific set of quantum gates with which one acts to construct the
desired state. In particular, the complexity would be expected to change by an overall multiplicative factor with different gate choices. This may then provide an explanation of the multiplicative factor relating the complexities found using the CA and CV dualities. That is, our holographic results suggest that the CA and CV dualities may both provide a consistent description of the complexity of holographic boundary states, however, the microscopic details of the algorithms used to define the complexity differs in each case.

As noted previously, the coefficients $k_d$ and $\tk_d$ both vanish for $d=2$. Hence in the case of two boundary dimensions, the complexity of formation is a fixed constant, independent of the temperature. Referring back to the discussion of MERA above (see also appendix \ref{lastA}), this result indicates that the complexity associated with constructing the layer of bridge tensors is essentially the same as for the IR portion of the vacuum network. This result is likely related to the recent discussion of MERA tensor networks in the context of kinematic space \cite{kinematic1,kinematic2,kinematic3}. In particular, it was found that for the special case of $d=2$, the bridge tensors can be constructed from the standard isometries and disentanglers appearing in the UV portion of the MERA \cite{kinematic2}. In the holographic context, the fact that $\Delta\mC$ is independent of the temperature is related to the fact that in three bulk dimensions, the BTZ black hole geometry is still locally the same as the vacuum AdS$_3$ space \cite{btz1,btz2}. Comparing to the Neveu-Schwarz vacuum in the boundary theory \cite{couscous}, our holographic results in eqs.~\reef{cc1} and \reef{eq:dcvBTZ} indicated that the complexity of formation is
\beq
d=2\,:\qquad\Delta\mC_A = -\frac{c}3 \qquad{\rm and}\qquad
\Delta\mC_V = +\frac{8\pi}3\,c\,,
\label{confused}
\eeq
where $c$ is the central charge of the boundary theory. Further, considering the Ramond vacuum instead, we find $\Delta\mC=0$ with both the CA and CV dualities. Naively, the difference in the sign of the two results in eq.~\reef{confused} might indicate some tension between the two holographic approaches to evaluating complexity. However, it seems more likely that the different microscopic details in the definition of the complexity, \eg the reference state, for the two approaches is simply producing different results at this fine-grained level. That is, the precise value (or even sign) of the complexity of formation is not robust against the ambiguities appearing in the definition of the complexity.

At this point, we note that, as discussed in \cite{LuisRob}, the boundary terms on the null boundary surfaces \reef{ActSurf2} and null joints \reef{ActJ2} are ambiguous. That is, evaluating the gravitational action for a particular spacetime geometry  generally produces different numerical values depending on different choices that can be made in constructing the boundary terms.  However, we show in appendix \ref{AppendixC} that our results for the complexity of formation are not effected by these ambiguities. The one exception to this statement is for the small hyperbolic black holes, \ie with $k=-1$ and $\omega^{d-2}<0$. In this case, we find that the complexity of formation is ambiguous due to the possibility of shifting $a$ in the joint contributions \reef{ActJ2} by an extra (arbitrary) function \cite{LuisRob}. In particular, $\Delta\mC$ is modified by such a shift through the joint terms where the null sheets from the opposite asymptotic boundaries meet between the inner and outer horizons  --- see appendix \ref{AppendixC} for a detailed discussion. Hence our results for the complexity of formation are not universal in this case. It is interesting that in this regime, we observed a discrepancy between the CA and CV approaches. Namely, $\Delta\mC_{V}$ diverges as $T$ approaches zero for $k=-1$, while $\Delta\mC_{A}$ remains finite. The CA duality also yields the curious result $d\mC_{A}/dt=0$ for small hyperbolic black holes --- see appendix \ref{app:Hypers}. All of these results highlight the exotic nature of these states, as was first commented by \cite{roberto}.

We also observe that the complexity of formation appears well-behaved, for small spherical black holes, \ie with $k=+1$ and $r_h<L$, as shown in figures \ref{CompareAdS5} and \ref{CompareAdS4}. However, these black holes are thermodynamically unstable \cite{HP,witten1} and the correct saddle point which dominates the bulk partition function is actually still the vacuum AdS space. Of course, with this saddle to represent the thermofield double, our calculations would yield a vanishing complexity of formation. However, this simply indicates that there is no leading order contribution to $\Delta\mC$ in the large central charge (or large $N$) expansion of the boundary CFT. That is, there is no contribution to $\Delta\mC$ of the magnitude of the central charge $C_T$, just as there is no entropy of this order. However, the thermofield double state still entangles the two CFTs  and it is simply that the entanglement entropy is an order one quantity. Hence we expect that the complexity of formation is also nonvanishing but only an order one quantity. It would be interesting to understand how to evaluate this contribution to $\Delta\mC$.

There are a variety of questions which we leave for future work. One interesting question is to compute the full time dependence of the complexity \cite{prep5}, which in a certain sense would interpolate between the complexity of formation (at $t_L=0=t_R$), considered here, and the late-time growth rate, considered in  \cite{Brown1,Brown2,LuisRob}. It would also be interesting to generalize the present calculations to charged and rotating black holes, where an analysis similar to that for the small hyperbolic black holes would apply. Finally, it would be instructive to investigate the complexity of formation for  higher curvature theories of gravity \cite{curv1,curv2}, although this would require first a better understanding of the boundary terms for the corresponding gravitational actions.

\section*{Acknowledgments}
We would like to thank Adam Brown, Dean Carmi, Bartek Czech, Lampros Lamprou, Djordje Radicevic, Dan Roberts, Sotaro Sugishita, Jamie Sully, Lenny Susskind, Brian Swingle, 
Guifr\'e Vidal and  Ying Zhao for useful comments and discussions. Research at Perimeter Institute is supported by the Government of Canada through Industry Canada and by the Province of Ontario through the Ministry of Research \& Innovation. SC acknowledges support from an Israeli Women in Science  Fellowship from the Israeli Council of Higher Education. RCM is supported by funding from the Natural Sciences and Engineering Research Council of Canada, from the Canadian Institute for Advanced Research and from the Simons Foundation through the ``It from Qubit'' collaboration.

\appendix

\section{Fefferman-Graham near Boundary Expansions} \label{App:FeGra}

The action on the WDW patch is divergent since this region of the bulk spacetime extends all the way up to spatial infinity.
In this appendix, we discuss how to regulate these divergences by introducing a UV cutoff surface at $r=r_{\mt{max}}$, following the standard approach in holographic calculations, see, \eg \cite{count,sken1,sken2}. To make meaningful comparison between the two different spacetimes (\ie black hole and vacuum AdS) we need to be able to relate the cutoffs in the two geometries. We do this by choosing $r_{\mt{max}}$ to correspond to the surface of $z=\delta$ in the asymptotic Fefferman Graham (FG) expansion for both cases.

\subsection{Relating the Cutoffs}
We begin by writing the metric \reef{HigherDMetric} in the FG form:
\begin{equation}\label{eq:FGMetric}
ds^2 = \frac{dr^2}{f(r)} - f(r) dt^2 +r^2 d \Sigma_{k,d-1}
=\frac{L^2}{z^2}\left(dz^2 +g_{ij}(z,x^i) dx^i dx^j\right)
\end{equation}
where
\begin{equation}\label{seriesappa}
z = \frac{L^2}{r}+\frac{c_1}{r^2} + \cdots + \frac{c_d}{r^{d+1}} +\frac{c_{d+1}}{r^{d+2}}+ \cdots
\end{equation}
goes to zero at the boundary and $g_{ij}(z\rightarrow 0, x^i)$ is the \emph{finite} boundary metric \reef{bound}. We fix the different coefficients $c_i$ to obtain:
\begin{equation}\label{good}
 \frac{L}{z}\, dz  = \frac{dr}{\sqrt{f(r)}}  \,,
\end{equation}
where, as is implicit in eq.~\reef{seriesappa}, we work in a series expansion for large $r$. We can invert the series \eqref{seriesappa} to obtain  $r_{\mt{max}}$ corresponding to the surface of constant $z=\delta$:
\begin{equation}
r_{\mt{max}} = \frac{L^2}{\delta}+ \tilde c_1 +\dots + \tilde c_d \delta^{d-1} +\tilde c_{d+1} \delta^{d}+ \dots\, .
\end{equation}
The first coefficient to depend on $r_h$ (and so, the first coefficient which differs from the FG expansion of vacuum AdS) is $c_d$ (or equivalently $\tilde c_d$). This follows immediately from the form of the blackening factor  in which the dependence on $r_h$ starts at the $d$-th subleading order in the boundary expansion.
In general, we can prove that
\begin{equation}
r^\mt{BH}_{\mt{max}}-r^\mt{\vac}_{\mt{max}}=  \frac{\omega^{d-2}  }{2 d L^{2 (d-2)}}\,\delta ^{d-1}\ +\ \mathcal{O}\!\left(\delta^{d+1}\right)\,,
\label{bounce}
\end{equation}
as given in eq.~\reef{differ}.
The proof goes as follows: Integrating eq.~\reef{good},  we obtain
\begin{equation}\label{feGra2}
- L \log \frac{\delta}{L^2} = \int^{r_{\mt{max}}} \frac{dr}{\sqrt{f(r)}}\, .
\end{equation}
The integration constant was fixed here by assuming that to leading order $z=\frac{L^2}{r}$. Further we understand the left-hand side of eq.~\eqref{feGra2} to be defined by a large-$r$ power series, \ie
\begin{equation}
\int^{r_{\mt{max}}}\frac{dr}{\sqrt{f(r)}} =
\int^{r_{\mt{max}}} dr \left(\frac{L}{r}+\sum_{n=2}^{\infty} \frac{a_n}{r^n} \right)
= L \log(r_{\mt{max}}) - \sum_{n=2}^{\infty} \frac{a_n}{(n-1)r_{\mt{max}}^{n-1}}\,.
\end{equation}
Now putting the UV cutoff surface at the $z=\delta$ surface in both the black hole and vacuum AdS backgrounds, we can subtract the corresponding equations to obtain
\begin{equation}
0 = \int^{r^{\BH}_{\mt{max}}} \frac{dr}{\sqrt{f(r)}}
- \int^{r^{\vac}_{\mt{max}}} \frac{dr}{\sqrt{f_0(r)}}\, ,
\end{equation}
where the upper limits are slightly different in the two integrals.
If we substitute $r^{\BH}_{\mt{max}}=r^{\vac}_{\mt{max}} +\delta r_\mt{max}$ into the first integral, the leading order contribution in the shift $\delta r_\mt{max}$ becomes
\begin{equation}
0 \simeq \frac{1}{\sqrt{f(r^{\vac}_{\mt{max}})}} (r^\mt{BH}_{\mt{max}}-r^\mt{\vac}_{\mt{max}}) + \int^{r^\mt{\vac}_{\mt{max}}} dr\left(\frac{1}{\sqrt{f(r)}}- \frac{1}{\sqrt{f_0(r)}}\right) \, .
\end{equation}
This integral is convergent and we can expand this expression for large $r$,
\begin{equation}
0 \simeq \frac{L}{r^\mt{\vac}_{\mt{max}}} (r^\mt{BH}_{\mt{max}}-r^\mt{\vac}_{\mt{max}}) + \int^{r^\mt{\vac}_{\mt{max}}} dr \, \frac{\omega^{d-2} L^3}{2\, r^{d+1}} \, .
\end{equation}
Integrating the last expression and using the leading order result $r_\mt{max}=L^2/\delta$ (which applies for both geometries), we recover eq.~\eqref{bounce}.

Finally, we compute $r_\mt{max}$ in the various vacuum AdS geometries by evaluating eq.~\eqref{feGra2} to obtain
\begin{equation}
z=\frac{2L^2}{r + \sqrt{k L^2 + r^2}}
\end{equation}
and
\begin{equation}
r_{\mt{max}}^{\mt{vac}} = \frac{L^2}{\delta} -\frac{k \delta }{4  }\, .
\end{equation}
As a consequence, eq.~\eqref{bounce} yields
\begin{equation}\label{appeq:rmaxBH}
r_{\mt{max}}^{\mt{BH}} = \frac{L^2}{\delta} -\frac{k \delta }{4  }+  \frac{\omega^{d-2} }{2 d L^{2 (d-2)}}\,\delta ^{d-1} \ +\ \mathcal{O}\!\left(\delta^{d+1}\right).
\end{equation}
Finally for the BTZ case, eq.~\eqref{feGra2} can be evaluated explicitly using the blackening factor \eqref{metricBTZ} which leads to:
\begin{equation}\label{zrRelBTZ}
r+\sqrt{r^2-r_h^2} =  \frac{2 L^2}{z}, \qquad
r = \frac{L^2}{z} + \frac{z r_h^2}{4L^2}\,,
\end{equation}
and
\begin{equation}\label{rBTZdel}
r_{\mt{max}}^\mt{BTZ} = \frac{L^2}{\delta} +  \frac{\delta r_h^2}{4L^2}\,.
\end{equation}

\subsection{Cutoff Independence of the Action}\label{app:Fegra2}

In this subsection we provide details for the various cancellations encountered in the main text when subtracting the vacuum AdS results from those of the black holes in the process of evaluating the action.
It will be useful in our discussion to use the following relation
\begin{equation}\label{differ4}
v_\infty -r^* (r_{\mt{max}}) = \delta +\cdots + w(r_h)\delta^{d+1} + \mathcal{O}(\delta^{d+2}).
\end{equation}
where due to the form of the blackening factor \eqref{BlackeningFactor} the $r_h$ dependence first appears at order $\delta^{d+1}$. Note that $v_\infty$ cancels a possible integration constant and the expansion therefore starts at order $\delta$. It is further possible to show\footnote{The notation $v'_\infty$ and $r_0^*(r)$ refers to the vacuum AdS geometry --- see appendix \ref{App:Vac}.}
\begin{equation}\label{deldelvp}
v_\infty -r^* (r^{\mt{BH}}_{\mt{max}}) -\left(v'_\infty -r_0^* (r_{\mt{max}}^{\mt{vac}}) \right) = w(r_h)- w(0) = \frac{ (d-1) }{2 d (d+1) }\,\frac{\omega^{d-2}}{L^{2 (d-1)}}\,
\delta ^{d+1} +
\ \mathcal{O}\!\left(\delta^{d+2}\right) .
\end{equation}
The arguments are similar to those in the previous section and the leading contribution reads
\begin{equation}
\begin{split}
v_\infty -r^* (r^{\BH}_{\mt{max}}) - & \left(v'_\infty -r_0^* (r_{\mt{max}}^{\vac}) \right) =
\int^{\infty}_{r^{\BH}_{\mt{max}}} \frac{dr}{f(r)}
- \int^{\infty}_{r^{\vac}_{\mt{max}}} \frac{dr}{f_0(r)}
\\
= &
\int_{r^{\vac}_{\mt{max}}}^{\infty} \frac{f_0(r) - f(r)}{f_0(r) f(r)} - \frac{1}{f(r^{\vac}_{\mt{max}})} (r_{\mt{max}}^{\BH} - r_{\mt{max}}^{\vac})
\\
= & \frac{ (d-1) }{2 d (d+1) }\,\frac{\omega^{d-2}}{L^{2 (d-1)}}\,
\delta ^{d+1} +
\ \mathcal{O}\!\left(\delta^{d+2}\right)\, ,
\end{split}
\end{equation}
where we have used eq.~\eqref{differ}.

With all this in hand, we are ready to prove some of the claims quoted in the main text regarding cancellations between vacuum and black hole contributions to the action of the WDW patch. The first claim is related to the bulk integrals. It explains why eq.~\eqref{ActionBulk} reduces to eq.~\eqref{Cform1a} after subtracting  the vacuum AdS contribution and why we can choose $r_{\mt{max}}$ in eq.~\eqref{Cform1a} to be either  of the two cutoffs.
We start with the difference of the bulk actions
\beqa
\Delta I_\mt{bulk} &=& - \frac{\Omega_{k,d-1} \, d}{2\pi G_N L^2} \, \int_0^{r_{\mt{max}}^{\BH}} dr\, r^{d-1}\, \Big({v_\infty -r^*(r)}\Big) \nonumber\\
&&\quad + \frac{\Omega_{k,d-1} \, d}{2\pi G_N L^2} \, \int_0^{r_{\mt{max}}^{\vac}} dr\, r^{d-1} \, \Big({v'_\infty -r_0^*(r)} \Big)\,.
\label{Cform1x}
\eeqa
 In particular, if we denote $r^{\BH}_{\mt{max}} = r_{\mt{max}}^{\vac} + X\,\delta^{d-1}$ where $X$ was given in eq.~\eqref{bounce} we recover eq.~\eqref{Cform1a}, with $r_{\mt{max}}$ being the vacuum AdS cutoff, plus an addition
\beq
\delta I_\mt{bulk,BH} = -  \frac{\Omega_{k,d-1} \, d}{2\pi G_N L^2} \, X\,\delta^{d-1} \Big[ r^{d-1} \, \Big({v_\infty -r^*(r)} \Big)\Big]_{r=r_{\mt{max}}}\,.
\eeq
Now to leading order, $r^{d-1}=r_{\mt{max}}^{d-1}=L^{2(d-1)}/\delta^{d-1}$ which is canceled by the factor $\delta^{d-1}$ in the pre-factor. But then $v'_\infty -r_0^*(L^2/\delta) \simeq \delta$ and so we find $\delta I_\mt{bulk,vac}\propto\delta$.

The second claim we want to prove is that the surface contributions at the cutoff surfaces cancel between the black hole and vacuum AdS backgrounds.
We find it convenient to define a function
\begin{equation}
g(r) =  \del_r f(r) + \frac{2(d-1)}{r}f(r),
\end{equation}
and a function $g_0(r)$ defined by replacing $f(r)$ by the vacuum blackening factor $f_0(r)$ in the expression above.
Using eq.~\eqref{extrinsic2} and subtracting the vacuum AdS result we obtain
\begin{equation}
\delta I_\mt{GHY}(r=r_{\mt{max}})  = \delta_1 + \delta_2 + \delta_3 + \dots
\end{equation}
where the ellipsis stands for higher orders in $\delta$ and we have defined the following expressions
\begin{equation}\label{deltacontr}
\begin{split}
\delta_1 & \equiv  \frac{\Omega_{k,d-1}\, \delta r_{\mt{max}}^{d-1}}{4 \pi G_N}   g_0(r_{\mt{max}}^{\vac}) \bigg( v'_\infty-r_0^*(r_{\mt{max}}^{\vac}) \bigg),
\\
\delta_2 & \equiv  \frac{\Omega_{k,d-1}\,(r_{\mt{max}}^{\vac})^{d-1}}{4 \pi G_N}   \delta g(r_{\mt{max}}) \bigg( v'_\infty-r_0^*(r_{\mt{max}}^{\vac}) \bigg),
\\
\delta_3 & \equiv  \frac{\Omega_{k,d-1}\,(r_{\mt{max}}^{\vac})^{d-1}}{4 \pi G_N}    g_0(r_{\mt{max}}^{\vac})
\left[v_\infty -r^* (r^{\BH}_{\mt{max}}) -  (v'_\infty -r_0^* (r_{\mt{max}}^{\vac}) \right],
\end{split}
\end{equation}
as well as
\begin{equation}\label{deltarmaxdm1}
\delta r_{\mt{max}}^{d-1}\equiv (r_{\mt{max}}^{\BH})^{d-1}-(r_{\mt{max}}^{\vac})^{d-1}
\simeq (d-1)(r_{\mt{max}}^{\vac})^{d-2} \left(r^\mt{BH}_{\mt{max}}-r^\mt{\vac}_{\mt{max}}\right) \simeq  \frac{d-1 }{2 d }\omega^{d-2}  \delta
\, ,
\end{equation}
and
\begin{equation}
\begin{split}\label{deltag}
\delta g(r_{\mt{max}})&\equiv g(r_{\mt{max}}^{\BH})- g_0(r_{\mt{max}}^{\vac})
=
g(r_{\mt{max}}^{\BH})- g(r_{\mt{max}}^{\vac})
+g(r_{\mt{max}}^{\vac})- g_0(r_{\mt{max}}^{\vac})
\\
&\simeq g'(r_{\mt{max}}^{\vac})\left(r^\mt{BH}_{\mt{max}}-r^\mt{\vac}_{\mt{max}}\right) - d \frac{\omega^{d-2}}{(r_{\mt{max}}^{\vac})^{d-1}}
\simeq -(d-1)
\frac{\omega^{d-2}  }{ L^{2 (d-1)}}\ \delta ^{d-1}
\, ,
\end{split}
\end{equation}
and simplified the expressions using eq.~\eqref{differ}. We can now show that all the contributions in eq.~\eqref{deltacontr} are of order $\delta$. For $\delta_1$ we use eq.~\eqref{deltarmaxdm1} together with
\begin{equation}
 g_0(r_{\mt{max}}^{\vac}) \simeq \frac{2 d r_{\mt{max}}^{\vac}}{L^2} \simeq \frac{2 d}{\delta}
\end{equation}
and  eq.~\eqref{differ4}. For $\delta_2$ we use again  eq.~\eqref{differ4} together with eq.~\eqref{deltag} and $(r_{\mt{max}}^{\vac})^{d-1}=L^{2(d-1)}/\delta^{d-1}$. For $\delta_3$ we use in addition eq.~\eqref{deldelvp}. Therefore, we conclude that $\delta I_\mt{GHY}(r=r_{\mt{max}})\propto\delta$.

The last claim which we want to show is that the joint terms at the cutoff surfaces cancel between the black hole background and vacuum AdS. We use eq.~\eqref{CornerCut} which we reiterate here for convenience:
\begin{equation}\tag{\ref{CornerCut}}
I_{\jnt,\text{cut}} =  \frac{\Omega_{k, d-1}}{4 \pi G_N}\,  r_{\mt{max}}^{d-1}\log{f(r_{\mt{max}})}\, .
\end{equation}
Dependence on $r_h$ comes either from the cutoff, as in eq.~\eqref{appeq:rmaxBH}, or from the explicit $\omega^{d-2}$ that appears in $f(r)$. Expanding the above expression for small $\delta$ we see  that the dependence on the horizon radius is at most of order $\delta \log{\delta}$. Therefore, the joint terms near the boundary cancel between the black hole background and vacuum AdS when we take the $\delta \rightarrow 0$ limit.

\section{Details for Vacuum AdS Actions}  \label{App:Vac}

As explained in section \ref{sec:genset}, there are subtle differences for the vacuum geometries corresponding to $k=\{+1,0,-1\}$. In particular, various caustics appear in the boundary of the WDW patch and there are also `mild' orbifold singularities in the planar and hyperbolic geometries. The WDW patches for the vacuum AdS geometries are shown in figure \ref{EmptyAdSDrawing}. In the following, we carefully consider these new singularities in calculating the WDW action but our conclusion is that they do not affect the final result. That is, the only nonvanishing contributions for the vacuum actions are those already calculated in section \ref{sec:genset}, \ie the bulk action \reef{ActionBulk}, as well as the GHY surface terms \reef{extrinsic2} and null joint terms \reef{CornerCut} associated with the UV cutoff surface $r=r_{\mt{max}}$. \vspace{.5em}

\begin{figure}
\centering
\includegraphics[scale=0.35]{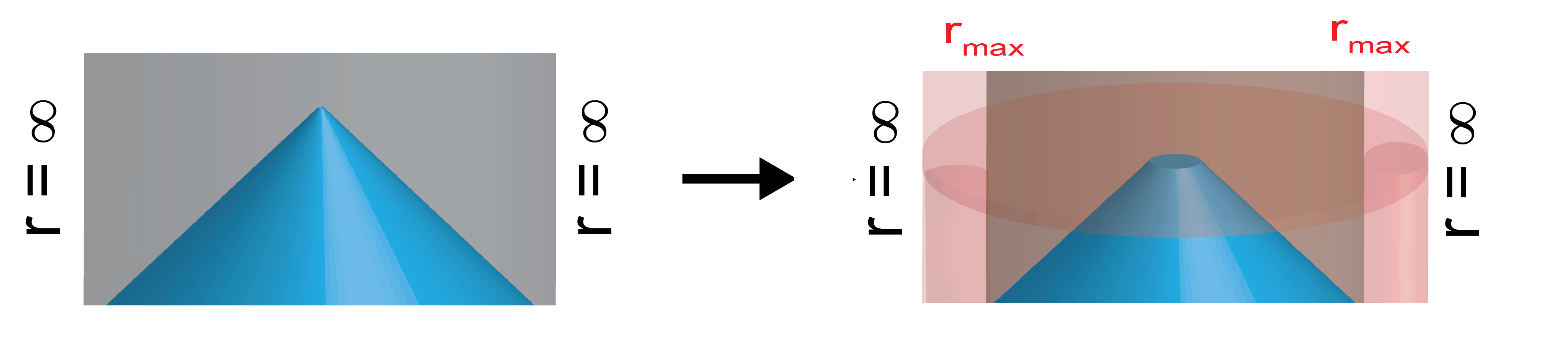}
\caption{The future caustic of the Wheeler-DeWitt patch in the vacuum global AdS geometry (left figure). The tip contribution can be effectively calculated by a regulator surface $t=v'_\infty-\epsilon_1$ (represented in the right figure) with well defined $\eps_1\to0$ limit.}
\label{CausalTip}
\end{figure}

\noindent{\bf a) Spherical geometry:} As noted before in section \ref{sec:genset}, the WDW patch terminates with a caustic at the past and future tips of the causal diamond shown in figure \ref{EmptyAdSSpherical} for $k=+1$. These points are located at $(t,r)=(\pm v_\infty,0)$. To determine the latter, we must first evaluate the tortoise coordinate \reef{tortoise} using $f_0(r)$, which yields
\begin{equation}\label{TortoiseEmpty}
r^{*}_0(r) = L \ \tan ^{-1}\! \left({r}/{L}\right) \, ,
\end{equation}
and eq.~\reef{eq:vinf} then gives
\begin{equation}\label{vinfsphempt}
v'_{\infty} =  {L\,\pi}/{2} \, .
\end{equation}
As mentioned in the previous section, boundary terms for such caustics were not considered in the recent discussion of \cite{LuisRob} but we will argue that in fact, they do not contribute to the gravitational action as follows: Focusing on the future tip, we introduce a regulator surface at $t=v'_\infty-\epsilon_1$, which cuts off the future tip and produces a flat cap on the WDW patch  --- see figure \ref{CausalTip}. The gravitational action can then be evaluated for this regulated geometry using the standard boundary terms, and the contribution of the caustic is recovered with the limit $\eps_1\to0$.

In evaluating the gravitational action with the new regulator surface, we must consider potential extra contributions of the GHY term \reef{ActSurf} integrated over this cap and of the joint term \reef{ActJ2} where this additional boundary intersects the null boundary $v=v'_\infty$. Hence, we introduce the (outward-pointing) unit normal to the regulator surface
\begin{equation}\label{VecEmpX}
{\bf t}'=t'_{\mu}\,dx^\mu =  \sqrt{f_0(r)}\ dt\,.
\end{equation}
Now one can easily verify that on this surface, the trace of the extrinsic curvature vanishes and so the GHY term \reef{ActSurf} makes no contribution. Next
we can combine eq.~\reef{VecEmpX} with eq.~\reef{niceNull} --- after replacing $f(r)$ by $f_0(r)$ --- to evaluate the corresponding joint term \reef{ActJ2},
\beq
I_{\mt{jnt},\mt{cap}} =\frac{\Omega_{1,d-1}}{16 \pi G_N}\, \epsilon_1^{d-1}\,\log\!\left(1+\frac{\epsilon_1^2}{L^2}\right)\, .
\label{vani2}
\eeq
From this expression, we can easily see that the joint contribution vanishes in the limit $\eps_1\to0$. Hence our conclusion is that the caustic at the future tip of the WDW patch does not contribute to the gravitational action and, of course, the same is true for the past tip by symmetry. In the presence of the regulator surface, the bulk contribution is also modified but of course, this change vanishes in the limit  $\eps_1\to0$. \vspace{.5em}

\noindent{\bf b) Planar geometry:} As described in section \ref{sec:genset} with $k=0$, a `conical' or orbifold singularity appears in the vacuum geometry along the Poincar\'e horizon, due to the compactification of the spatial geometry. To carefully evaluate the corresponding gravitational action, we introduce a timelike regulator surface at $r=\eps_0$, as shown in figure \ref{EmptyAdSPlanar}.\footnote{An alternative approach is to introduce spacelike regulator surfaces at $t=v'_{\infty}-r_0^*(\eps_1)$ and $t=u'_{\infty}+r_0^*(\eps_1)$.  We have confirmed that one arrives at the same conclusion with this approach. That is, there are no additional contributions to the gravitational action coming from the orbifold singularity at $r=0$.}

We evaluate the contributions due to this regulator to the gravitational action and demonstrate that they vanish in the $\epsilon_0 \rightarrow 0$ limit.
For the vacuum planar AdS space, we have $f_0(r) = {r^2}/{L^2}$ and the corresponding tortoise coordinate \reef{tortoise} is simply
\begin{equation}
r_0^*(r) = -{L^2}/{r}\,.
\end{equation}
In turn, using eq.~\reef{eq:vinf}, the future null boundary of the WDW patch is given by $v=v'_\infty=0$.

In evaluating the gravitational action, the new surface term  on the regulator surface takes the form in eq.~\eqref{extrinsic1}, with the appropriate substitutions of $f_0(r)$, $v'_\infty$ and $r^*_0(r)$, which yields
\begin{equation}
I_\mt{GHY}  = - \frac{d  \Omega_{0,d-1}}{2 \pi G_N}\, \epsilon_0^{d-1}\,.
\end{equation}
Similarly, the joint contribution where the null boundary intersects the regulator  surface takes the form in the first line of eq.~\eqref{CornerSingSP}, with $f(r)$ replaced by $f_0(r)$, which yields
\beq \label{CornerSingX}
I_{\mt{jnt},\mt{sing}} =  - \frac{\Omega_{0, d-1}}{4 \pi G_N}\,  \eps_0^{d-1}  \log(\eps_0^2/L^2)\,.
\eeq
Of course, both of these expressions vanish in the limit $\epsilon_0 \rightarrow 0$.
Further, the change in the bulk action produced by cutting off the radial integral at $r=\eps_0$ is proportional to $\eps_0^{d-1}$, which again vanishes in the limit  $\eps_0\to0$.  \vspace{.5em}

\begin{figure}
\centering
\includegraphics[scale=0.25]{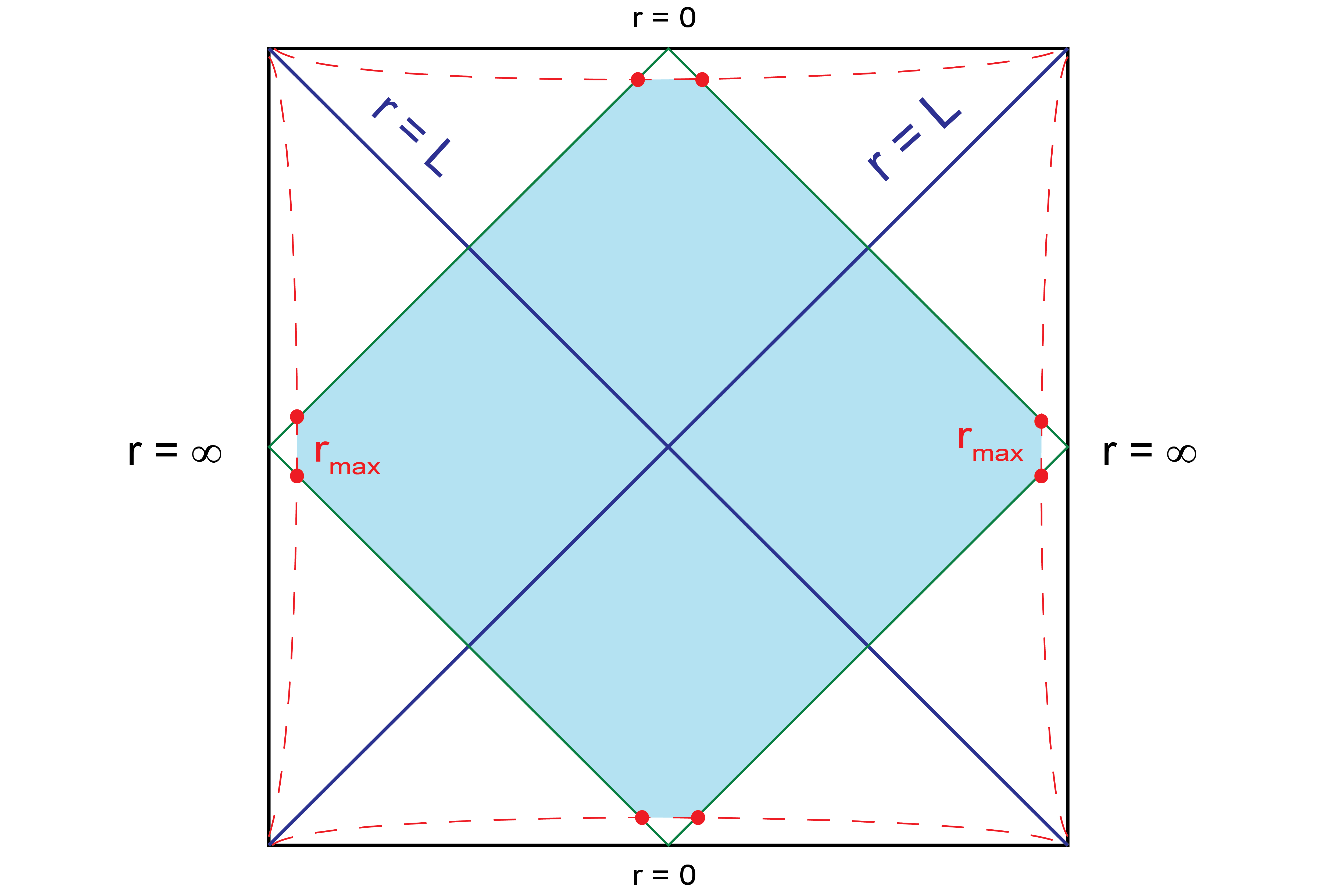}
\caption{Penrose diagram for the vacuum hyperbolic AdS space.}
\label{EmptyHyper}
\end{figure}

\noindent{\bf c) Hyperbolic geometry:} Recall that the $k=-1$ vacuum actually describes an entangled state of two copies of the CFT on a hyperbolic geometry. With $f_0(r)={r^2}/{L^2}-1$, there is a horizon at $r_h=L$ and even though locally the geometry corresponds to that of vacuum AdS space, the Penrose diagram looks essentially the same as for the black hole metric --- see figure \ref{EmptyHyper}. In particular, because we have compactified the hyperbolic geometry, there is an orbifold singularity at $r=0$. The tortoise coordinate \reef{tortoise} becomes
\begin{equation}\label{TortoiseHypEmp}
r^{*}_0(r) = \frac{L}{2} \log  \frac{  |r-L| }{r+L} \, ,
\end{equation}
and from eqs.~\reef{eq:uinf} and \reef{eq:vinf}, we see that the boundaries of the WDW patch are given by $v'_\infty=0=u'_\infty$.
One can readily confirm that the null surfaces future boundaries on the left and right sides of the Penrose diagram meet at precisely $r=0$.\footnote{As usual the tortoise coordinate \reef{TortoiseHypEmp} diverges  at the horizon (\ie $r^{*}_0(r\to L)\to-\infty$), but then returns zero at $r=0$. Hence the null-ray $v=0$ hits the singularity $r=0$ at precisely $t=0$.} This joint is a particularly singular feature in the boundary of the WDW patch and so as before, we regulate the boundary geometry by cutting it off on the spacelike surface $r=\eps_0$, as shown in figure \ref{EmptyHyper}. The gravitational action then receives extra contributions from the GHY term \reef{ActSurf} integrated along this regulator surface and from the joint term \reef{ActJ2} where this new boundary intersects the null boundary $v=v'_\infty=0$. However, following calculations identical to those given previously, we find that both of these contributions vanish in the limit $\eps_1\to0$. Essentially the size of the relevant integration region shrinks to zero as $\eps_1^{d-1}$ and there is no compensating singularity in the other geometric factors --- consider eqs.~\reef{extrinsic1} and \reef{CornerSingSP}, with the appropriate substitutions of $f_0(r)$, $v'_\infty$ and $r^*_0(r)$.  As in the previous cases, introducing the regulator surface changes the lower limit in the radial integration in the bulk contribution to produce a small modification, $\delta I_\mt{bulk}\propto \eps_1^{d+1}$, which vanishes in the limit  $\eps_1\to0$. \vspace{.5em}

In addition, it is possible to write general expressions for the bulk contribution to the WDW action in general dimension. For $d$ even we obtain:
\begin{equation} 
2I^{\vac}_{\textrm{bulk}} = - \frac{\Omega_{1,d-1} \, d}{2 \pi G_N} \left( \frac{r_{\mt{max}}^{d-1}}{(d-1)} + \sum_{n=1}^{\frac{d-2}{2}} (-k)^{n} \frac{L^{2 n} r_{\mt{max}}^{d-1-2 n}}{(2 n +1)(d-1-2 n)} + \delta_{k,1} (-1)^{\frac{d}{2}} \frac{ \pi L^{d-1}}{2 d} \right)  \, ,
\end{equation}
while for $d$ odd:
\begin{align}
\begin{split}
&2I^{\vac}_{\textrm{bulk}} = - \frac{\Omega_{1,d-1} \, d}{2 \pi G_N} \left( \frac{r_{\mt{max}}^{d-1}}{(d-1)} + \sum_{n=1}^{\frac{d-3}{2}} (-k)^{n} \frac{L^{2 n} r_{\mt{max}}^{d-1-2 n}}{(2 n +1)(d-1-2 n)}
\right.
\\
&~~~~~~~~~~~~~~~~~~~~~~~~~~~~~~~~~~~~~
\left.
 +(-k)^{\frac{d-1}{2}} \frac{L^{d-1}}{d^{2}} \left(1+ d \log{\frac{r_{\mt{max}}}{L}} \right) \right) \, .
 \end{split}
\end{align}
Note, that for odd dimensions, there is a logarithmic divergence while for even dimensions, there is an additional constant term for the spherical geometry. This is the origin of the  $\delta_{k,1}$ term in the bulk action appearing in eq.~\eqref{BulkActiond4k}.

\section{Small Hyperbolic Black Holes}\label{app:Hypers}

\begin{figure}
\centering
\includegraphics[scale=0.4]{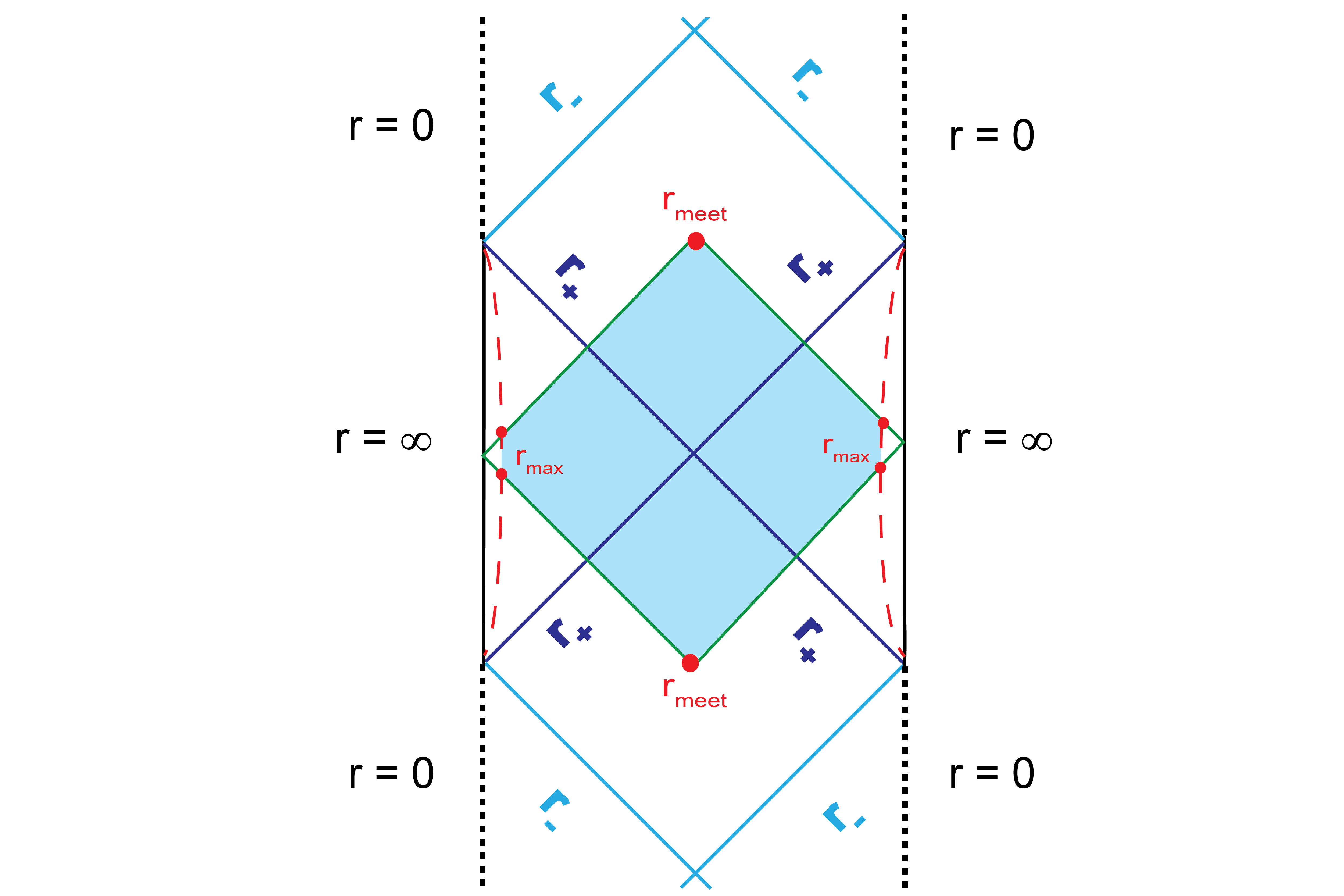}
\caption{Penrose diagram for small hyperbolic black holes with $r_h < L$. The outer horizon is drawn in dark blue and labeled $r_{+}$ and the inner horizon is drawn  in lighter blue and labeled $r_{-}$. The two ingoing null rays meet in the region between the inner and outer horizon, and the joint term between them has a non vanishing contribution to the action.}
\label{PenroseInOut}
\end{figure}

The `small' hyperbolic black holes (\ie $k=-1$ and $r_h<L$) have a causal structure similar to that encountered for  charged AdS black holes \cite{Chamblin:1999tk}. In this case, the blackening factor $f(r)$ in eq.~\eqref{BlackeningFactor} admits two positive real roots. This means that the black hole has two horizons, an inner one which we will denote by $r=r_-$ and an outer one with $r=r_h=r_+$. The relevant Penrose diagram is  shown in figure \ref{PenroseInOut}. One feature which can be noticed right away is that the null surfaces bounding the WDW patch do not fall into the singularity. Instead, they meet at some point between the two horizons which we will denote by $r=r_{\text{meet}}$ (of course $r_{-}< r_{\text{meet}} < r_{+}$). For this reason, instead of the surface term that we have encountered in the previous cases discussed in this paper we will have  two new joint contributions. The joint contributions can be computed according to the rules of \cite{LuisRob}.\footnote{In this case, we have a null-null joint for which the function $a$ is given in eq.~\reef{ball}.} The total contribution from the two joints at $r=r_{\text{meet}}$ becomes
\begin{equation}\label{jointrmeet}
I_{\text{jnt}} = - \frac{\Omega_{-1,d-1}}{4 \,\pi \, G_N} \, r_{\text{meet}}^{d-1} \, \log{|f(r_{\text{meet}})|} \, .
\end{equation}
The point in which the null rays meet $r=r_{\text{meet}}$ can be calculated from the following equation for the tortoise coordinate \eqref{tortoise}:
\begin{equation}\label{eq:rmeethyper}
r^{*}(r_{\text{meet}}) = \frac{v_{\infty} - u_{\infty}}{2} = v_{\infty} \, .
\end{equation}
We will have to solve for $r_{\text{meet}}$ numerically since this equation is usually transcendental.
In addition, the bulk contribution for the small hyperbolic black holes is modified, since the volume integral only goes as low as  $r_{\text{meet}}$ now:
\begin{equation}
\begin{split}\label{eq:bulkappC}
 \Delta I_{\bulk}= & - \frac{\Omega_{-1,d-1} \, d}{2\pi G_N L^2} \, \left[\int_{r_{\text{meet}}}^{r_{\mt{max}}} dr\, r^{d-1} \,\Big[v_\infty -r^*(r)\Big]-\int_0^{r_{\mt{max}}} dr\, r^{d-1} \Big[v'_\infty-r_0^*(r)\Big]\right]\, .
\end{split}
\end{equation}
Combining these results, we obtain the complexity of formation for small hyperbolic black holes:
\begin{equation}
\qquad\Delta\mC =  \frac{1}{\pi}\big[\,\Delta I_{\bulk} + I_{\text{jnt}} \,\big]\,.\tag{\ref{Cform2}}
\end{equation}
In the following, we study the cases of $d=3$ and $d=4$ in more detail.
We also chose to include in this appendix the results for hyperbolic black holes in $d=3$ with $L<r_h<\frac{2 L}{\sqrt{3}}$ since these black holes have a blackening factor with three real roots, analogously to the case of small hyperbolic black holes, and so parts of the computation overlap. Note however that out of these three roots, only one is positive in this case, and so we have a single horizon and the same causal structure as the one for large hyperbolic black holes.

\subsection{d=4}
\begin{figure}
\centering
\includegraphics[scale=0.65]{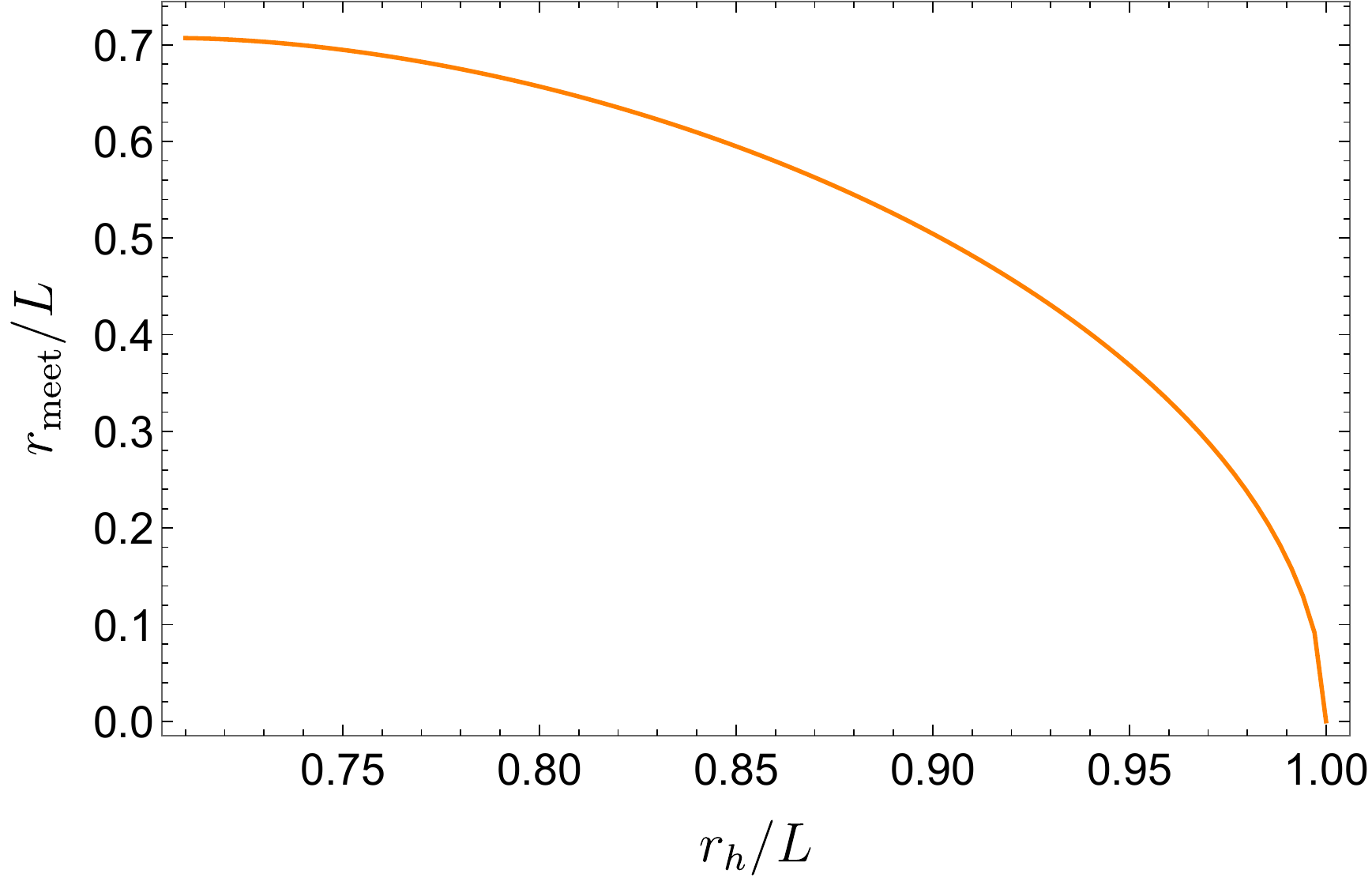}
\caption{The meeting point of the two ingoing null rays as a function of $r_h$. The meeting point is always between the inner and outer horizons. In particular, when the black hole becomes extremal the rays meet at $r_{\text{meet}} =r_h= L/\sqrt{2}$.}
\label{figureRays}
\end{figure}

In $d=4$, the blackening factor \eqref{BlackeningFactor},
\begin{equation}
f(r) = \frac{r^2}{L^2} - 1-
\frac{r_h^{2}}{r^{2}}\left(\frac{r_h^2}{L^2} -1 \right)\, ,
\end{equation}
has two  positive real roots:
\begin{equation}\label{eq:2radii5d}
r_1=r_h, \qquad r_2=\sqrt{L^2 - r_h^2}\, .
\end{equation}
For $r_h =L$ we should recover the results of hyperbolic vacuum AdS.  The problem is completely symmetric under the redefinition $\tilde r_h = \sqrt{L^2-r_h^2}$ and so no loss of generality is involved in assuming
$r_h>r_2$.
For the special case
$r_h = L/\sqrt{2}$ the two horizons become degenerate which results in an extremal black hole.
To obtain the tortoise coordinate one has to factorize the inverse blackening factor with respect to both roots. After integration we obtain:
\begin{equation}\label{TortoiseSmallHyp}
r^{*} (r) = \frac{L^2}{2 (r_h^2-r_2^2)}
\left(
 r_h
  \log \left[\frac{| r-r_h | }{r+r_h}\right]
  - r_2 \,
 \log \left[\frac{| r-r_2| }{r+r_2}\right]
\right)\, .
\end{equation}
The point where the ingoing null rays meet inside the black hole can be calculated numerically using eq.~\eqref{TortoiseSmallHyp} and the meeting condition \eqref{eq:rmeethyper} which reads in this case
\begin{equation}
r^{*}(r_{\text{meet}})
= 0 \, .
\end{equation}
Since the rays meet between the two horizons one has to choose the appropriate branches of the logs in eq.~\eqref{TortoiseSmallHyp} when solving this equation.
We show the result for $r_\text{meet}$ in figure  \ref{figureRays}.

For the bulk integral, we evaluate eq.~\eqref{eq:bulkappC} and obtain:
\begin{align}\label{bulkhypsmall}
\begin{split}
\Delta I_{\text{bulk}} = &\frac{ \,  \Omega_{-1,3}}{12 \pi G_N (r_h^2-r_2^2)}
\left[
3 r_2 \left(r_{\text{meet}}^4- r_2^4 \right) \log \left[\frac{r_{\text{meet}}-r_2}{r_{\text{meet}}+r_2}\right]
\right.
\\
&\left.
+2 r_{\text{meet}} \left(r_h^2-r_2^2\right) \left(3 L^2+r_{\text{meet}}^2\right)
+3 r_h (r_h^4- r_{\text{meet}}^4) \log \left[\frac{r_h-r_{\text{meet}}}{r_h+r_{\text{meet}}}\right]\right]\, ,
\end{split}
\end{align}
and the joint contribution \eqref{jointrmeet}:
\begin{equation}\label{cornerhypsmall}
I_{\text{jnt}} = - \frac{\Omega_{-1,3}}{4 \,\pi \, G_N} \, r_{\text{meet}}^{3} \, \log{|f(r_{\text{meet}})|} \, .
\end{equation}
Substituting the numerical solution for $r_\text{meet}$, we obtain the result plotted in orange in figure \ref{CompareAdS5}.

\subsection{d=3}
In $d=3$ when $r_h<\frac{2}{\sqrt{3}} L$ the blackening factor \eqref{BlackeningFactor},
\begin{equation}
f(r) = \frac{r^2}{L^2} - 1-
\frac{r_h}{r}\left(\frac{r_h^2}{L^2} -1 \right)\, ,
\end{equation}
has three real roots which we denote by
\begin{equation}
r_1 = r_h, \qquad r_2 = \frac{1}{2} \left(\sqrt{4 L^2-3 r_h^2}-r_h\right),
\qquad
r_3 = - \frac{1}{2} \left(\sqrt{4 L^2-3 r_h^2}+r_h\right).
\end{equation}
For small hyperbolic black holes ($r_h<L$), $r_2$ becomes positive and there are two event horizons (see figure \ref{PenroseInOut}). The root $r_3$ is always negative and so does not indicate the presence of a third horizon. We study in this subsection both the case of $r_2<0$, $L<r_h<\frac{2}{\sqrt{3}} L$ , \ie large hyperbolic black holes and the case of $r_2>0$, $r_h<L$, \ie  small hyperbolic black holes.
Once again for $r_h=L$, we recover vacuum AdS while for $r_h = L/\sqrt{3}$, our two horizons become degenerate.
As before, we can assume without loss of generality that $r_h>r_2$.
The inverse blackening factor can be decomposed as follows:
\begin{equation}
\frac{1}{f(r)}= \frac{L^2}{(r_h - r_2)(r_h-r_3)(r_2 - r_3)}
\left[ \frac{ r_h (r_2-r_3)}{(r-r_h)}
-\frac{r_2 (r_h-r_3)}{ (r-r_2)}
+\frac{r_3 (r_h-r_2)}{  (r-r_3)}
\right]
\end{equation}
leading to the following tortoise coordinate:
\begin{equation}\label{rstar4hyper}
\begin{split}
r^{*}(r)= \frac{L^2}{(r_h-r_2)(r_h-r_3)(r_2-r_3)}
&\left[
r_3 r_2 \log\frac{|r-r_2|}{|r-r_3|}
 +r_h r_2  \log\frac{|r-r_h|}{|r-r_2|}
+r_3 r_h \log\frac{|r-r_3|}{|r-r_h|}
\right]
\end{split}
\end{equation}
which implies $v_{\infty} = 0$.

\subsubsection*{$r_h<L$}
\begin{figure}
\centering
\includegraphics[scale=0.7]{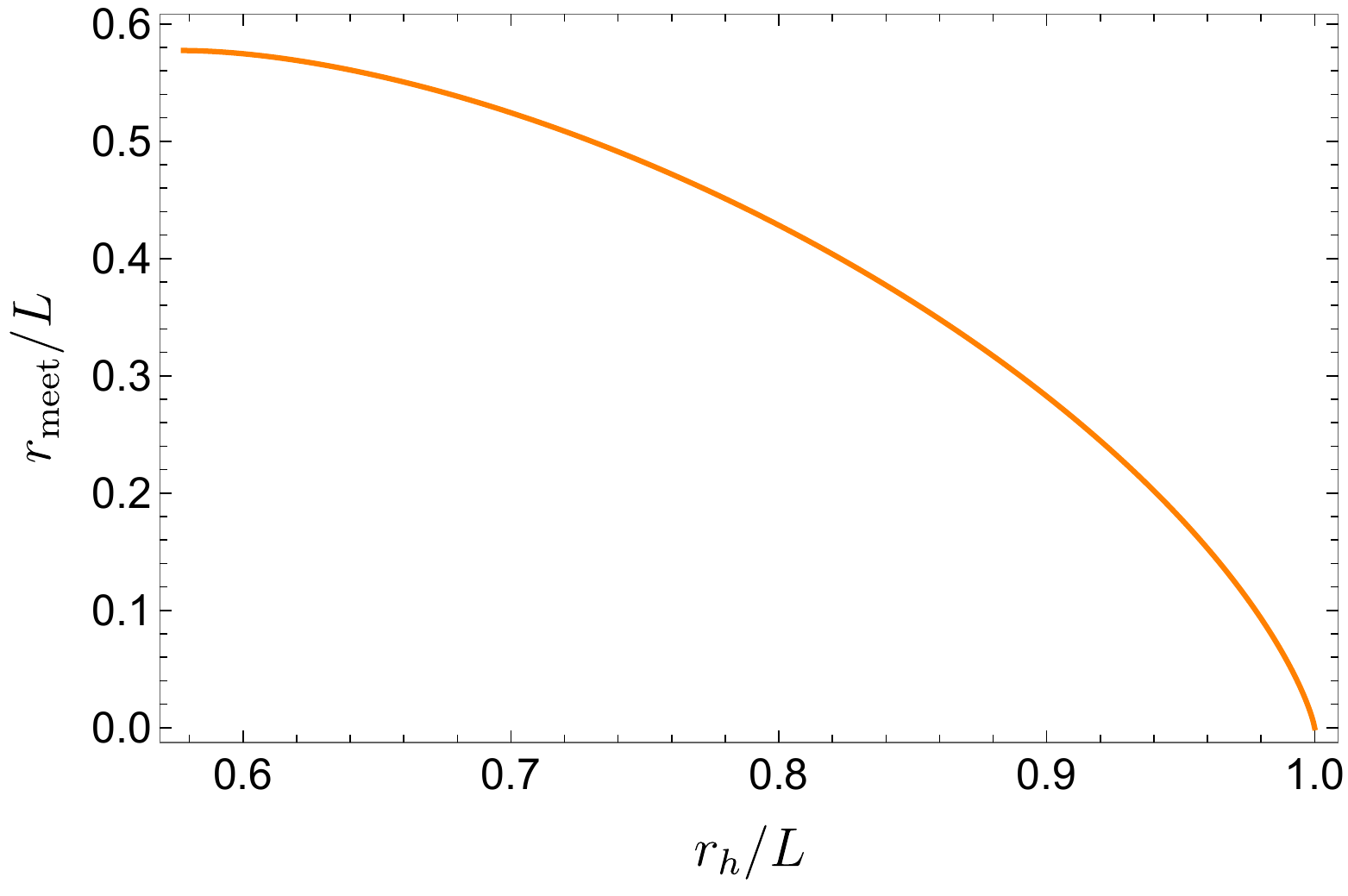}
\caption{Meeting point of the two ingoing null rays for hyperbolic black holes in AdS$_4$ as a function of the horizon radius. Orange plot (right) indicates the region in which $r_h>r_2$.}
\label{figureRays4}
\end{figure}

For this range of parameters, we have that $r_2>0$ and the black hole has two horizons. The absolute values inside the logarithms differentiates the regions inside, outside, and between the inner and outer horizons. Evaluating eq.~\eqref{eq:bulkappC}, we obtain:
\begin{equation}\label{bulkhyper4small}
\begin{split}
&\Delta 	I_{\bulk}=
-\frac{\Omega_{-1,2}  }{4 \pi  G_N (r_2-r_3) (r_h-r_2) (r_h-r_3)}
\left[ - 2 r_2 \left(r_{\text{meet}}^3-r_2^3\right) (r_h-r_3) \log \left(\frac{r_{\text{meet}}-r_2}{L}\right)
\right.
\\
&\left.
+2 r_3 (r_h-r_2) \left(r_{\text{meet}}^3-r_3^3\right) \log \left(\frac{r_{\text{meet}}-r_3}{L}\right)
- 2 r_h (r_2-r_3) \left(r_h^3-r_{\text{meet}}^3\right) \log \left(\frac{r_h-r_{\text{meet}}}{L}\right) \right]
\\
&
+\frac{\Omega_{-1,2} (2 r_2+2r_3+2r_h+r_{\text{meet}})  r_{\text{meet}} }{4 \pi  G_N }\, ,
\end{split}
\end{equation}
where $r_{\text{meet}}$ can again be computed numerically using eq.~\eqref{rstar4hyper} and the condition \eqref{eq:rmeethyper} which reads in this case
\begin{equation}
r^{*}(r_{\text{meet}})
= 0 \, .
\end{equation}
Since the null sheets meet between the two horizons one has to choose the appropriate branches of the logarithms in eq.~\eqref{rstar4hyper} when solving this equation.
We show the result for $r_{\text{meet}}$ in figure  \ref{figureRays4}.
The joint contribution \eqref{jointrmeet} gives:
\begin{equation}\label{cornerhypsmall4}
I_{\text{jnt}} = - \frac{\Omega_{-1,2}}{4 \,\pi \, G_N} \, r_{\text{meet}}^{2} \, \log{|f(r_{\text{meet}})|} \, .
\end{equation}
The total action is the sum of the bulk \eqref{bulkhyper4small} and joint  \eqref{cornerhypsmall4} terms. Substituting the numerical solution for $r_\text{meet}$ we obtain the result plotted in orange in figure  \ref{CompareAdS4}.

\subsubsection{$L<r_h <\frac{2 L}{\sqrt{3}}$}
The black holes for this region of parameter space only have one positive root, since $r_2<0$. Therefore, we should use the tortoise coordinate \eqref{rstar4hyper} to evaluate the contributions to the action \eqref{Cform1}. We obtain:
\begin{align}
\begin{split}
\Delta I_{\bulk} =&
\frac{\Omega_{-1,2}}{2 \pi  G_N (r_2-r_3) (r_h-r_2) (r_h-r_3)}
\left[- r_2^4 (r_h-r_3) \log \left(-\frac{r_2}{L}\right)
\right.
\\
&
\left.
+r_3^4 (r_h-r_2) \log \left(-\frac{r_3}{L}\right)+r_h^4 (r_2-r_3) \log \left(\frac{r_h}{L}\right)\right]
\end{split}
\end{align}
and
\begin{equation}
\begin{split}
I_{\mt{GHY}}^{\BHx} = -\frac{3 r_h \Omega_{-1,2}  \left(r_h^2-L^2\right) }{4 \pi  G_N (r_2-r_3) (r_h-r_2) (r_h-r_3)}
\left[ - r_2(r_h-r_3) \log \left(-\frac{r_2}{L}\right)
\right.
\\
\left.
+r_3 (r_h-r_2) \log \left(-\frac{r_3}{L}\right)+r_h (r_2-r_3) \log \left(\frac{r_h}{L}\right)\right]\, .
\end{split}
\end{equation}
Combining these results, we recover the last missing part of figure \ref{CompareAdS4}.

\subsection{Late-Time Growth of Complexity}

This subsection is a small aside in which we extend the results of \cite{LuisRob} for the rate of growth of complexity at late times to account for the possibility of small black holes with hyperbolic horizons. In this case the blackening factor \eqref{BlackeningFactor} reads:
\begin{equation}\label{gld}
f(r) = \frac{r^2}{L^2}-1 - \frac{r_h^{d-2}}{r^{d-2}}\left( \frac{r_h^2}{L^2}-1\right)\, ,
\end{equation}
and as mentioned before, it has two positive real roots representing two  horizons with radii which we labeled $r_+$ and $r_{-}$ respectively.\footnote{By design, $r_h$ in eq.~\reef{gld} is one of the horizon radii and we set $r_+=r_h$.}

The computation parallels strongly the computation for a charged black hole described in \cite{LuisRob} and we will not repeat the full details here. The contribution from the volume integral is most easily computed in the $(r,v)$ and $(r,u)$ coordinates and is given by:
\begin{equation}
\begin{split}\label{bcrate}
I_{\bulk} =& \frac{1}{16 \pi G_N}\int (R-2\Lambda) \sqrt{-g} d^{d+1} x = - \frac{d \,\Omega_{k,d-1}}{8 \pi G_N L^2} \int dv dr  r^{d-1}
\\
= &- \frac{\Omega_{k,d-1}}{8 \pi G_NL^2} \, \delta t\,    r^{d} \biggr{|}^{r_{+}}_{r_{-}}\, .
\end{split}
\end{equation}
The joint terms are given by equations (3.36) of \cite{LuisRob} without any modification:
\begin{equation}\label{jcrate}
I_{\jnt} = \frac{\Omega_{k,d-1}}{16 \pi G_N} \left(r^{d-1} \frac{df}{dr}\right)\biggr{|}^{r_{+}}_{r_{-}} \delta t\, .
\end{equation}
Plugging the blackening factor and summing together eq.~\eqref{bcrate} and eq.~\eqref{jcrate} leads to:
\begin{equation}
\frac{\delta \mathcal{C}}{\delta t} = \frac{1}{\pi} \left(\frac{\delta I}{\delta t} \right)= 0
\end{equation}
which implies that the growth rate vanishes for $r_h<L$. We would like to point out that this does not provide evidence for a discontinuity in the complexity growth rate as $r_h$ approaches $L$ since in this limit the black hole mass vanishes --- see eq.~\eqref{eq:BHMass}.

\section{Ambiguities in the Action Calculations} \label{AppendixC}

\begin{figure}
\centering
\includegraphics[scale=0.19]{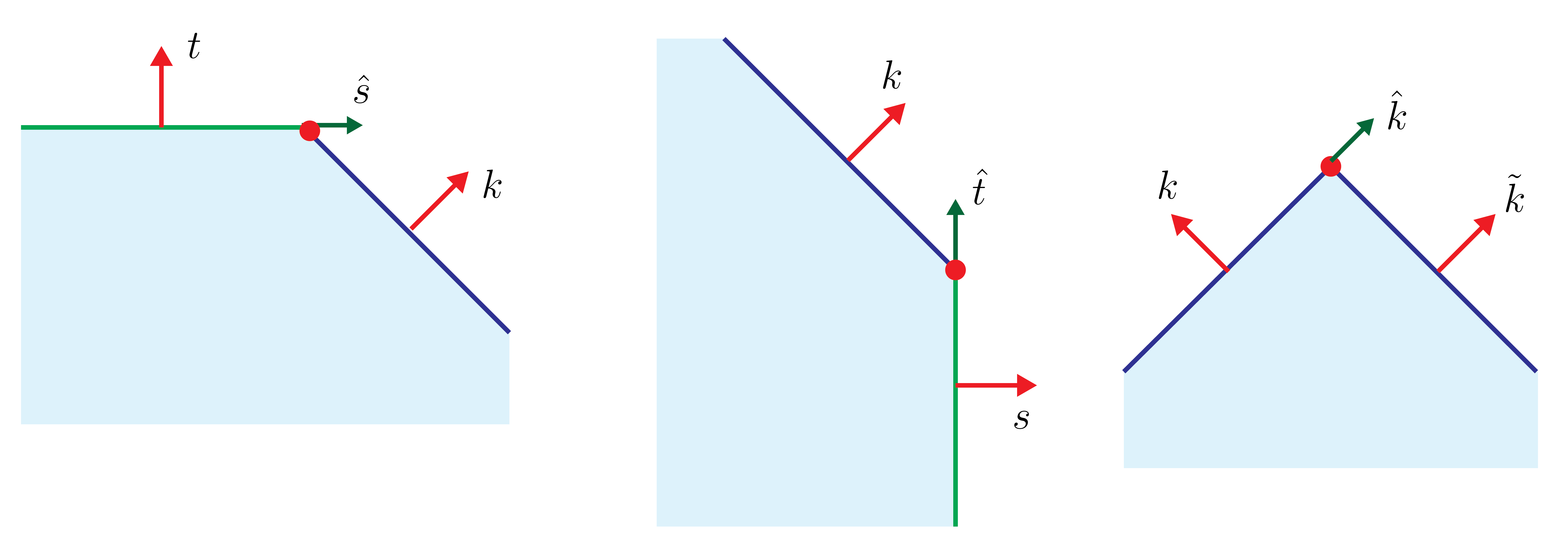}
\caption{Various possible null junctions appearing in our action calculations. We show $k$, $t$ and $s$  as outward-directed one-forms, following the convention of \cite{Pratik}.}
\label{cornerdrawing}
\end{figure}

In this appendix, we recall that, as discussed in \cite{LuisRob}, the boundary terms on the null boundary surfaces \reef{ActSurf2} and null joints \reef{ActJ2} may introduce some ambiguities in the gravitational action. By construction, the variation of these boundary terms is well-defined and cancels the corresponding total derivative terms coming from the variation of the bulk action. However, evaluating the gravitational action for a particular spacetime geometry will generally yield different numerical values depending on different choices that can be made in constructing the boundary terms.   In particular, $\kappa$ in eq.~\reef{ActSurf2} depends on an arbitrary choice
of parameterization for the null generators.  Further, in eq.~\reef{ActJ2}, $a$
depends on the arbitrary normalization of the null tangent $k^\alpha$ and in principle, we could add an additional function $a_0$ to $a$ in eq.~ \eqref{ActJ2}, which remains fixed when the action is varied. For convenience, we reiterate the expressions used to evaluate $a$:\footnote{Here we use the conventions introduced in \cite{Pratik}.}
\begin{equation}\label{ball}
a=
\begin{cases}
 \epsilon \log{ |k \cdot t| } \qquad\text{for  spacelike-null joint with }\epsilon = -\mbox{sign}(k \cdot t)\, \mbox{sign}(k \cdot \hat s)\,,
 \\
\epsilon \log{| k \cdot s |}\qquad \text{for  timelike-null joint with }\epsilon = -\mbox{sign}(k \cdot s) \, \mbox{sign}(k \cdot \hat t)\,,
\\
 \epsilon \log{| k \cdot \tilde k/2 |} \qquad \text{for null-null joint with }\epsilon = -\mbox{sign}(k \cdot \tilde k)\mbox{sign}(\hat k \cdot \tilde k)
 \,.
\end{cases}
\end{equation}
In the equation above, $\hat n$ and $\hat s$ ($\hat k$) are unit \emph{vectors} (null \emph{vector}) that are in the tangent space of the appropriate
boundary region, orthogonal to the junction and pointing outward from the boundary
region --- see figure \ref{cornerdrawing}.
In this appendix, we will examine the influence of all these  ambiguities on our results and show that except for the small hyperbolic black holes, our results are not effected by the different possible choices.

\subsection{Redefinition of the Function Defining the Null Hypersurface}\label{app:a0amb}
It was argued in \cite{LuisRob} that it is possible to introduce an ambiguity in the joint terms without redefining the null normal  $k_{\alpha}$. This is done by modifying the function $\Phi$ that describes the hypersurface (\ie with $\Phi(x)=0$).
In general $k_{\alpha}= \mu \, \partial_{\alpha} \Phi$ and therefore $a$ in the joint action depends on both $\mu$ and $\Phi$.
We can however redefine $\Phi \rightarrow \bar \Phi (\Phi)$ (where also $\bar \Phi$ is required to vanish on the hypersurface) and choose $\bar \mu \equiv \mu\, d \Phi/d \bar \Phi$ in such a way that our normal vector is left unchanged:
\begin{equation}
k_{\alpha}= \mu \, \partial_{\alpha} \Phi = \, \bar \mu \partial_{\alpha} \bar \Phi \, .
\end{equation}
This implies that $a$ is modified as follows:
\begin{equation}
a_{\text{new}} = a + a_0 = a + \log{\bigg[ \frac{d \Phi}{d \bar \Phi} \bigg] }.
\end{equation}
In principle, there is no reason that $a_0$ should be the same on all the joints nor does it have to be a constant over a given joint.\footnote{In principle, $a_0$ could be any scalar function with vanishing variation. Note however, that to maintain the additive character of the action, there are some restrictions that need to be imposed on $a_0$ for different kinds of joints --- see section II.H of \cite{LuisRob}.} We will consider including a fixed constant $a_0$ for all joints as a simple test case. We will check how this addition influences our results.

We start from the joint contributions at infinity, eq.~\eqref{CornerCut} will be modified by:
\begin{equation}
\Delta I_{\jnt,\text{cut}} =
I_{\jnt,\text{cut}} =  a_0 \, \frac{\Omega_{k, d-1}}{2 \pi G_N}\,  r_{\mt{max}}^{d-1}\, .
\end{equation}
As discussed earlier (see eq.~\eqref{differ}), the difference in the cutoffs between vacuum AdS and the black hole background is of order $\delta^{d-1}$. The leading order divergence near infinity is $r_{\mt{max}} = L^2/\delta +\cdots$, which implies that the subtraction of this term between the black hole and vacuum AdS spacetimes will result in an order $\delta$ contribution. Of course, we are assuming here that the same $a_0$ appears for both spacetimes.

Moving to the joint terms near the singularity we have that these are modified by
\begin{equation}
\Delta I_{\jnt,\text{sing}} =
I_{\jnt,\text{cut}} =  a_0 \, \frac{\Omega_{k, d-1}}{2 \pi G_N}\,  \epsilon_0^{d-1}\, ,
\end{equation}
which vanishes in the limit $\epsilon_0 \rightarrow 0$.
This implies already that most of the results presented in this paper are left unchanged under  such a modification of the joint terms.

\begin{figure}
\centering
\includegraphics[scale=0.6]{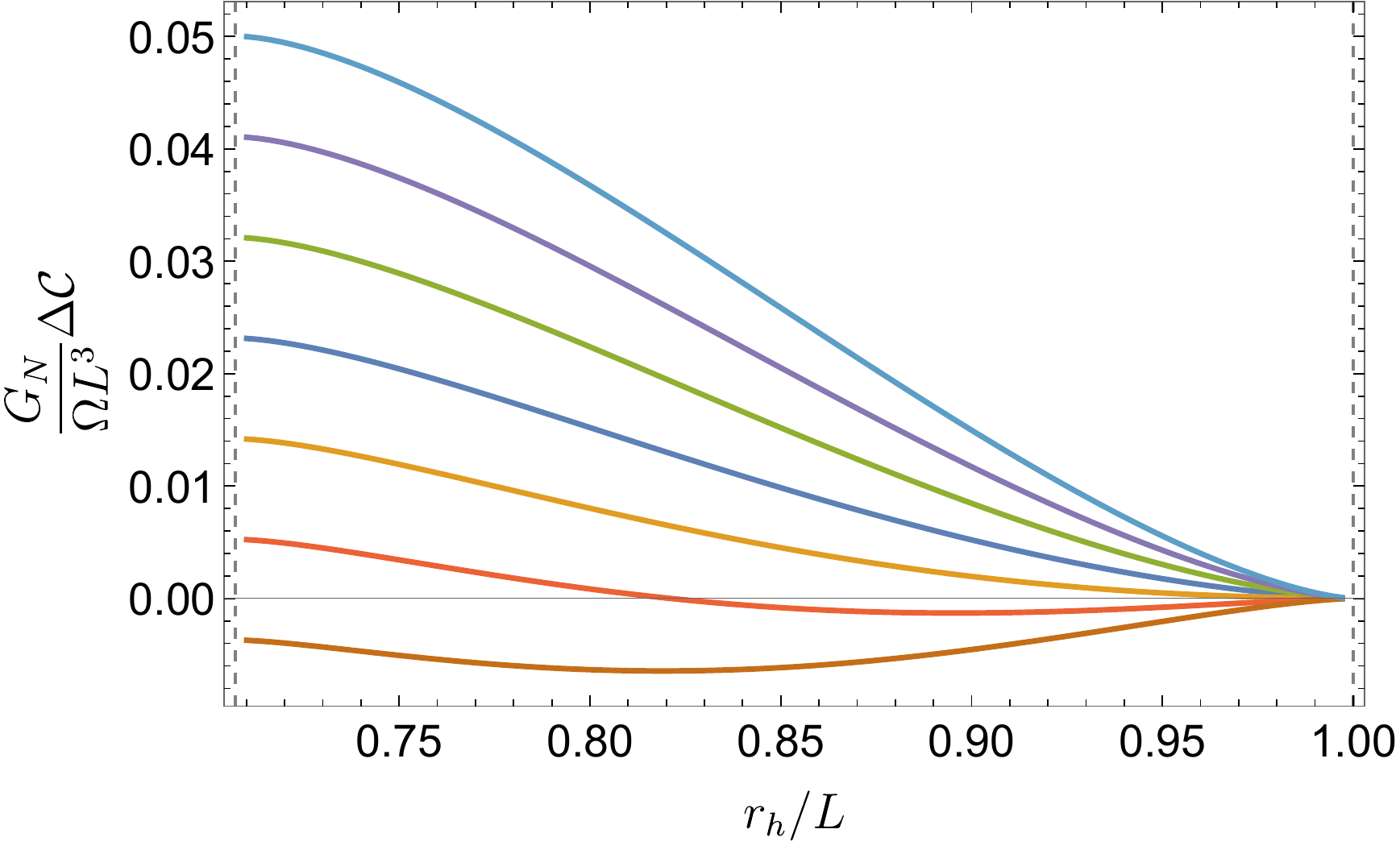}
\caption{Complexity of formation for the small hyperbolic black holes in $d=4$ for different choices of $a_0$ ranging from $a_0=-3$ (lowest line) to $a_0=3$ (highest line) in jumps of 1. We can observe that for certain values of $a_0$ the complexity becomes negative.}
\label{a0amb}
\end{figure}

However, the small hyperbolic black holes are once again an exception. In this case we have also the joint terms at $r_{\text{meet}}$ --- see eq.~\eqref{jointrmeet}. These will be modified by
\begin{equation}
\Delta I_{\jnt,\text{meet}} = a_0 \frac{\Omega_{-1,d-1}}{4\pi G_N} r_{\text{meet}}^{d-1}\, .
\end{equation}
There is no equivalent contribution in vacuum AdS and so we will be left with a finite modification of the complexity of formation. The result for the small hyperbolic black holes in $d=4$ with $a_0=\{-3,-2,-1,0,1,2,3\}$ is shown in figure \ref{a0amb}.

\subsection{Reparameterizations}
In this subsection we demonstrate that our complexity of formation is invariant under a certain class of reparameterizations of the null generators $\lambda \rightarrow \bar\lambda(\lambda,\theta^A)$.
The behavior of the various geometric quantities under reparameterizations was already studied in section II.C.4 of \cite{LuisRob} and we briefly review it here. We use $\lambda$ to denote the parameter along null generators and $\bar \lambda$ for the one associated to the new parameterization. We also define $e^{-\beta}=\del \bar\lambda/\del \lambda$. The normal vectors can be defined using the surface embedding functional $x^\alpha=x^\alpha(\lambda,\theta^A)$ where $\theta^A$ are the other intrinsic coordinates:
\begin{equation}
k^\alpha = \frac{\del x^\alpha}{\del \lambda}.
\end{equation}
Under a reparameterization, we have:
\begin{equation}\label{reparam}
\bar k^{\alpha} = e^\beta k^\alpha,
\qquad
\bar \kappa = e^\beta (\kappa+\del_\lambda\beta).
\end{equation}
Let us look at the upper right normal in figure \ref{PenroseBHa}. In this case, the parametric representation of the null sheet reads $(t,r,\theta^A) =  (v_\infty - r^{*}(r), r, \theta^A)$ where $\lambda =  -r$ is the affine parameter along the null generators (oriented towards the future) and the null normal is given by the expression in eq.~\eqref{niceNull}. We choose to look at a reparameterization with $e^\beta = g(r)$ for the upper right null sheet in figure \ref{PenroseBHa}. Applying the relations \eqref{reparam}, we obtain:
\begin{equation}\label{reparam2}
\bar k_{\mu}dx^\mu = g(r)\, k_{\mu}dx^\mu=g(r) \, \left(dt+  \frac{dr}{f(r)}\right) \, ,\qquad \bar \kappa = - \frac{d \, g(r)}{d \, r}\,.
\end{equation}
Recall that we require our normal forms to be pointing outwards and so we will assume $g(r)>0$ to maintain this condition.
The other relevant normal forms are given by eqs.~\eqref{VecBH} and \eqref{VecEmpX}.
The change in the action for the upper right null hypersurface of figure \ref{PenroseBHa} is then:
\begin{align}
\Delta I_\mt{null surface} =- \frac{\Omega_{k,d-1}}{8\pi G_N}  \int \, r^{d-1} \, \bar \kappa \, d \bar \lambda  =  \frac{\Omega_{k, d-1}}{8 \pi G_N}\, \int_{0}^{r_{\mt{max} }} \, r^{d-1} \, \partial_{r} \log{g(r)} \, ,
\end{align}
where the integral in the first equality is taken with integration limits from past ($r=r_{\mt{max}}$) to future ($r=0$).
The difference in the action due to joint terms
is:
\begin{equation}\label{CornerNulls}
\Delta I_{\jnt}=-\frac{\Omega_{k, d-1}}{8 \pi G_N}  \,  r^{d-1} \log{g(r)} \bigg|^{r_{\mt{max}}}_{0} \, .
\end{equation}
Summing the two contributions together, multiplying by four (assuming that we rescaled all our null normals with the same function $g(r)$) and using integration by parts we obtain:
\begin{equation}\label{ActionDiffKappa}
\Delta I = - \frac{(d-1) \, \Omega_{k, d-1}}{2 \pi G_N} \, \int_{0}^{r_{\mt{max}}} r^{d-2} \, \log{g(r)} \, dr\, .
\end{equation}
We recall that from this difference we are to subtract that of vacuum AdS, and so it will be enough for our purposes to demonstrate that the difference is independent of $r_h$ in the limit $\delta \rightarrow 0$.

Let us now focus on the example of $\bar \kappa$ constant. This is achieved by choosing:
\begin{equation}
 g(r) = 1 + \bar \kappa (r_{\mt{max}}-r) \, .
 \end{equation}
where we chose an additive constant  to guarantee that the normalization condition $\bar k \cdot \hat t = \pm 1$ at the cutoff surface is maintained.

If we are to compare the different spacetimes, the nontrivial difference is in the $r_{\mt{max}}$ dependence on the horizon radius. Since $g(r)$ is a known function, we can integrate eq.~(\ref{ActionDiffKappa}),
\begin{align}
&\bar I_{\bar \kappa} - I_\kappa =- \frac{\Omega_{k,d-1} r^{d-1}}{2\pi G_N \, d(d +1)( \bar \kappa r_{\mt{max}}+1)^{2}} \bigg[ d \, r^2 \bar \kappa^2 \, _2F_1\left(1,d+1;d+2;\frac{\bar \kappa  r}{\bar \kappa  r_{\mt{max}} +1}\right) + \nonumber \\
&(d+1)(\bar \kappa r_{\mt{max}}+1) (\bar \kappa r + d (\bar \kappa r_{\mt{max}}+1) \log{(1 + \bar \kappa (r_{\mt{max}}-r))} \bigg] \bigg|_{0}^{r_{\mt{max}}} \, .
\end{align}
This integral has a simple analytic expression for several dimensions.
Since we know from eq.~\eqref{differ} that the  difference in $r_{\mt{max}}$ between the black hole and vacuum AdS is of order $\delta^{d-1}$, the subtraction between the two spacetimes is of order $\delta \, \log \delta$,
\begin{equation}
\Delta I_{\mt{BH}} - \Delta I_{\mt{vac}} = \mathcal{O}(\delta \, \log \delta) \, .
\end{equation}
Therefore, if the surface gravity changes to a constant value, the difference in the action between the two spacetimes can still be evaluated and the result in independent of $\bar \kappa$ in the limit $\delta \to 0$.\footnote{For small hyperbolic black holes, however, we have again an ambiguous result.}

\subsection{Changing the Normalization Condition at the Boundary}
The last source of ambiguity which we chose to explore is the possibility to normalize differently the null normals at the cutoff surface.
Throughout the paper we have normalized the null normal, \eg in eq.~\eqref{niceNull}, at the asymptotic AdS boundary such that ${\bf k} \cdot {\bf{\hat t}} = 1$ where ${\bf \hat t}=\del_t$. However, as suggested in \cite{LuisRob},  ${\bf k} \cdot {\bf{\hat t}} = c$ with $c$ a positive constant would be an equally natural choice. Let us explore the consequences of choosing such a constant $c>0$. This will lead to a rescaling of the null normal similar to the one in the previous subsection eq.~\eqref{reparam2}. However, since in this case $g(r)$ is a constant, the surface gravity remains zero and the only contribution is of the form \eqref{CornerNulls} with $g(r)=c$:
\begin{equation}\label{CornerNulls2}
\Delta I_{\jnt}=-\frac{\Omega_{k, d-1}}{8 \pi G_N}  \,  r_{\mt{max}}^{d-1} \log{c} \, .
\end{equation}
We can now use again the fact that from eq.~\eqref{differ} the  difference in $r_{\mt{max}}$ between the black hole and vacuum AdS is of order $\delta^{d-1}$. The expression \eqref{CornerNulls2} is therefore of order $\delta$ and vanishes when subtracting the two backgrounds. The small hyperbolic black holes are again different and will suffer ambiguities similar to those already discussed in subsection \ref{app:a0amb}.

\subsection{A Comment on the Cutoff Choice}\label{app:cutoff choice}

\begin{figure}
\centering
\includegraphics[scale=0.3]{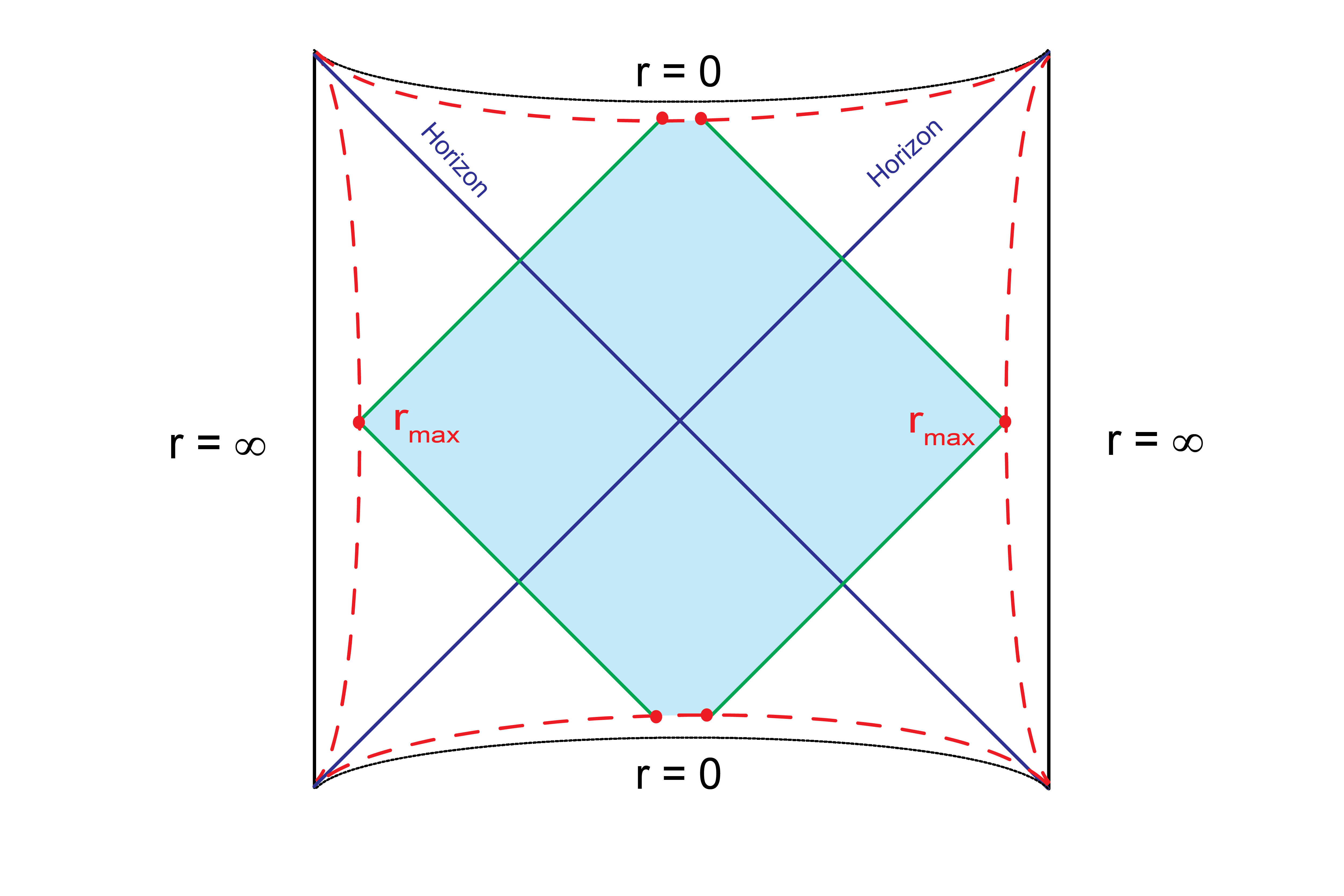}
\caption{Penrose diagram of a black hole in $d>2$ representing a different way of regularizing the WDW patch. In this case we shoot the null rays from $r=r_{\mt{max}}$ and not from the boundary as in figure \ref{PenroseBHa}.}
\label{PenroseBHadc}
\end{figure}

We close this appendix with another  nontrivial test of our results which is to check that they would not change for a certain modified regularization scheme. Suppose that instead of regularizing our action by a cutoff surface at $r=r_{\mt{max}}$  as we did in the text, we would bound our region by null rays which are emitted at $t=0$ at $r=r_{\mt{max}}$, as depicted in figure \ref{PenroseBHadc}. In this appendix we demonstrate that our results do not change for such a choice.
 To be more precise, we show that the change introduced in the black hole action under our modified regularization scheme does not depend on $r_h$ when taking the limit $\delta \rightarrow 0$ and so will cancel against that of vacuum AdS.

Let us start with the bulk contribution to the action. The sole modification to the integral \eqref{ActionBulk} is that $v_\infty$ would be replaced by:
\begin{equation}
\tilde v_\infty = r^{*}(r_{\mt{max}})
\end{equation}
which results in the following change of the bulk action:
\begin{equation}\label{RegBulk}
\, \Delta I_{\bulk} = - \frac{d\Omega_{k,d-1}}{2\pi G_N L^2}  \int_0^{r_{\mt{max}}} (\tilde v_{\infty}- v_\infty) \, r^{d-1} \, dr =
- \frac{\Omega_{k,d-1} r_{\mt{max}}^d}{2\pi G_N L^2}   (\tilde v_{\infty}- v_\infty) \,  .
\end{equation}
Using the definition \reef{tortoise} of the tortoise coordinate:
\begin{equation}\label{delv}
\tilde v_{\infty}- v_\infty = - \int_{r_{\mt{max}}}^{\infty} \frac{dr}{f(r)}
= - \frac{L^2}{r_{\mt{max}}} +\cdots.
\end{equation}
This means that the bulk integral has a leading contribution proportional to $r_{\mt{max}}^{d-1}$ when $r_{\mt{max}}$ is large. Using again eq.~\eqref{differ} for the difference between the cutoffs of the black hole and vacuum AdS, we see that the $r_h$ dependence of this expression is of order $\delta$ and vanishes in the limit  $\delta \rightarrow 0$.

A similar argument holds for the surface contribution near the singularity \eqref{extrinsic1a} which is modified by:
\begin{align}\label{RegSurf}
\, \Delta I_{\mt{GHY}} = \frac{d\,\Omega_{k,d-1}\,\omega^{d-2}}{4 \pi G_N} \left(\tilde v_{\infty}- v_{\infty}\right) .
\end{align}
From eq.~\eqref{delv} we see that the difference $\left(\tilde v_{\infty}- v_{\infty}\right)$ has a leading $1/r_{\mt{max}}$  contribution and goes to zero when we take the $\delta \rightarrow 0$ limit.

One last contribution which has to be considered is that of the right joint near the boundary (between the two null sheets):
\begin{equation}
\begin{split}
I_{\text{jnt}}&=\frac{r_{\mt{max}}^{d-1}\Omega_{k,d-1}}{8\pi G_N}\log f(r_{\mt{max}})\, .
\end{split}
\end{equation}
But notice that this (after a factor of 2 accounting for the two sides of the black hole) is exactly the same as in eq.~\eqref{CornerCut} and so $\Delta I_{\text{jnt}}=0$.
This completes our argument that this different regularization scheme would give the same result for the complexity of formation.

\section{Insights from MERA} \label{lastA}

We found that for high temperatures, the leading contribution to $\Delta\mC$  is proportional to the entropy. This result arose for the complexity=action duality in eq.~\reef{PlanarGenerald}, but also using the complexity=volume approach in eq.~\reef{VolSd}, with a different pre-factor. Hence our holographic calculations indicated that at least for high temperatures, the additional complexity required in preparing the entangled thermofield double state (compared to preparing each of the CFTs in its vacuum state) is proportional to the entanglement entropy between the two boundary CFTs in this state --- or the thermal entropy of the corresponding mixed state (produced by tracing over one of the CFTs).  Further, in the following it will be useful to recall that in a CFT, the leading contribution to the entropy takes the form 
\beq
S=C_\mt{S}\,\mV\,T^{d-1}\,,
\label{entro}
\eeq
where $\mV$ is the spatial volume and $C_\mt{S}$ is a parameter characterizing the number of degrees of freedom in the CFT, \eg $C_\mt{S}\sim C_\mt{T}$ in a holographic CFT dual to Einstein gravity, \eg see \cite{fthem}. Of course, in the above expression, we have neglected the possibility of curvature corrections to this leading result.
\begin{figure}
        \centering
       \begin{subfigure}[(a)]{.4\textwidth}
       \includegraphics[scale=0.6]{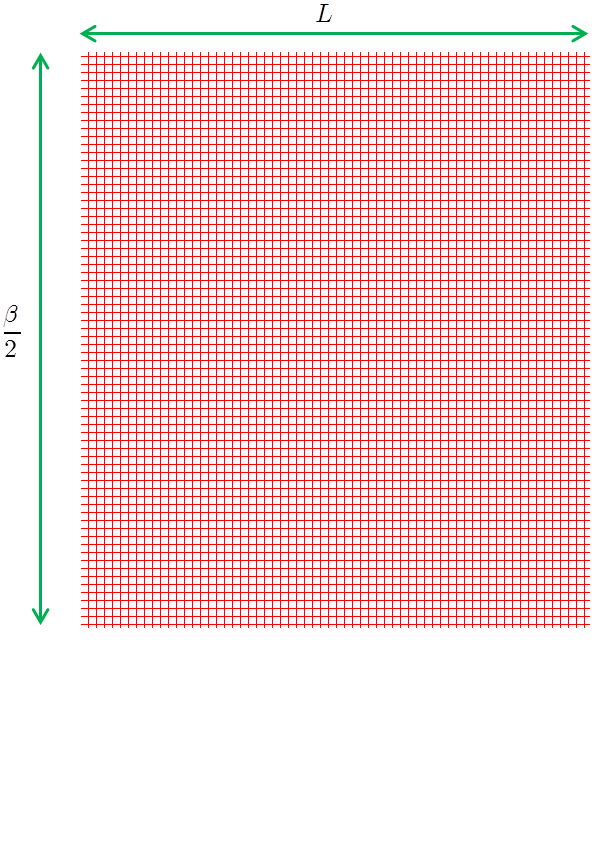}
                \caption{Thermofield double state}\label{gooa}
        \end{subfigure}
                \hskip2cm
        \begin{subfigure}[(b)]{.4\textwidth}
       \includegraphics[scale=0.6]{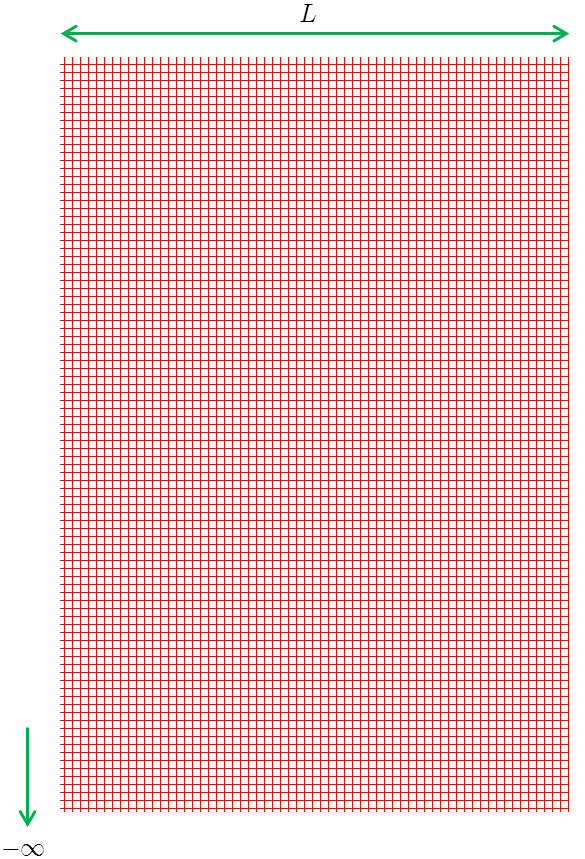}
                \caption{Vacuum state}\label{goob}
        \end{subfigure}
        \caption{Tensor network representations of the Euclidean path integral giving (a) the thermofield double state and (b) the vacuum state. In panel (a), both the top and bottom of the tensor network correspond to open indices, while in panel (b), only the top of the tensor network has open indices.} \label{goo}
\end{figure}
\begin{figure}
        \centering
        \begin{subfigure}[(a)]{.34\textwidth}
       \includegraphics[scale=0.6]{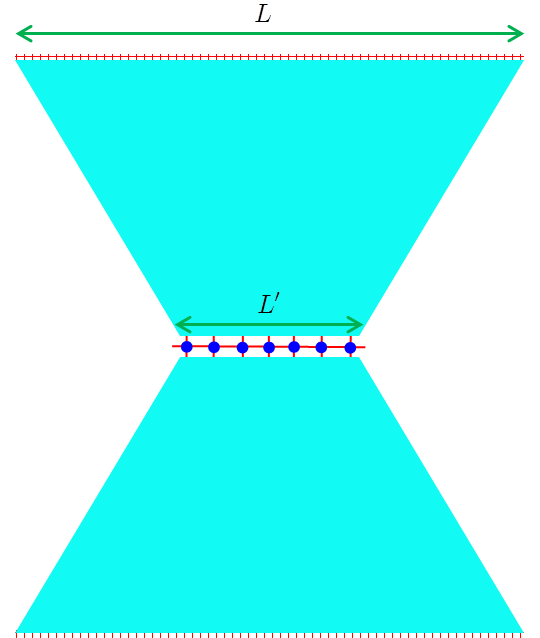}
                \caption{Thermofield double state}\label{googooa}
        \end{subfigure}
                \hskip2cm
         \begin{subfigure}[(b)]{.34\textwidth}
       \includegraphics[scale=0.6]{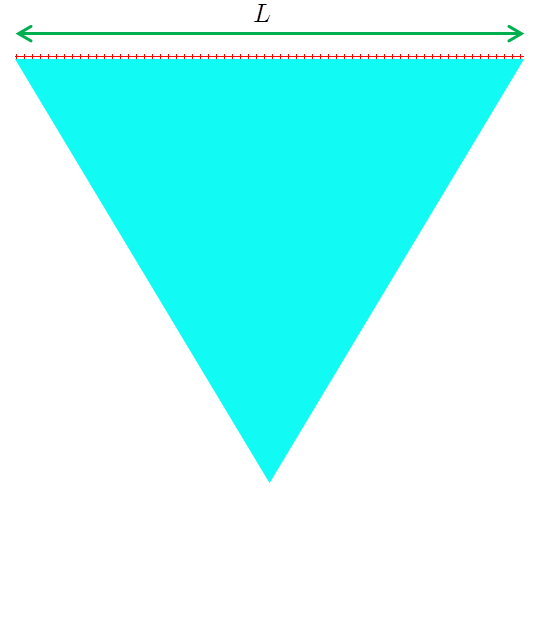}
                \caption{Vacuum state}\label{googoob}
        \end{subfigure}
        \caption{MERA network representations of (a) the thermofield double state and (b) the vacuum state. The aqua regions are composed of disentanglers and isometries. In panel (a), there remains a single layer of fixed point tensors at the center of the tensor network.} \label{googoo}
\end{figure}

In this appendix, we would like get some insight into this result from the description of analogous CFT states with MERA tensor networks \cite{mera1,mera2,mera3}.\footnote{We thank Guifre Vidal for discussions on the material presented in this appendix. The interested reader may also look at appendix E of \cite{prep9}, which suggests some ways of bounding complexity in thermal states. While the discussion there considers systems defined by an ensemble of Hamiltonians, it may still be interesting to gain further intuition for the complexity of formation.} Of course, the key underlying assumption here will be that the MERA network is comparable to the optimal circuit of universal gates for preparing the desired states and so the complexity is proportional to the number of gates in the MERA circuit. 
Let us first begin with a tensor network representation of the Euclidean path integral giving the thermofield double (TFD) state and the vacuum state \cite{tnr2,tnr1}, as shown in figure \ref{goo}. While the details will not be important here, these networks are constructed with a Suzuki-Trotter decomposition to approximate the Euclidean time evolution operator $e^{-\beta H}$ and here we will also assume that tensor network renormalization (TNR) has been applied to identify the fixed point tensors, which implicitly appear at each of the vertices in figure \ref{goo} --- see further discussion below.  Let us note that in the TFD network, there are open indices both at the top and bottom of the tensor network, while the network representing the vacuum only has open indices on the top. For simplicity, we assume periodic boundary conditions and so indices on the left and right sides of both networks are contracted with each other. Finally, although our figures are two-dimensional, we can think of them as illustrating the network for a $d$-dimensional CFT if each horizontal layer is an $L^{d-1}$ array of tensors and so the tensor network corresponds to a $d$-dimensional Euclidean path integral.

Next, we apply TNR as in \cite{tnr2} to produce the corresponding MERA networks, as shown in figure \ref{googoo}. Again, the TFD network has open indices on the top and bottom while the vacuum circuit only has open indices on the top. In the figure, the angled (aqua) regions represent networks constructed from the usual disentanglers and isometries found in the MERA network \cite{mera1,mera2,mera3}. Hence for the vacuum state in panel (b) of figure \ref{googoo}, the entire tensor network in figure \ref{goob} has been replaced by these new tensors and we can think of the resulting network as a quantum circuit of unitaries which maps a trivial state to the CFT vacuum state \cite{mera1,mera2,mera3}. Now in contrast to the vacuum MERA, not all of the fixed point tensors are eliminated in the TFD state in figure \ref{googooa}. In particular, there is a  single layer of the fixed point tensors at the center of the network which bridges between two MERA circuits above and below. We emphasize that this bridge layer is constructed from the same tensors used to construct the original tensor network in figure \ref{goob} and we recall that the latter are not unitary gates.

\begin{figure}
\centering
\includegraphics[scale=0.65]{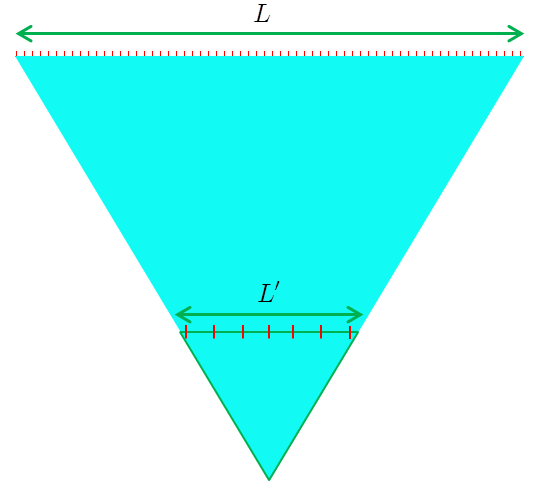} 
  \caption{MERA network representing the vacuum state. The IR portion of the MERA denoted by the green triangle gives a coarse-grained version of the vacuum state. The coarse graining is suggested by the sparsity of vertical red bonds running through the horizontal cut of length $L'$.} \label{jointX}
\end{figure}
Now towards evaluating the complexity of formation, we realize in comparing the MERA circuits for the TFD state and for (two copies of) the vacuum state that the UV portions of the circuits are identical. That is, the circuit represented by the trapezoid in the top half of figure \ref{googooa} is identical to that represented by the trapezoid in the upper part of the vacuum MERA shown in figure \ref{jointX}. In the latter figure, the horizontal line denoted by the length $L'$ is the same size as the central layer in the TFD circuit. That is, there are just as many (vertical) bonds in this horizontal cut through the vacuum circuit as run between the bridge layer and either of the MERA portions of the TFD circuit. We can think of the IR portion of the MERA denoted by the bottom (green) triangle in figure \ref{jointX} as giving a coarse-grained version of the vacuum state. 

Now again, our key assumption is that the MERA circuits give some approximation of the optimal preparation of the desired states and so the complexity is proportional to the number of gates in the MERA circuit. It follows that the complexity of the vacuum state can be separated as 
\beq
\mC(vac) = \mC_\mt{UV} + \mC_\mt{c-g}(vac)
\label{cold}
\eeq
where $\mC_\mt{UV}$ and $\mC_\mt{c-g}$ denote the complexities of the UV (top) and IR (bottom) parts of the circuit in figure \ref{jointX} --- the subscript `c-g' denotes coarse-grained. For the TFD state, we make the further assumption that there is a similar addition of complexities, \ie
\beq
\mC({\rm TFD}) = 2\,\mC_\mt{UV} + \mC_\mt{c-g}({\rm TFD})
\label{hot}
\eeq
where $\mC_\mt{UV}$ is precisely the same quantity as appears in eq.~\reef{cold} and $\mC_\mt{c-g}({\rm TFD})$ is the complexity associated with the bridge layer of the TFD circuit --- which we will argue in a moment is simply a coarse-grained version of the TFD state. Hence with these assumptions, the complexity of formation would be given by
\beq
\Delta\mC = \mC({\rm TFD}) - 2\,\mC(vac)=\mC_\mt{c-g}({\rm TFD})-2\,\mC_\mt{c-g}(vac)\,.
\label{form}\eeq
The important point being, of course, that the complexities associated with the UV portions of the circuits have precisely canceled in this difference.
\begin{figure}
        \centering
           \begin{subfigure}[(a)]{.85\textwidth}
       \includegraphics[scale=0.55]{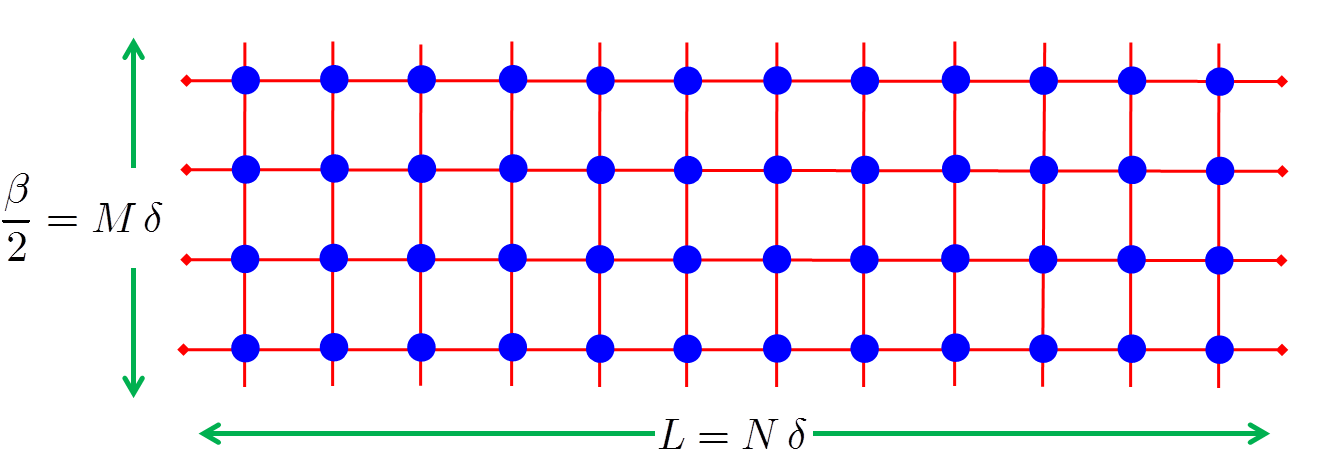}
                \caption{Thermofield double state}\label{gogoa}
        \end{subfigure}
             \vskip0.5cm
        \begin{subfigure}[(b)]{0.35\textwidth}
       \includegraphics[scale=0.55]{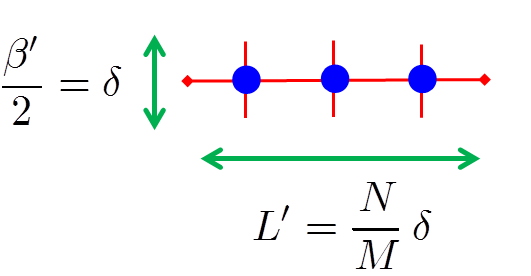}
                \caption{`Coarse-grained' TFD}\label{gogob}
        \end{subfigure}
        \caption{Two different versions of the thermofield double state, shown in figure \ref{gooa}, with different coarse-graining. The blue dots represent the fixed point tensors for the CFT.} \label{gogo}
\end{figure}

A natural question is: what is the (linear) size of this middle layer in the TFD circuit?
Of course, by construction, this is also the size of the coarse-grained vacuum circuit in figure \ref{jointX}. To begin, we consider the dimensions of the original circuit in figure \ref{gooa} --- see figure \ref{gogo}. We assign the (linear) spatial dimension  $L=N\, \delta$ where $\delta$ is some lattice spacing between sites. This tensor network describes a path integral in a CFT and so this lattice spacing is arbitrarily chosen and then corresponds to our resolution of the wavefunction in the spatial direction. Again, for a $d$-dimensional CFT, each layer has the spatial volume $L^{d-1}$ and, \eg there are $N^{d-1}$ open indices at the top and bottom of the network. The size of the Euclidean time direction is $\beta/2=M\,\delta$. We are considering the high temperature regime and so we have $L\gg \beta$ or $N\gg M$. Now let us assume that each step of the TNR reduces each of the dimensions by a factor of 2. That is, acting on the networks in figures \ref{goo} or \ref{gogo}, we group the tensor network into blocks of $2^d$ tensors and applying one step of TNR, then reduces each of these blocks to a single tensor. With our assumption that we have already identified the fixed point tensors for the CFT, the tensors appearing in the network before and after this step are the same and this simply produces a coarse-graining of the original network.\footnote{For example, evaluating the partition function with either of the circuits in figure \ref{gogo} will give the same result -- up to small numerical errors from properly representing the fixed point tensors.} Now if we assume that $M=2^m$, then after $m$ steps the Euclidean time direction is reduced to one layer, \ie $\beta'/2=\delta$. Similarly, the spatial dimension is reduced by the same factor and so $L' = N'\,\delta$ with $N'=N/M=L/(\beta/2)$. That is, the number of tensors in the TFD bridge is $N'{}^{d-1}=(2 L/\beta)^{d-1}\simeq {\cal V}\, T^{d-1}$. In other words, the number of fixed point tensors in the bridge layer is proportional to the entropy \reef{entro} of the thermofield double state!

Now to evaluate the complexity of formation \reef{form}, let us begin with $\mC_\mt{c-g}(vac)$. By construction, the volume of the top layer of this coarse-grained vacuum MERA is also $L'{}^{d-1} = \tilde N\,\delta^{d-1}$ with $\tilde N=N'{}^{d-1}\simeq \mV\,T^{d-1}$. One of the remarkable features of MERA is that the total number of gates in the entire circuit is also proportional to $\tilde N$. Hence we expect that the complexity satisfies $\mC_\mt{c-g}(vac)\propto \mV\,T^{d-1}$. One might go further to argue for a factor of $C_\mt{T}$ in the limit of a large central charge. In this case, we expect that there are $C_\mt{T}$ degrees of freedom associated with each bond and so the complexity of the individual gates in the MERA circuit should be proportional to the central charge. Then we arrive at
\beq
\mC_\mt{c-g}(vac)= \tilde k_0(d)\,C_\mt{T}\, \mV\,T^{d-1} = k_0(d)\,S\,,
\label{vacC}
\eeq
where $S$ is the thermal entropy in eq.~\reef{entro} and $k_0(d)$ is some order one factor, which can depend on the spacetime dimension $d$.

\begin{figure}
\centering
\includegraphics[scale=0.6]{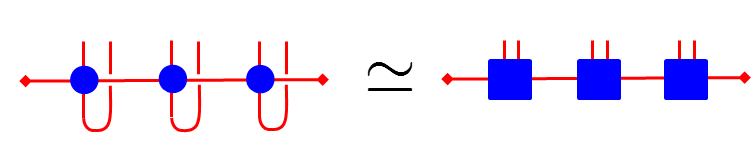} 
  \caption{The coarse-grained TFD state is similar to an MPS description of a gapped ground state, in that both exhibit only short-range correlations.} \label{joint}
\end{figure}
Now we would like to consider $\mC_\mt{c-g}({\rm TFD})$, the complexity of the coarse-grained TFD state. Above, we found that the total number of tensors in this network was $\tilde N=N'{}^{d-1}\simeq \mV\,T^{d-1}$ and we would like to argue that the complexity is proportional to this number. Recall that here the individual fixed point tensors are not unitaries. However, looking back to figure \ref{gogob}, we can think of this circuit as describing a high temperature TFD state with temperature $T\simeq 1/\delta$ and so it will only have short-range correlations in the spatial direction. In this regard, this tensor network is like a matrix product state \cite{orus} describing the ground state of a massive or gapped theory --- see figure \ref{joint}. Therefore it is natural to think that only a circuit with a fixed finite depth (independent of $L'$) is needed to prepare this state.\footnote{In higher dimensions ($d>2$), gapped states may be topologically ordered and this statement would not apply for such states. However, the assumption here is that (the  purification of) the thermal state for the CFT only contains short-range correlations and is not topologically ordered.} Of course, this implies $\mC_\mt{c-g}({\rm TFD})\propto \mV\,T^{d-1}$ as desired. As above, in the limit of a large central charge, we can also argue for a factor of $C_\mt{T}$. That is, we expect that there are $C_\mt{T}$ degrees of freedom associated with each bond and so the complexity of the individual fixed point tensors should be proportional to the central charge. Then we conclude that
\beq
\mC_\mt{c-g}({\rm TFD})= \tilde k_T(d)\,C_\mt{T}\, \mV\,T^{d-1} = k_T(d)\,S\,,
\label{TC}
\eeq
where again $S$ is the thermal entropy in eq.~\reef{entro} and $k_T(d)$ is some new order one factor, which again can depend on the spacetime dimension $d$.

Substituting these results into eq.~\reef{form} now yields
\beq
\Delta\mC = \left(k_T(d)-2\,k_0(d)\right)\,S\,.
\label{form2}
\eeq
That is, the complexity of formation is proportional to the thermal entropy, as in our holographic result \reef{PGd}, and the pre-factor is simply given by the difference of the numerical factors appearing in eqs.~\reef{vacC} and \reef{TC}. Our holographic calculations produce a positive pre-factor for $d>2$, which then suggests that $k_T(d)>2\,k_0(d)$ at least for those holographic CFTs. 
Of course, in our holographic calculations, the coefficients $k_d$ and $\tk_d$ both vanished for $d=2$ and $\Delta\mC$ became a fixed constant. Hence in this special case, $k_T(d=2)=2\,k_0(d=2)$. Now it was found that for the special case of $d=2$, the central layer of fixed point tensors can be replaced by a network involving the standard isometries and disentanglers appearing in the UV portion of the MERA \cite{kinematic2}. However, we should note that this new circuit does not respect the `time flow' of the rest of the circuit, which is usually a part of the definition of complexity. Further it appears that a naive comparison yields $\Delta\mC$ which grows as the size of the horizon.\footnote{We would like to thank Bartek Czech and Jamie Sully for discussions on this point.} However, it still seems that this construction may be useful in understanding $k_T(d=2)=2\,k_0(d=2)$. Of course, the construction in \cite{kinematic2} relies on the {\it local} conformal invariance of the underlying $d=2$ CFT and so it will not apply in higher dimensions, which is consistent with the holographic result  $k_T(d)>2\,k_0(d)$ for $d>2$.

To close, a few more words on that central layer of the TFD state: As mentioned above, it can be regarded as the tensor network describing a high temperature state (with $T\simeq 1/\delta$) with only short-range spatial correlations. One suggestion \cite{swingle1,swingle2} was to replace the bridge layer with trivial identity tensors, raising the temperature to infinity and eliminating the spatial correlations altogether --- see also \cite{don2}. While this picture provides useful intuition, it seems to be an oversimplification when considering the complexity, \eg a consequence would be $\Delta\mC<0$. As emphasized above, this central layer is just a coarse-grained version of the original TFD state. It is constructed from the same fixed point tensors used to build the original tensor networks in figure \ref{goo} and contains much of the same information. Hence there are very specific correlations controlled by the conformal dimensions of the primary operators --- and up to an overall rescaling (and numerical errors), one reproduces the correlation functions in the original TFD state for which the spatial separation is chosen as some multiple of $\beta$. Hence the correlations established by the central bridge of the TFD state are short-range but they are definitely not trivial.

\end{document}